\documentclass[fleqn,usenatbib]{mnras}

\usepackage[T1]{fontenc}
\usepackage{ae,aecompl}

\usepackage{graphicx}	
\usepackage{amsmath}	
\usepackage{amssymb}	
 
\usepackage[dvipsnames]{xcolor}

\usepackage{xspace}
\usepackage{xfrac} 

\usepackage{subcaption}
\usepackage[percent]{overpic}

\newcommand{\dd}{\mathrm{d}}

\newcommand{\pow}[1]{\ifmmode{}^{#1}\else ${}^{#1}$\fi}
\newcommand{\HI}{{\text{H\MakeUppercase{\romannumeral 1}}}\xspace}

\newcommand{\Lya}{\ifmmode{\mathrm{Ly}\alpha}\else Ly$\alpha$\xspace\fi}

\def\gsim{\;\rlap{\lower 2.5pt
 \hbox{$\sim$}}\raise 1.5pt\hbox{$>$}\;}
\def\lsim{\;\rlap{\lower 2.5pt
   \hbox{$\sim$}}\raise 1.5pt\hbox{$<$}\;}

\newcommand{\cm}{\,\ifmmode{{\mathrm{cm}}}\else cm\fi}

\newcommand{\ergps}{\,{\rm erg}\,{\rm s}\ifmmode{}^{-1}\else${}^{-1}$\fi}
\newcommand{\Mpch}{\,{\rm Mpc}\,\ifmmode h^{-1}\else $h^{-1}$\fi}
\newcommand{\snru}{\,\ifmmode{\mathrm{Myr}^{-1}}\else Myr${}^{-1}$\fi}
\newcommand{\kms}{\,\ifmmode{\mathrm{km}\,\mathrm{s}^{-1}}\else km\,s${}^{-1}$\fi\xspace}

\newcommand{\cl}{\mathrm{cl}}
\newcommand{\mytime}[1]{\ifmmode{{t_{\mathrm{#1}}}}\else $t_{\mathrm{#1}}$\xspace\fi}
\newcommand{\tcc}{\mytime{cc}}
\newcommand{\tcool}[1]{\mytime{cool,#1}}
\newcommand{\tcoolrat}[1]{\tcool{#1}/\tcc}

\defcitealias{Gronke2018}{Paper I}

\title[Cold gas transport \& growth]{How cold gas continuously entrains mass and momentum from a hot wind}

\author[M. Gronke \& S. P. Oh]{
  Max Gronke\thanks{E-mail: maxbg@ucsb.edu, Hubble fellow}
  and S. Peng Oh
\\
Department of Physics, University of California, Santa Barbara, CA 93106, USA
}

\date{Draft from \today}

\pubyear{2019}

\begin{document}
\label{firstpage}
\pagerange{\pageref{firstpage}--\pageref{lastpage}}
\maketitle

\begin{abstract}
  
  The existence of fast moving, cold gas ubiquitously observed in galactic winds
  is theoretically puzzling, since the destruction time of cold gas is much
  smaller than its acceleration time. In previous work, we showed that cold
  gas can accelerate to wind speeds and grow in mass if the radiative cooling
  time of \textit{mixed} gas is shorter than the cloud destruction time. Here,
  we study this process in much more detail, and find remarkably robust cloud
  acceleration and growth in a wide variety of scenarios. Radiative cooling,
  rather than the Kelvin-Helmholtz instability, enables self-sustaining
  entrainment of hot gas onto the cloud via cooling-induced pressure gradients.
  Indeed, growth peaks when the cloud is almost co-moving. The entrainment
  velocity is of order the cold gas sound speed, and growth is accompanied by
  cloud pulsations. Growth is also robust to the background wind and initial
  cloud geometry. In an adiabatic Chevalier-Clegg type wind, for instance, the
  mass growth rate is constant. Although growth rates are similar with magnetic
  fields, cloud morphology changes dramatically, with low density, magnetically
  supported filaments which have a small mass fraction but dominate by volume.
  This could bias absorption line observations. Cloud growth from entraining and
  cooling hot gas can potentially account for the cold gas content of the CGM.
  It can also fuel star formation in the disk as cold gas recycled in a galactic
  fountain accretes and cools halo gas. We speculate that galaxy-scale
  simulations should converge in cold gas mass once cloud column densities of
  ${\rm N} \sim 10^{18} \ {\rm cm^{-2}}$ are resolved.
\end{abstract}


\section{Introduction}
\label{sec:intro}
Galactic winds are an essential building block of our theory of galaxy formation and evolution \citep[for recent reviews, see][]{Veilleux2005,Rupke2018,Zhang2018}. They expel gas out of the potential well, thus, regulating star-formation and affecting the future fate of the galaxy. This process leaves a clear imprint on multiple observables, from the galaxy luminosity function 
to the morphology of galaxies.
Another important effect of galactic winds is on the chemical and structural evolution of the surroundings of galaxies, and the IGM. Metals have been detected in low-column densities absorbers in the IGM \citep[e.g.,][]{1997ApJ...487..482H} where they have likely been deposited by galactic winds.

Observations show that galactic outflows are common throughout cosmic time. For instance,  `down the barrel' absorption lines are commonly blue-shifted with respect to the host-galaxy's systemic redshift, and emission lines also show effects indicative of outflowing gold gas \citep[e.g.,][]{Veilleux2005,Steidel2010a,2017MNRAS.469.4831C}.

Multi-wavelength observations of nearby galaxies allow us to study the structure of galactic winds. They show that galactic winds are multiphase, with cold $\sim 10^4\,$K gas co-spatial with hot $\gtrsim 10^6\,$K gas \citep{Heckman1990,Strickland2009,Rubin2014}, and moving outward at speeds comparable to virial and escape velocities.
In our own galaxy, a large population of fast moving, cold gas clouds -- dubbed `high velocity clouds' (HVCs) -- have been detected \citep[see review by][]{Wakker1997}. They seem to be distributed throughout the hot galactic halo \citep[e.g.,][]{Putman2002,DiTeodoro2018,2019ApJ...884...53F}.
At higher redshifts, absorption line studies paint a similar picture \citep[e.g.,][]{Rudie2019}. \\

Multi-phase gas is often poorly resolved in galactic scale simulations, and  modeled in a `sub-grid' fashion \citep[e.g.][]{Springel2002}.
One key question which requires high-resolution idealized simulations is whether hydrodynamic ram pressure from a hot wind can accelerate dense clouds to the observed velocities \citep[e.g.][]{Klein1994,Mellema2002,Pittard2005,Vieser2007,Cooper2009,Scannapieco2015a,Bruggen2016,Schneider2016,Gronnow2018}. This has proven to be remarkably challenging. 

The central problem associated with the acceleration of cold gas by a hot wind can be illustrated by a simple timescale argument. On the one hand, the destruction timescale of a blob of cold gas with radius $r_{\cl}$, and density $\rho_{\cl}$ which is impinged by a wind with velocity $v_{\mathrm{wind}}$ and density $\rho_{\mathrm{wind}}$ is
\begin{equation}
  \label{eq:tcc}
  \tcc \sim \chi^{1/2}\frac{r_\cl}{v_{\mathrm{wind}}}
\end{equation}
where $\chi\equiv \rho_{\mathrm{cl}} / \rho_{\mathrm{wind}}$ is the overdensity of the cloud. This `cloud-crushing' time \tcc is approximately equal to the Kelvin-Helmholtz, Rayleigh-Taylor, and shock-crossing time of the system. In adiabatic hydrodynamic simulations, the cold gas is usually destroyed within a few $\tcc$ \citep[e.g.,][]{Klein1994}. On the other hand, the acceleration time-scale of the cold gas is
\begin{equation}
  \label{eq:tdrag}
  t_{\mathrm{drag}}\sim \chi\frac{r_\cl}{v_{\mathrm{wind}}}\sim \chi^{1/2}\tcc\;.
\end{equation}
Thus, for the overdensities $\chi \sim 100-1000$ associated with cold clouds embedded  in a hot wind, $\tcc \ll t_{\mathrm{drag}}$. The cold gas will be destroyed before it can be accelerated.

Several solutions out of this dilemma have been suggested. Early on, \citet{Klein1994} mentioned the potentially important role of cooling which might stabilize the cold cloud, and thus extend its lifetime. \citet{Mellema2002} later supported this conjecture using two-dimensional hydrodynamical simulations of a spherical blob of cold gas surrounded by a moving, hot wind which included radiative cooling. While they found that the lifetime is somewhat extended, they did not follow the evolution of the system long enough in order to come to a firm conclusion. 

More recently, \citet{Cooper2009}, \citet{Scannapieco2015a}, and \citet{Schneider2016} used modern, three-dimensional hydrodynamical simulations with radiative cooling to cover a wide range of the parameter space of the problem.
They found that the lifetime of the cloud can be extended by a factor of a few due to radiative cooling. However, they all concluded that eventually the cold gas will be mixed away, and cannot be entrained, i.e., the cold gas disappears before the velocity difference between the phases becomes $\sim$ zero.

These studies inspired \citet{Zhang2015a} to compare observations of galactic winds with the results of the recent hydrodynamical simulations. Specifically, they checked whether the prolonged destruction time from radiative cooling as parametrized by \citet{Scannapieco2015a} can explain the fast-moving cold gas as seen in observations of M82 \citep[e.g.][]{Strickland2009}, dwarf starburst \citep[e.g.][]{2004ApJ...610..201S}, and (U)LIRGs \citep{2014MNRAS.437.1698M}. They concluded that it could not. This cast strong doubts that the cold gas seen could have been accelerated hydrodynamically. 

Multiple potential solutions to this puzzle have  been put forward. One is that once the hot wind gas cools adiabatically to temperatures at the peak of the cooling curve, radiative cooling will allow cold gas clouds to form via thermal instability. Thus, the cold clouds are born comoving out of the hot wind gas, and not subject to hydrodynamic instabilities \citep{Wang1995,Thompson2015,Schneider2018}. This is an attractive solution, but it is unlikely to be universal. For instance, it is difficult to explain the prevalence of dust in the outskirts of galaxies (e.g., \citealt{menard10,peek15}), if it was not transported there by outflows where the cold gas survives. Alternatively, the cold gas is accelerated by non-thermal rather than hydrodynamic forces, which then decouples the relation between acceleration and the development of hydrodynamic instabilities (Eqs.~\eqref{eq:tcc} and \eqref{eq:tdrag}). If radiation pressure is responsible for acceleration, the cloud survival time is prolonged, but the cloud still continuously loses mass and disrupts \citep{Zhang2018-dusty-cloud}. Cosmic rays might be a potentially promising mechanism \citep[e.g.,][]{Wiener2017,Wiener2019}, but current simulations are highly simplified, and much more work is needed. Yet another possibility is that magnetic  field draping across the cloud suppresses Kelvin-Helmholtz instability via magnetic tension, and promotes acceleration via magnetic drag, enabling cloud survival \citep{McCourt2015}. However, as we discuss  in \S\ref{sec:magnetic_fields}, this only works at low overdensities for typical wind conditions. For clouds with $\chi \sim 100-1000$ to survive, the wind has to be magnetically dominated $\beta < 1$.

In \citet{Gronke2018} (henceforth: \citetalias{Gronke2018}), we re-visited this classical `entrainment problem'\footnote{A clarification: the customary use of the term `entrainment' in this context refers to the case when a cold cloud becomes comoving with a wind. Our work shows that this takes place because the cold cloud {\it entrains} the hot gas onto the cloud via cooling-induced pressure gradients, acquiring the mass and momentum of the hot gas. We use both senses of the word `entrainment' in this paper; the meaning should be clear from the context.}, and found a quantitative criterion for cloud survival due to cooling. The key timescale is the cooling time of the \textit{mixed} gas $t_{\mathrm{cool,mix}}$ which depends on $T_{\mathrm{mix}}\sim \sqrt{T_\cl T_{\mathrm{wind}}}$, and $n_{\mathrm{mix}}\sim \sqrt{n_\cl n_{\mathrm{wind}}}$ \citep{Begelman1990}. This renders $t_{\mathrm{cool,mix}}$ approximately an order of magnitude larger than the cooling time of the cold gas. If it is smaller than \tcc, the mixed gas can cool as quickly as it is produced by mixing, increasing the total cold gas mass. The inequality $\tcc < \tcool{mix}$ can be re-arranged in order to give a criterion that clouds exceed a critical cloud size $r> r_{\mathrm{cl,crit}}$ \citep{Gronke2018} to escape destruction:
\begin{equation}
r_{\mathrm{cl,crit}} \sim \frac{v_{\rm wind}\tcool{mix}}{\chi^{1/2}} \approx 2 \, {\rm pc} \ \frac{T_{\rm cl,4}^{5/2} \mathcal{M}_{\mathrm{w}}}{P_{3} \Lambda_{\rm mix,-21.4}} \frac{\chi}{100}\;.
\label{eq:rcrit}
\end{equation}
where $T_{\rm cl,4} \equiv (T_{\rm cl}/10^{4} \, {\rm K})$, $P_{3} \equiv nT/(10^{3} \, {\rm cm^{-3} \, K})$, $\Lambda_{\rm mix,-21.4} \equiv \Lambda(T_{\rm mix})/(10^{-21.4} \, {\rm erg \, cm^{3} \, s^{-1}})$, $\mathcal{M}_{\mathrm{w}}$ is the Mach number of the wind, and we write $v_{\mathrm{wind}} = c_{\mathrm{s,wind}} \mathcal{M}_{\rm wind} \sim c_{\mathrm{s,cl}}\mathcal{M}_{\rm wind}\chi^{1/2}$. We have assumed thermal pressure balance. Most of the mass accumulates in a long tail, and a sufficiently long simulation box is required to capture this. 
In some previous studies, the cloud size was simply too small, while in others the simulation domain was too short to capture tail formation effects.

Interestingly, while all this work was motivated by galactic winds, in an adjacent community motivated by different considerations -- the survival of high velocity clouds in the Galactic fountain of the Milky Way -- another group found cloud survival and even slight growth in some of their simulations   \citep{Marinacci2010,Armillotta2016,Armillotta2017}\footnote{Note that by Eq.~\eqref{eq:rcrit}, clouds which should survive by our criterion are nonetheless disrupted in the simulations of \citet{Armillotta2017}. We attribute this to the difference between 2D and 3D simulations: disruption is easier in 2D and continues at cloud sizes which survive in 3D simulations (see \S~\ref{sec:appendix_2d}).}  (see section 5.1 in \citetalias{Gronke2018} for a comparison to previous work). These were the first simulations to see such behavior. They also noted that larger clouds tend to survive longer. However, they did not quantify the required physical scale for survival and growth.

The findings in \citetalias{Gronke2018} showed that clouds larger than $r_{\mathrm{cl,crit}}$ will not only survive the acceleration phase but also grow significantly in mass, by more than an order of magnitude in some of our simulations.
This mass growth is the main subject of this paper. 
In particular, we aim to develop a quantitative understanding of the rate of mass growth and the relevant physical parameters. We also want to understand when and how this growth finally saturates, and thus the final cold gas mass in the wind. We will consider the influence of magnetic fields, initial cloud geometry, and background wind on mass growth rates. Understanding this is essential in order to compare galactic models to observations. One has to keep in mind that $r_{\mathrm{cl,crit}}$ in the wind and CGM of galaxies is at present unresolved in most hydrodynamic simulations (see \S\ref{sec:convergence}). In this paper, we therefore aim to find scaling relations describing the cold mass acceleration and growth which can be used to guide larger-scale simulations, and to compare our findings to observations.

This paper is structured as follows. In Sec.~\ref{sec:analytics}, we first introduce the problem, and lay out some analytical considerations in order to set the stage. In Sec.~\ref{sec:method}, we describe our numerical setup, and we present our numerical results in Sec.~\ref{sec:results} and Sec.~\ref{sec:result_robustness}. Specifically, we focus on the origin of the mass growth in Sec.~\ref{sec:results}, and show its dependence on the physical conditions in Sec.~\ref{sec:result_robustness}. We discuss our findings in a broader context in Sec.~\ref{sec:discussion} before we conclude in Sec.~\ref{sec:conclusion}.

Videos visualizing our numerical results are available at \url{http://max.lyman-alpha.com/cloud-crushing}.

\section{Analytic considerations}
\label{sec:analytics}

\subsection{Entrainment and mass growth}
\label{sec:entrainment}
As described in \citetalias{Gronke2018} and recapitulated in Sec.~\ref{sec:intro}, a cloud of cold gas will become entrained in a hot wind if $\tcool{mix} / \tcc < 1$. In addition to surviving, the cold gas mass will grow by up to $\sim $ an order of magnitude \citepalias[cf. left panel of figure~1 in][]{Gronke2018}. This raises the question of what sets the cold gas  mass growth rate and the final mass of the cloud. Globally, this sets the cold gas mass loading of the wind. Answering these questions will be the main content of this paper. In this section, we sketch some analytic considerations which provide a framework for understanding and interpreting out numerical results.  

Simply on dimensional grounds, we can write the mass growth rate as: 
\begin{equation}
  \label{eq:mdot_general}
  \dot m \sim A \rho_{\mathrm{wind}} v_{\mathrm{mix}},
\end{equation}
where $A$ is the surface area of the cloud, and $v_{\mathrm{mix}}$ the characteristic mixing velocity of the process. This equation is strictly true from mass conservation if mass only flows  one way (from hot to cold), if  one imagines drawing a Gaussian surface of area A around the cloud. Of course, there can be another contribution to mass flow from cold to hot; equation \ref{eq:mdot_general} becomes an increasingly good approximation  as growth dominates.

What is $v_{\mathrm{mix}}$? Naively, one could assume that the dominant process funneling in new, hot gas into the mixing layer is the Kelvin-Helmholtz instability. This would imply that $v_{\mathrm{mix}}$ scales with the velocity difference between the hot and the cold medium $\Delta v$. As a consequence, once the cloud is entrained ($\Delta v \ll v_{\mathrm{wind}}$) and the shear between the hot and the cold medium is negligible, the mixing and hence the mass growth halts.

However, another possibility is that the mixing is instead dominated by cooling. If mixed gas cools on timescales shorter than the sound crossing time, it sets up pressure gradients which funnels new, hot gas into the boundary layer. With this effect, the mass growth rate would not depend on the velocity difference $\Delta v$ between the hot and the cold gas, but instead on the cooling rate. 

Recently, \citet{Ji2018} performed high resolution hydrodynamic and MHD simulations of radiatively cooling, turbulent mixing layers in plane-parallel geometry, where all cooling lengths are resolved by many cells. They found that the latter effect is dominant: radiative cooling, rather than the Kelvin-Helmholz instability, is responsible for the entrainment of hot gas into the mixing layer. This agrees with our findings from \citetalias{Gronke2018} where we found mass growth even towards the end of the simulations, when $\Delta v$ is decreased substantially. In the remainder of this section, we will therefore focus on this `cooling induced mixing' effect.\footnote{In \S~\ref{sec:mass_growth_cause}, we will numerically investigate what the contributions of the two channels of mass growth are.}\\

What are reasonable choices for the mixing velocity $v_{\rm mix}$ and surface area $A$? Cooling induced mixing is driven by pressure gradients which arise when gas cools on timescales shorter than a sound crossing time.  
Thus, the characteristic mixing velocity is of order the sound speed of gas which dominates cooling. In practice, due to the steep nature of the cooling function,
this is only somewhat larger than the cloud temperature. This is indeed what was found in \citet{Ji2018}, who found mixing speeds of order $v \sim 3-10 \, {\rm km \, s^{-1}}$, and that thermal pressure deficits (and turbulent velocities) were maximized at $T_{\rm mix} \sim {\rm few} \times 10^{4}$K. Thus, $v_{\rm mix} \sim c_{\rm s,cl}$ to a good approximation. However, this ignores the dependence on the cooling time, which we now derive. 

In radiative turbulent mixing layers, cooling is balanced by radiative heat diffusion. We can characterize the turbulent heat diffusion coefficient as $\kappa \sim \alpha^{2} r_{\rm cl} c_{\rm s,cl}$, where $r_{\rm cl}$ and $c_{\rm s,cl}$ are characteristic size and velocity scales for the cloud (and set scales for when pressure balance with the surroundings -- which then shuts off mixing --- is restored), and $\alpha^{2}$ is a fudge factor  which has to be obtained numerically. Over some characteristic lengthscale $H$,  the diffusion time is equal to the cooling time,  i.e. $t_{\rm diffuse} \sim H^{2}/\kappa \sim t_{\rm cool}$, which gives the diffusion length: 
\begin{equation}
H \sim \sqrt{\kappa t_{\rm cool}} \sim \alpha  (r_{\rm cl} c_{\rm s} t_{\rm cool})^{1/2}
\end{equation} 
the geometric mean of the cloud size and cooling length\footnote{This is generic for diffusive processes. For instance, the Field length is the geometric mean of the elastic and inelastic electron mean free paths \citep{Field1977}, which characterize the competing processes of  thermal  conduction and radiative cooling.}. 
In the plane parallel simulations of \citet{Ji2018}, the size of the mixing layer indeed scales as $H \propto t_{\rm cool}^{1/2}$.  

If the inflow is driven by radiative cooling, the inflow velocity should depend on the cooling time. Direct examination of  the fluid equations gives a Bernoulli like constraint (see \citealp{Ji2018} for more detailed justification and numerical verification in high resolution simulations of radiative mixing layers):
\begin{equation}
P + \rho v_{\rm mix}^{2} \approx \ {\rm const}.
\label{eq:total_pressure} 
\end{equation}
In steady state, the competition between cooling (which increases pressure fluctuations at a rate $\delta \dot{P} \sim P/t_{\rm cool}$) and sound waves (which damp pressure fluctuations at a rate  $\delta \dot{P} \sim - \delta P/t_{\rm sc,H}$) across the mixing layers balance one another $P/t_{\rm cool} \sim \delta P/t_{\rm sc,H}$ to give: 
\begin{equation}
\frac{\delta P}{P} \sim \left(\frac{t_{\rm sc,H}}{t_{\rm cool}} \right) \sim  \alpha \left( \frac{r_{\rm cl}}{c_{\rm s} t_{\rm cool}}  \right)^{1/2} = \alpha \left( \frac{t_{\rm sc,cl}}{t_{\rm cool,cl}} \right)^{1/2}. 
\label{eq:deltaP}
\end{equation}
Since the term in brackets is larger than unity, cooling can remain quasi-isobaric $\delta P/P \ll 1$, only if $\alpha \ll (t_{\rm cool,cl}/t_{\rm sc,cl})^{1/2}$. We shall see that this is true in our simulations,  where we find $\alpha \approx 0.04$. Combining Eqs.~\eqref{eq:total_pressure} and \eqref{eq:deltaP}, we obtain
\begin{equation}
  v_{\rm mix} \approx \alpha^{1/2} c_{\rm s,cl} \left( \frac{t_{\rm cool,cl}}{t_{\rm sc,cl}} \right)^{-1/4}
  \label{eq:vmix_analytic}
\end{equation}
i.e. $v_{\rm mix}$ is  of order the {\it cold}  gas sound speed,  rather than the hot gas sound speed as one might naively expect. We shall see that this is true in our numerical simulations, where we also verify the $v_{\rm mix} \propto t_{\rm cool}^{-1/4}$ scaling\footnote{Note an important distinction between $v_{\rm z} \propto t_{\rm cool}^{-1/2}$ defined in \citet{Ji2018} and $v_{\rm mix} \propto t_{\rm cool}^{-1/4}$ discussed here. The former refers to advective inflow in the frame of the mixing layer (and includes a component due to growth of the cloud), whereas $v_{\rm mix}$ here refers to advective inflow in the frame of the cloud. If we take the simulations of \citet{Ji2018} and measure the advective inflow velocities $v_{\rm mix}$ from the box boundary, consistent with  the definition used here, we recover $v_{\rm mix} \propto t_{\rm cool}^{-1/4}$.}. 
 
As for the cloud surface area $A(t)$, we expect its evolution to be characterized by two phases: initially, the `tail formation phase', when the cloud is not yet entrained, and tail growth dominates the change in surface area. 
Subsequently, during the `entrained phase, when $\Delta v \sim 0$, the areal growth is isotropic and slower. A simple estimate for the former phase, when the largest increase in surface area takes place, is:
\begin{equation}
  \label{eq:Atail}
  A(t) \sim r_\cl v_{\mathrm{wind}} t \sim r_{\cl}^{2} \chi^{1/2} (t / \tcc) \qquad \text{ for } t \lesssim t_{\mathrm{ent}}
\end{equation}
where $t_{\mathrm{ent}}$ is the entrainment time. Due to the momentum transfer caused by the addition of cooled wind gas to the cloud, we found in \citetalias{Gronke2018} that $t_{\mathrm{ent}}\lesssim t_{\mathrm{drag}}$. Note that 
\begin{equation}
t_{\rm cc} \sim \frac{r_{\rm cl}}{\chi^{1/2} v_{\rm wind}} \sim \frac{r_{\rm cl}}{\chi^{1/2} c_{\rm s,w} \mathcal{M}_{\mathrm{w}}} \sim \frac{r_{\rm cl}}{c_{\rm s,cl} \mathcal{M}_{\mathrm{w}}} \sim t_{\rm sc}
\label{eq:tcc_tsc}
\end{equation}
for a transonic wind with $\mathcal{M}_{\mathrm{w}} = v_{\rm wind}/c_{\rm s,w} \sim 1$, i.e. the cloud crushing time is of order the cloud sound crossing time. Thus, whenever we normalize to $\tcc$, it is equivalent to normalizing by $t_{\rm sc}$.

We can use this to find the rate of cloud growth after entrainment. If we set $t_{\rm ent}/t_{\rm cc} \sim \chi^{1/2}$ in Eq.~\ref{eq:Atail}, then after entrainment $A\sim r_{\rm cl}^{2} \chi \rho_{\rm w}$. We obtain for the mass growth rate:
\begin{equation}
\dot{m}  \sim \rho_{w}r_{cl}^{2} \chi c_{s,cl} \left(\frac{m_{cl,0}}{\rho_{cl} r_{\rm cl}^{3}} \right) \sim \frac{m_{\rm cl,0}}{t_{sc}}
\label{eq:growth_rate_cc} 
\end{equation}
where we multiply and divide by $m_{\rm cl,0}$, the initial cloud mass. We will compare this expectation to simulation results. Note that this does {\it  not} imply that the cloud grows on a sound-crossing time in the entrained phase, since $t_{\rm grow} \sim m/\dot{m}  \sim t_{\rm sc} (m/m_{i}) \gg t_{\rm sc}$.
The cloud growth time can be  written as:  
\begin{equation}
t_{\rm grow} = \frac{m}{\dot{m}} \approx \frac{ \rho_{cl} r_{cl} A}{\rho_{w} v_{\rm mix} A} \approx \chi \frac{r_{\rm cl}}{v_{\rm mix}} \sim \chi t_{\rm sc} \left(\frac{t_{\rm cool,cl}}{t_{\rm sc,cl}} \right)^{1/4}
\label{eq:t_grow} 
\end{equation}
In our numerical simulations, we shall see that this has a physical interpretation: the cloud pulsates on a timescale of order the sound crossing time $t_{\rm sc}$, thus sweeping up hot gas with a volume of order the cloud size, and thus a mass $\sim m_{\rm cl}/\chi$. Thus, the cloud grows on a timescale $\chi$ times longer than the pulsation frequency.

\begin{figure*}
  \centering
  \includegraphics[width=.9 \textwidth]{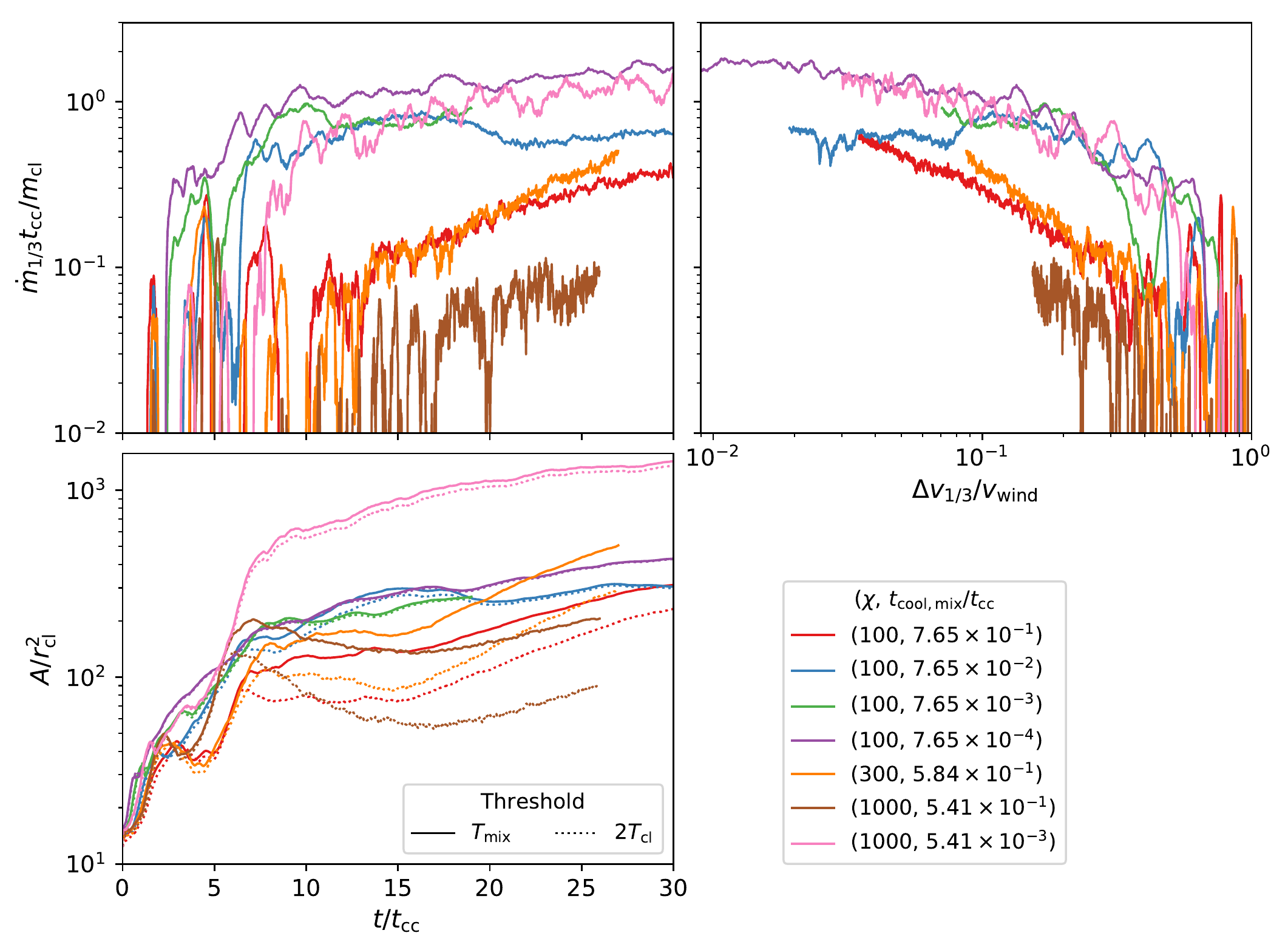}
  \caption{Evolution of the mass growth rate for different cloud sizes, which is equivalent to different $t_{\rm cool, mix}/t_{\rm cc}$. The mass growth converges to a non-negligible asymptotic value (\textit{upper left panel}) although the clouds are almost entrained and the velocity difference between the hot and the cold gas drops (\textit{upper right panel}; note that the velocity difference decreases toward the left in  this figure). This suggests that cooling, rather than the Kelvin Helmholtz instability, dominates the mixing process. The growth rate closely tracks the surface area of the stretched cloud (\textit{lower left panel}). The runs shown here had a resolution of $l_{\mathrm{cell}} / r_\cl = 8$, and an overdensity of $\chi=100$.}
  \label{fig:mass_growth_new_setup_multiplot}
\end{figure*}

\subsection{Mass growth in a galactic wind}
\label{sec:analytic_wind}
In a realistic setting, the background properties are not constant but change with radius from the galaxy. 
Let us consider an adiabatic wind as 
encapsulated in the \citet{Chevalier1985} (CC85) solution. In this framework,
mass and energy are injected at a constant rate $\dot{m}_*$ and $\dot{E}_*$ respectively in a hot `superbubble'. From energy conservation, this must drive a hot wind with asymptotic velocity $v_\infty \sim (2 \dot  E_* / \dot m_*)^{1/2}$. From mass conservation, $\rho v_\infty r^{2}=\,$const, this implies $\rho \propto r^{-2}$ and $P \propto \rho^{\gamma} \propto r^{-10/3}$.

In order to understand the evolution of an ejected cold gas cloud with $t_{\mathrm{cool,mix}}/\tcc < 1$ out to $\sim$kpc radius and beyond, it is important to take this background evolution into account.
The mixing speed drops as $v_{\mathrm{mix}}\propto t_{\mathrm{cool}}^{-1/4} \sim (P / n^{2} \Lambda(T_{\mathrm{cl}}))^{-1/4} \sim (k_{\mathrm{B}}^{2} T_{\cl}^{2}/P \Lambda(T_{\cl})^{-1/4} \propto  P^{1/4} \propto r^{-5/6}$ (cf. Eq.~\eqref{eq:vmix_analytic}).
Here we have assumed that $T_{\rm cl} \sim 10^{4}$K is constant due to photoionization by the UV background. On the other hand, the surface area grows due to the decrease in pressure, $A\propto (m_{\cl} / \rho_{\cl})^{2/3}\propto (m_{cl} T_{\rm cl}/P)^{2/3} \propto m^{2/3}r^{20/9}$. If we also use $\rho_{\rm w} \propto r^{-2}$, we obtain for the mass growth in an adiabatically expanding background
\begin{equation}
  \label{eq:mdot_adiabatic_wind}
  \dot m \propto v_{\rm mix} \rho_{\rm w} A \propto r^{-11/18} m^{2/3} \approx r^{-2/3} m^{2/3}.
\end{equation}
When the cloud is entrained, it is moving at the wind speed which is $\sim$ constant at larger radii for the \citet{Chevalier1985} solution. Thus, writing $\dot{m} \sim \dot{r} \, (dm/dr)$ and using the fact that $v_{\infty}\sim$const at large radii,
\begin{equation}
  \label{eq:mdot_adiabatic_wind2}
  \dot m \approx v_{\infty} \frac{\dd m}{\dd r} \propto r^{-2/3} m^{2/3}
\end{equation}
which implies
\begin{equation}
\dot{m} \sim \, {\rm const}, \ \ \ m \propto r
\end{equation}
in the entrained phase. The increase in area and decrease in wind density and cooling rates cancel out to give a constant mass growth in the entrained phase. If we combine this with Eq.~\eqref{eq:growth_rate_cc}, we obtain: 
\begin{equation}
m \sim m_{0} \left(\frac{t}{t_{\mathrm{sc},0}} \right) \sim m_{0} \left(\frac{r}{r_{\rm cl}} \right) \frac{1}{\mathcal{M}_{\mathrm{w}} \chi^{1/2}} 
\end{equation} 
where $r\sim v_{\rm w} t$ is the distances the cloud reaches in the halo. Thus,  very  large growth factors $m/m_{0} \sim 10-100$ are possible. The rapid increase in cold gas fraction suggests that the asymptotic cold gas mass can rise to be of order the wind mass, {\it even if the wind itself does not radiatively cool, and the cold gas fraction is small at the launch radius.} We discuss this further in  \S\ref{sec:disc_cgm}.

Our discussion above ignores the interaction of the wind with the CGM, and applies only out to distances where the constant entropy Chevalier-Clegg solution is a good approximation. In galaxy scale simulations, this appears true out to $\gtrsim 50\,$kpc \citep[e.g.,][]{Fielding2017}, though note that the latter simulations show a slow increase of entropy with radius, $K=T/n^{2/3} \propto r^{0.4}$. 
Cloud growth likely slows down once interaction with the CGM becomes important. A reverse shock will heat the wind to higher entropy as it encounters CGM gas;  the wind can also mix with high-entropy CGM gas. Higher entropy gas is more strongly pressurized and inhibits cloud expansion. If we  write  
\begin{equation}
\dot{m} \propto m^{2/3}  P^{-5/12}  \rho_{\rm  w} \propto m^{2/3} K_{\mathrm{w}}^{-5/12} \rho_{\rm w}^{11/36}
\label{eq:mdot_entropy}
\end{equation}
then it is clear that if the background / wind entropy rises with radius, the growth rate slows. For instance, consider an isothermal wind $P \propto \rho_{w} \propto r^{-2}$ (which may arise in a situation where heat input such as Compton heating is important), where $K_{w} \propto r^{4/3}$.  We study this numerically in \S\ref{sec:adiabatic_wind}. Then $\dot{m} \propto m^{2/3} r^{-7/6}$, which implies that asymptotically, $\dot{m} \rightarrow 0$, $m \sim$const with radius. Similar results may apply if the wind shocks or mixes with high entropy CGM gas. For $\rho\propto^{-2}$ and constant wind velocity, then for $K \propto r^{\beta}$, then $\beta \gsim 1$ is roughly the boundary at which cloud growth is shut off. However, the density and velocity profile is likely to be strongly affected by interaction with the CGM; feedback and turbulent mixing can also change the underlying entropy profile. Such questions are best handled by numerical simulations.

\begin{figure}
  \centering
  \includegraphics[width=\linewidth]{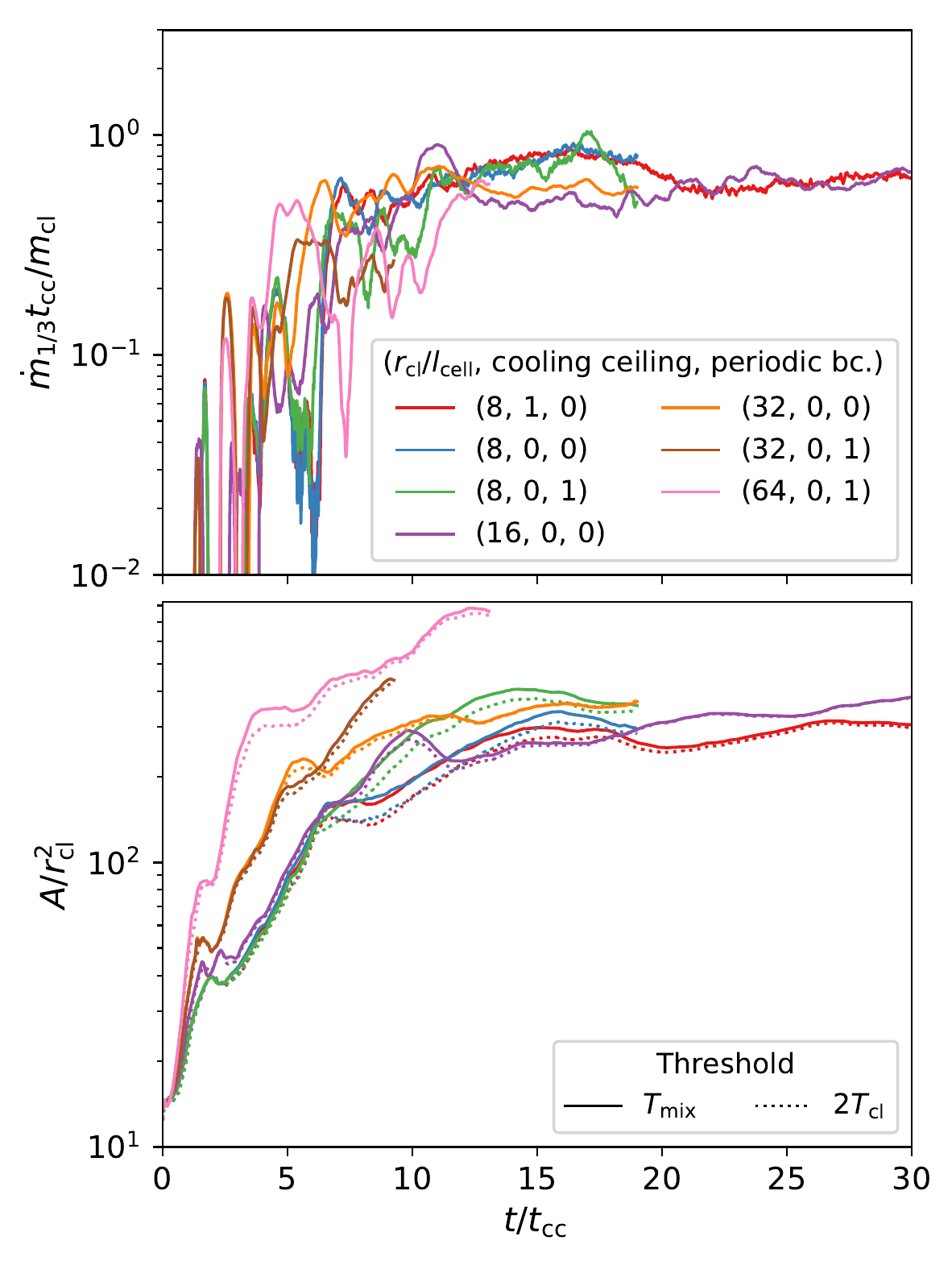}
  \caption{Growth rate \textit{(upper panel)} and surface area \textit{(lower panel)} of clouds with $\chi=100$, and $\tcoolrat{mix} \sim 8\times 10^{-2}$ with different resolutions, boundary conditions, and cooling setups. While the surface area is not converged, mass growth is independent of resolution.}
  \label{fig:mass_growth_d100_convergence}
\end{figure}

\section{Numerical Methods}
\label{sec:method}

We used \texttt{Athena} $4.0$ \citep{Stone2008} to run a suite of simulations on a regular, three-dimensional, Cartesian grid with grid-spacing $d_{\mathrm{cell}}$. For the hydro-only runs we used the HLLC Riemann solver, second-order reconstruction with slope limiters in the primitive variables, and the van Leer unsplit integrator \citep{Gardiner2008} while for the MHD runs we used the HLLD solver and a third-order reconstruction method. We use resolutions ranging from $r_\cl/d_{\mathrm{cell}}=8$ to $128$.

In \texttt{Athena} we have implemented the `exact' \citet{Townsend2009} radiative cooling algorithm, which  allows for fast exact solutions of radiative cooling  for a piecewise power-law cooling function. For simplicity, we use a seven-piece powerlaw fit to the \citet{Sutherland1993} $Z= Z_\odot$ cooling curve \citepalias[see][for the impact of the variation of the cooling curve on our results]{Gronke2018}.

Our simulation setup is mostly identical to \citetalias{Gronke2018}. We place a cloud in a hot wind entering the simulation domain from the $x\rightarrow -\infty$ direction. We use inflowing boundary conditions in the negative $x$-direction, and outflowing boundary conditions otherwise. Specifically, we enforce the wind solution and only allow for $v_x \ge 0$ in the ghost cells of the left boundary. 
Furthermore, we use a cloud-tracking scheme which means we continuously change the reference frame to follow the center of mass of the cold gas. This allows us to reduce the required box size, thus lowering the computational cost. Finally, to prevent cooling  of the wind over long timescales, we turn off cooling for $T > 0.6 T_{\rm wind}$. We previously showed in \citet{Gronke2018} that our results are robust to this; also see Fig. \ref{fig:mass_growth_d100_convergence} of this paper. Note that we do not take this step in the expanding wind solution described below. 

Our baseline simulations are purely hydrodynamic and consider an initially spherical cloud. Unless clearly indicated otherwise, this is the setup used in this paper. However, we also study the impact of non-spherical cloud geometries (\S~\ref{sec:nosphere}) for which we extracted a region from a turbulently stirred simulation, and use this as initial conditions for the cloud.

We also consider the impact of magnetic fields. For the MHD simulations, we follow \citet{McCourt2015}, and initialize a tangled, approximately force-free magnetic field inside the cloud with magnetic coherence length $r_\cl / 10$, and strength $\beta_\cl \equiv 8 \pi P / B_\cl^2$. In the wind, we use a horizontal magnetic field in the $z$-direction with strength $\beta_{\mathrm{w}}=8 \pi P / B_{\mathrm{w}}^2$.

In order to check the effect of a varying background as in the \citet{Chevalier1985} solution, we implemented a scaling routine in \texttt{Athena} \citep[similarly to what was done in][]{Scannapieco2017}. For a spherically symmetric, adiabatically expanding wind, we scale the density $\rho$, the pressure $P$, and the velocities orthogonal to the wind $v_{y,z}$ at every time step as
\begin{align}
  \label{eq:scalings_adiabatic}
  \rho \propto a^{-2},\quad  P \propto  a^{-10/3},\text{ and}\quad v_{y,z} \propto  a^{-1}
\end{align}
where $a(t)\equiv r(t) / r_0$ is the `scale factor' defined  through the cloud's position $r(t)$, and the starting position of the cloud $r_0\equiv r(t=0)$. The first two relations are standard for CC85 (as described in \S~\ref{sec:analytic_wind}), and the  third represents the usual decay of peculiar  velocities in an expanding background. We also consider an isothermal wind for which $P\propto a^{-2}$ instead. Note that this assumes that the cloud is in sonic contact with the hot medium.

\begin{figure}
  \centering
  \includegraphics[width=.95\linewidth]{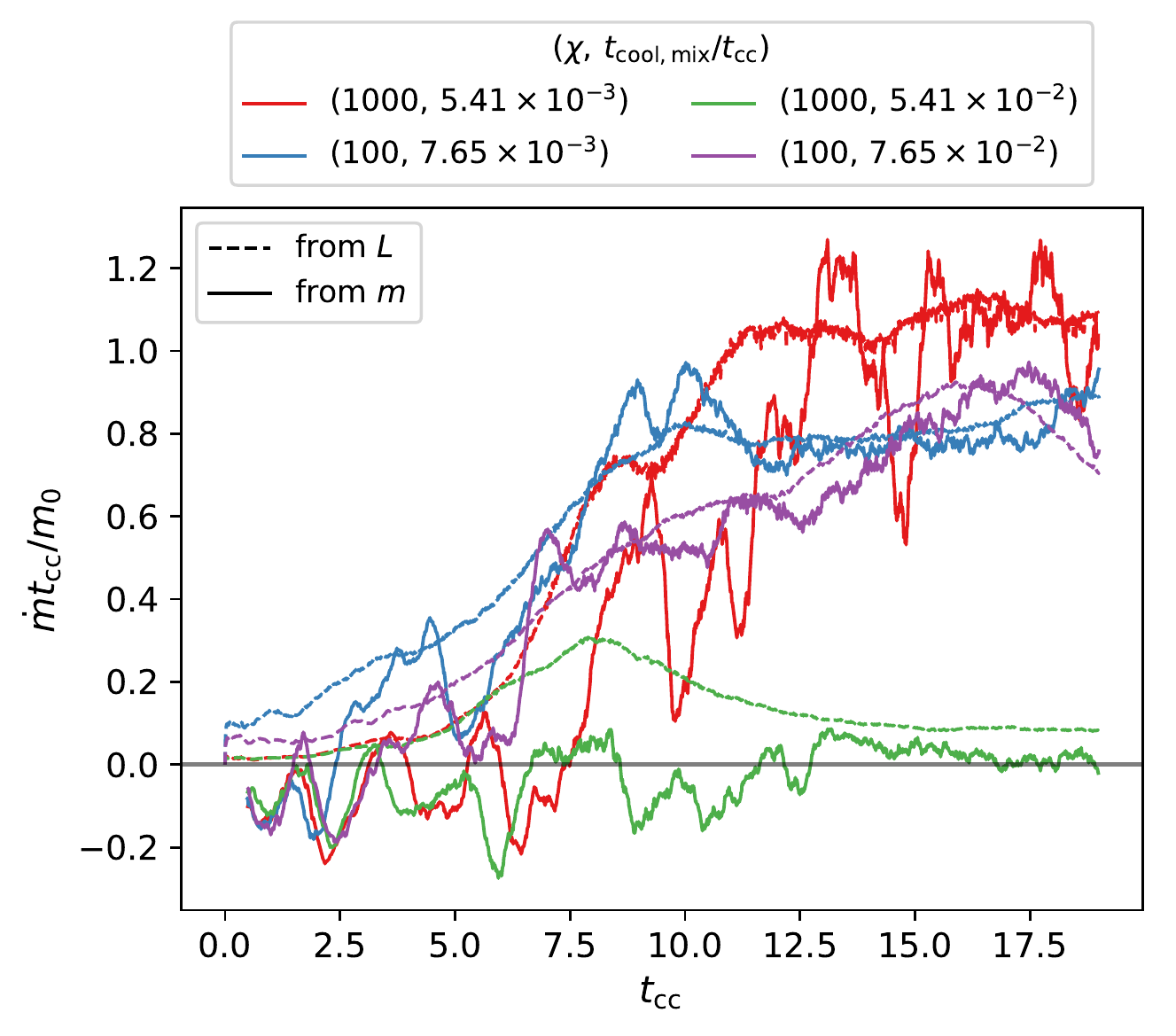}
  \caption{Cold gas mass growth computed directly from $m_{1/3}$ and the total luminosity using Eq.~\eqref{eq:isobaric_cooling}. The two agree well, showing that quasi-isobaric cooling of gas from the hot wind is a good approximation. The simulations have a resolution of $r_\cl / d_{\mathrm{cell}}=8$.}
  \label{fig:mass_luminosity}
\end{figure}

\section{Results: The Nature of the Cold Gas Mass Growth}
\label{sec:results}
In this section, we investigate numerically the mass growth for an initially spherical cloud in  a time-steady wind of $\mathcal{M}_{\mathrm{w}}=1.5$.
First, we study the cause of the mass growth (\S\ref{sec:mass_growth_cause}), how it depends on the luminosity of the cloud (\S~\ref{sec:luminosity_mdot}), and where this mass growth occurs (\S~\ref{sec:location_mass_growth}). We then analyze its components, namely the areal growth and mixing velocity (\S~\ref{sec:surface_area} and \S~\ref{sec:vmix}, respectively).

\begin{figure*}
  \centering
  \input{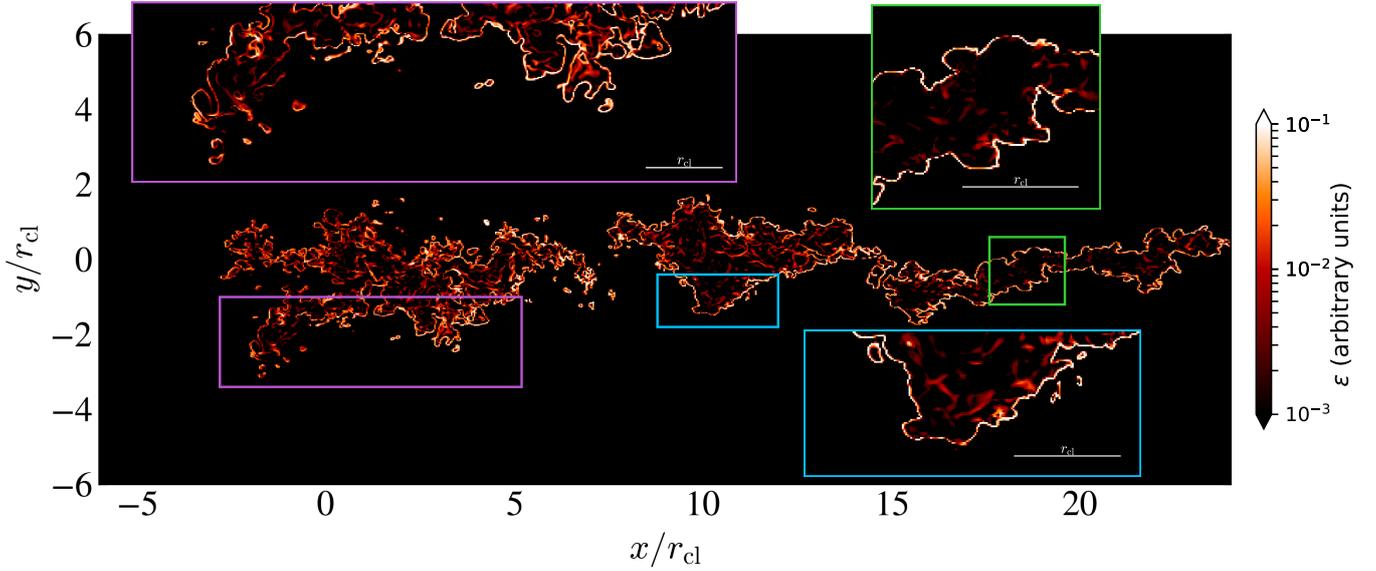}
  \vspace{-0.5cm}
  \caption{Slice through the $z=0$ plane at $t\sim 13\tcc$ for our $\chi=100$, $l_{\mathrm{cell}}/r_\cl=64$ run showing the emissivity with enlarged regions marked in the same color. Clearly, regions with highest emissivity are at the boundary between hot and cold gas. See \S~\ref{sec:location_mass_growth} for a census of the total luminosity.}
  \label{fig:2d_emissivty_slice}
\end{figure*}

\subsection{What causes the mass growth?}
\label{sec:mass_growth_cause}
In Fig.~\ref{fig:mass_growth_new_setup_multiplot}, we show the mass growth evolution (top left panel; normalized by $m_\cl / \tcc$, where $m_\cl, \tcc$ as evaluated for the {\it initial} cloud) for differently sized clouds (all with $\tcoolrat{mix} < 1$), and two overdensities ($\chi=100$ and $\chi=1000$). We measure the mass growth $\dot m_{1/3}$ of  all gas with $\rho > \rho_\cl / 3$.  After an initial steep rise, the growth flattens at $t \sim 10\,\tcc$.

In the upper right panel of the same figure, we show the same mass growth but this time as a function of $\Delta v_{1/3}$ (i.e., the difference between the mass-weighted velocity of gas with $\rho > \rho_\cl / 3$ and $v_{\mathrm{wind}}$; hence, $\Delta v_{1/3}(t = 0) = v_{\mathrm{wind}}$ and decreasing for larger $t$) first thing to note is that $\Delta v_{1/3}$ is \textit{anti-}correlated with $\dot m_{1/3}$, contrary to what might be expected if Kelvin-Helmholtz mixing causes mass growth. The anti-correlation suggests that rather than fueling mass growth, the Kelvin-Helmholz instability is a competing procress which destroys cold gas. Secondly, one can note from the top right panel of Fig.~\ref{fig:mass_growth_new_setup_multiplot} that even for basically entrained cold gas, i.e., $\Delta v_{1/3} < v_{\mathrm{wind}} /10$, the mass growth does not cease but seems to instead converge to a value of $\dot{m} \sim m_{\cl,0} / \tcc \sim m_{\cl,0} /t_{\rm sc}$ (consistent with Eq.~\eqref{eq:growth_rate_cc}). In Appendix \ref{sec:static_setups}, we confirm this directly by simulating near static setups.  

These two indications point towards the cooling dominated mass growth discussed in Sec.~\ref{sec:analytics}. A third thing to check is that in this case $v_{\mathrm{mix}}\sim \mathrm{const.}$, i.e.,  $\dot m$ is primarily set by the increase of the surface area of the cloud. Using a marching cubes algorithm \citep{MarchingCubes}\footnote{The marching cubes algorithm analyzes $8$ grid cells at a time, and identifies the surface which separates the scalar value at these points above and below some defined threshold. After this is carried out for all points in the grid, the individual surfaces are merged.}, we computed the surface area of the cold gas at each snapshot. Here, we defined `cold gas' as $T < T_{\mathrm{mix}}$ and as an alternative used threshold $T < 2 T_\cl$. The lower right panel of Fig.~\ref{fig:mass_growth_new_setup_multiplot} shows the evolution of this measured surface area, and indeed qualitatively the shape of $A(t)$ agrees with the corresponding $\dot m(t)$ curve in the panel above. We find numerically that $A\propto m^{\alpha}$ with $\alpha\sim \sfrac{1}{2}-\sfrac{2}{3}$ which is slightly lower than the $\alpha=\sfrac{2}{3}$ used in \S~\ref{sec:analytics}. This is due to the late time contraction of the cloud's tail and coagulation of individual clumps. For the simulations with  short cooling times  ($\tcoolrat{mix}\lesssim 0.1$) the  surface areas defined via the different temperature thresholds agree well. However, for larger cooling times the cold  gas surface area is not as well defined.

A concern regarding the cold gas surface area -- which seems to set the mass growth rate -- is that due to a fractal-like boundary, $A(t)$ strongly depends on the simulation resolution. While the simulations presented in Fig.~\ref{fig:mass_growth_new_setup_multiplot} were run with a rather low resolution of $l_{\mathrm{cell}} / r_\cl = 8$, we show in Fig.~\ref{fig:mass_growth_d100_convergence} results for a resolution up to $l_{\mathrm{cell}} / r_\cl = 64$.
The mass growth rate is roughly independent of the resolution. The area seems to be larger, however, in the high-resolution simulations.
We discuss the different convergence properties of mass and area further in \S\ref{sec:shattering}. Since the area $A$ and hence the mixing velocities $v_{\rm mix} \sim \dot{m}/\rho_{\rm w} A$ we derive are resolution dependent, their absolute values are not robust. However, the trends they describe are meaningful. The `coarse-grained' area at low resolution, which corresponds roughly to the projected area, is the `effective' area which seems to set the mass entrainment rate.  In Fig.~\ref{fig:mass_growth_d100_convergence}, we also show that the shutoff of cooling for $T > 0.6 \, T_{\mathrm{wind}}$ has negligible effects (also see section 5.2 in \citetalias{Gronke2018}).

Overall, it is clear from the results presented in this section that: \textit{(i)} the mass growth does not cease even with negligible $\Delta v$, and \textit{(ii)} it in fact converges to approximately $\dot m \sim m_{\cl,i} / \tcc$.

\begin{figure}
  \centering
  \includegraphics[width=\linewidth]{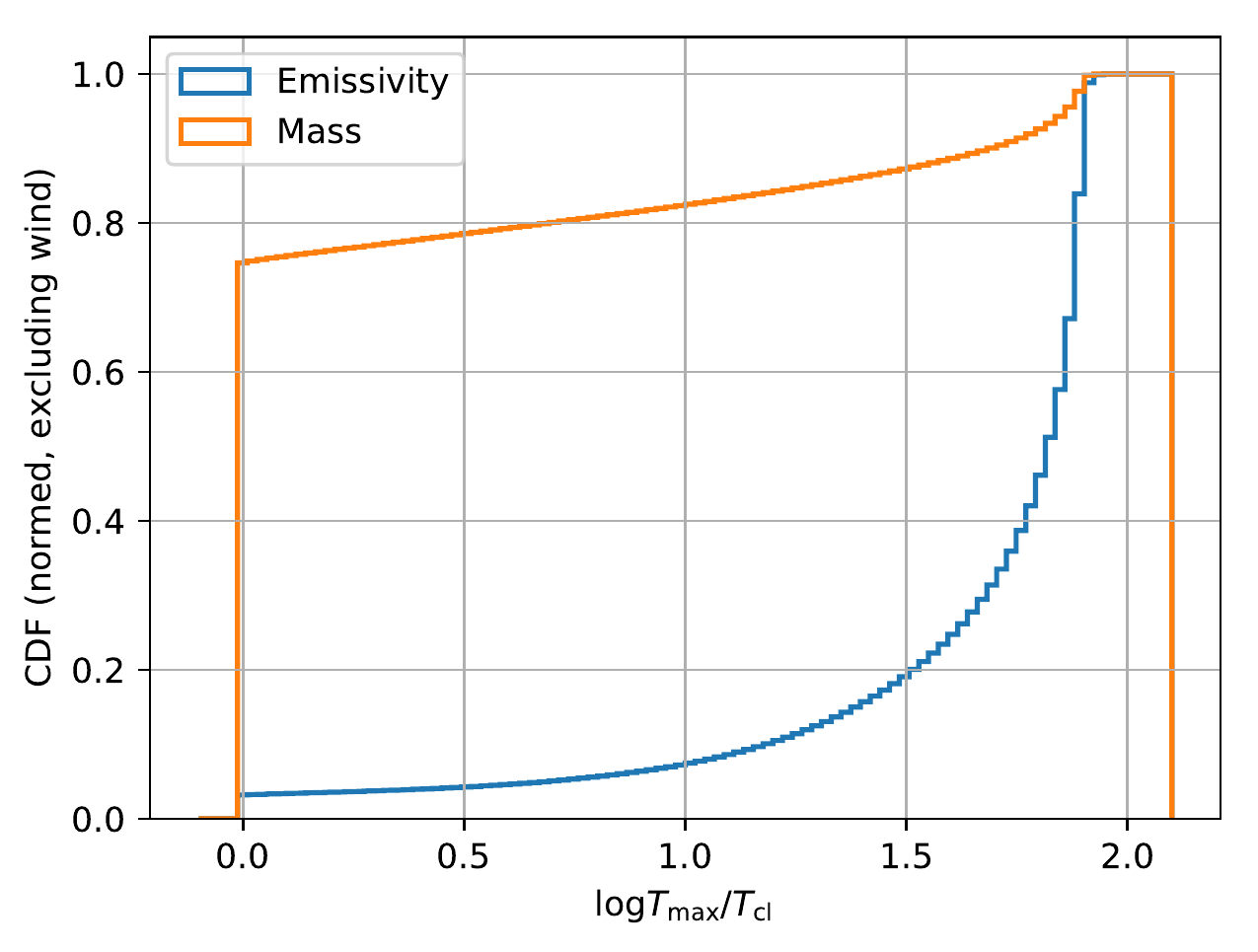}
  \caption{Normalized cumulative distribution function (CDF) of the emissivity and the density as a function of the maximum temperature of the neighboring cells $T_{\rm max}$ for the snapshot shown in Fig.~\ref{fig:2d_emissivty_slice}. Note that we excluded the wind cells for which $\rho < 2 \rho_{\rm wind}$ to focus on the cold gas evolution. While most of the gas mass lies in the cloud interior, most of the emissivity is in the cloud boundary.}
  \label{fig:cdf_Tmax}
\end{figure}

\begin{figure}
  \centering
  \includegraphics[width=\linewidth]{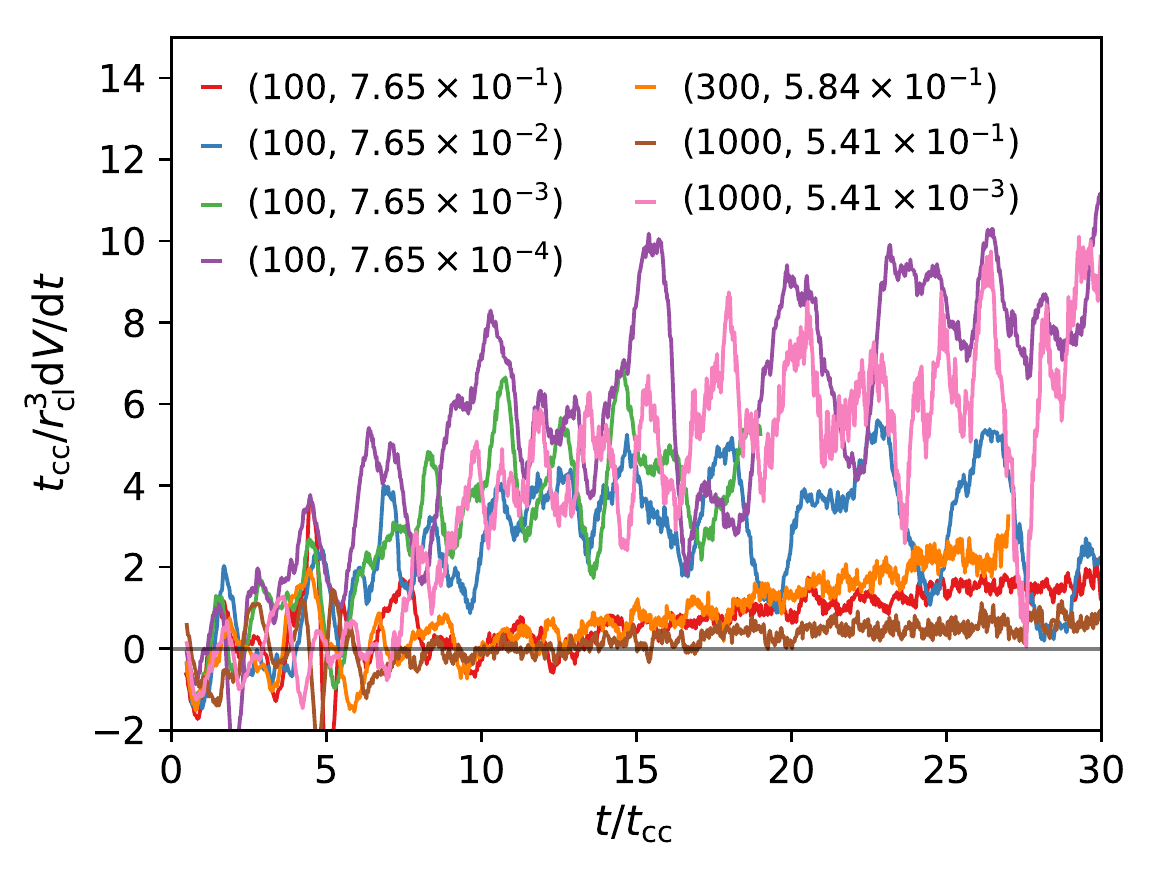}
  \caption{Rate of change in dense gas volume with $\rho > \rho_{\rm cl} / 3$ versus time.  There are clear pulsations in cloud volume which do not decay.}
  \label{fig:pulsations_evolution}
\end{figure}

\subsection{Mass growth and luminosity}
\label{sec:luminosity_mdot}

If mass accretion onto the cold cloud is indeed powered by quasi-isobaric  cooling, in the entrained phase the cloud should obey the standard relationship for isobaric cooling \citep{Fabian1994}: 
\begin{equation}
L= \frac{5}{2} \frac{\dot{m}}{\mu m_{\rm p}} k_{\rm B} T_{\rm w} \left( 1 + \mathcal{M}^2 \right),  
\label{eq:isobaric_cooling} 
\end{equation}
where the factor of $\left( 1 + \mathcal{M}^2 \right)$ accounts for the fact that the wind gas gives up both thermal and kinetic energy when it is entrained onto the cloud. Figure~\ref{fig:mass_luminosity} shows the mass growth in some simulations calculated via the numerical derivative of the cold gas mass $m_{1/3}$, and via the total luminosity $L$ using Eq.~\eqref{eq:isobaric_cooling}, assuming $\mathcal{M} \sim 0$, as appropriate for the entrained phase. Here, $L$ was obtained directly from the simulations by outputting how much energy was lost due to cooling.
The two measures agree well, showing that Eq.~\eqref{eq:isobaric_cooling} is roughly valid. Note that the agreement is reached only at late times when the cloud is entrained. Earlier on, not all emission corresponds to an increase in cold gas mass, due to the efficiency of Kelvin-Helmholtz mixing, which results in some intermediate temperature gas not surviving but getting mixed back into the hot wind. In addition, our assumption of $\mathcal{M} \sim 0$ is no longer valid. Thus, $\dot{m}$ calculated from the luminosity is  noisier and above the true $\dot{m}$. Agreement is reached later for higher overdensity clouds because they entrain more slowly.

\begin{figure}
  \centering
  \includegraphics[width=.95\linewidth]{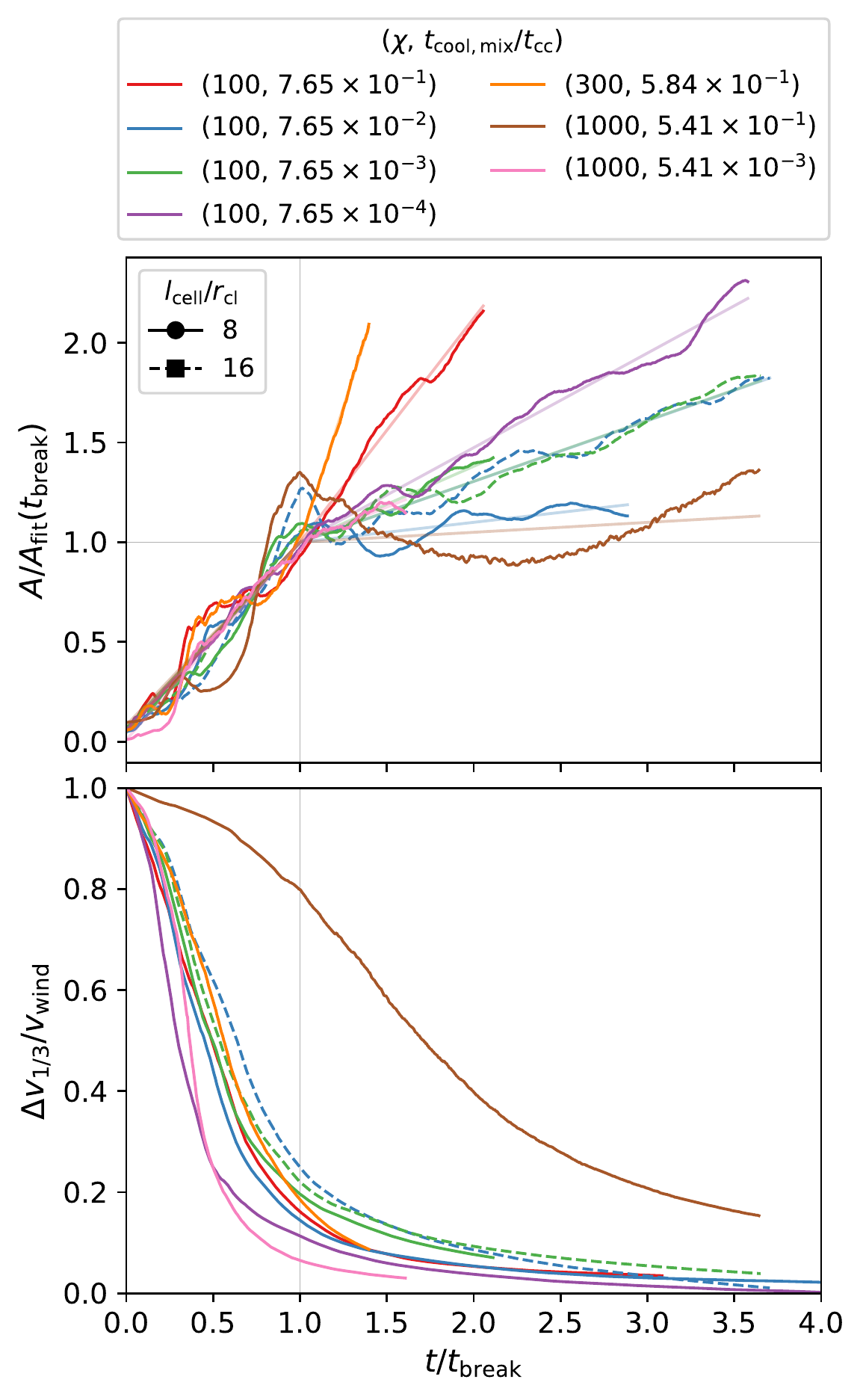}
  \caption{\textit{Upper panel:} evolution of the cold gas surface area (defined as $T < T_{\mathrm{mix}}$) with times and area normalized to values at the break Fits to Eq.~\ref{eq:A_fit} are shown with semi-transparent curves in the same color. \textit{Lower panel:} Velocity difference between the hot and the cold gas on the same time scale. 
  }
  \label{fig:evolution_area}
\end{figure}

\subsection{Location of the mass growth}
\label{sec:location_mass_growth}

Figure~\ref{fig:2d_emissivty_slice} shows an emissivity slice through our $r_\cl/d_{\mathrm{cell}}=64$ simulation at $t\sim 13\tcc$ (see \S~\ref{sec:luminosity_mdot} for the connection between luminosity and $\dot m$). The emissivity shown is outputted from the simulation and averaged over one time-step ($\sim 10^{-4}\,\tcc$). The figure clearly shows an order of magnitude higher emissivity $\epsilon$ at the boundary layer compared to the interior. In spite of the larger volume of the latter, the total luminosity is also dominated by the boundary layer with $\gtrsim 3/4$ of it stemming from regions with $\log\epsilon\gtrsim -1.5$ in the snapshot shown in Fig.~\ref{fig:2d_emissivty_slice} (which is $\gtrsim 40\times$ the median emissivity of the cloud).

In order to investigate further whether the majority of cooling occurs within the cloud or in the  boundary layer, we show in Fig.~\ref{fig:cdf_Tmax} the emissivity as a function of the maximum temperature of the neighboring $26$ cells $T_{\mathrm{max}}$. While in the interior of the cloud $T_{\mathrm{max}}\sim T_\cl$, at the  boundary this value is significantly larger and approaches $T_{\mathrm{wind}}$.
From Fig.~\ref{fig:cdf_Tmax} we can see that indeed, the majority of the emission stems from cells with $T_{\mathrm{max}}\gtrsim 10^{1.5}T_\cl$, i.e., from the mixing layer between the hot and the cold gas.

Figure~\ref{fig:2d_emissivty_slice} also shows emitting regions within the cloud. These regions are highly dynamic (as suggested by their wave-like pattern). We quantify these oscillations in Fig.~\ref{fig:pulsations_evolution} where we show the time evolution of the change in dense gas volume (defined, as before, as $\rho > \rho_\cl /3 $). Note that the oscillations are not damped, which means they are not caused by the initial cloud-crushing shock. These pulsations are likely overstable sound waves driven by loss of pressure balance due to cooling; they likely help drive mixing. We will study them in detail in future work.

\subsection{Components of the mass growth}

In \S~\ref{sec:analytics}, we wrote $\dot m$ in a form which depends on the cold gas mass surface area $A(t)$, and the characteristic mixing velocity $v_{\mathrm{mix}}$ (cf. Eq.~\eqref{eq:mdot_general}). In this section, we  investigate both components numerically.

\subsubsection{Cold gas surface area}
\label{sec:surface_area}

Figure~\ref{fig:evolution_area} shows in the top panel, the evolution of the cold mass surface area versus time for a range of cloud sizes, overdensities (denoted by colors), and resolutions (marked with different linestyles). Here, we compute the surface area using the threshold $T = T_{\mathrm{mix}}$. 
For each curve, we fit the piecewise function
\begin{equation}
  \label{eq:A_fit}
  A(t) = r_\cl^2 (4 \pi + s_i t / \tcc)
\end{equation}
with $s_i=s_1$ for $t < t_{\mathrm{break}}$ and $s_i=s_2$ otherwise. This fit is motivated by two stages of area growth (`tail formation' and `entrained') described in section \ref{sec:entrainment}. This leaves us with the three fit parameters $s_1$, $s_2$, and $t_{\mathrm{break}}$.

The upper panel of Fig.~\ref{fig:evolution_area} illustrates that $A(t)$ can be fairly well reproduced by fits to Eq.~\eqref{eq:A_fit}. Furthermore, in the lower panel of Fig.~\ref{fig:evolution_area} we show the velocity evolution on the same time scale. This illustrates that -- apart from the $\chi=1000$ run with $\tcoolrat{mix} > 0.5$, the time $t_{\mathrm{break}}$ corresponds to the point when $\Delta v_{1/3}$ dropped below $\sim 0.2 v_{\mathrm{wind}}$; agreeing well with the description in Sec.~\ref{sec:analytics} where we predicted $t_{\mathrm{break}}\lesssim t_{\mathrm{drag}}\sim \chi^{1/2}\tcc$. The values obtained from our numerical fits match this expectation fairly well (as we identified from our numerical fits shown in Fig.~\ref{fig:evolution_area}): the $\chi=100$ points cluster around $t_{\mathrm{break}}\sim 10\tcc$, whereas the $\chi \sim 1000$ simulation with $\tcoolrat{mix}\ll 1$ yields $t_{\mathrm{break}}\sim 23$. Similarly, from analytic considerations we expect $s_1\sim \chi^{1/2}$ (cf. Eq.~\eqref{eq:Atail}) which matches the values up to a factor of $\sim 2$.

In the period $t > t_{\mathrm{break}}$, when the cloud is entrained, the areal growth is slower\footnote{An exception are the runs with $\tcoolrat{mix}\gtrsim 0.1$ and $\chi = 100,300$ where $s_2\gtrsim s_1$. In these cases the cooling is just strong enough to compensate for destructive cloud-crushing instabilities which causes the original cloud to break into many small droplets at $t \sim t_{\rm break}$, causing an increase in surface area. At this point, they are close to fully entrained and thus no longer susceptible to shear-driven instabilities.}, with $s_2 < s_1$.  We obtain larger values of $s_2$ for a smaller $\tcoolrat{mix}$. 

In summary, we find that the most rapid area growth occurs in the `tail growth' phase, with $A\sim \chi r_{\rm cl}^{2}$ during the entrained phase (i.e., $t \gsim t_{\rm break}$), a property we used in Eq.~\eqref{eq:growth_rate_cc}. The area evolution can be fit by equation \eqref{eq:A_fit} where $s_1 \sim 2 \chi^{1/2}$, $s_2 \sim \chi^{1/2}$, and $t_{\rm break} \sim \chi^{1/2} t_{\rm cc}$.  

\begin{figure}
  \centering
  \includegraphics[width=\linewidth]{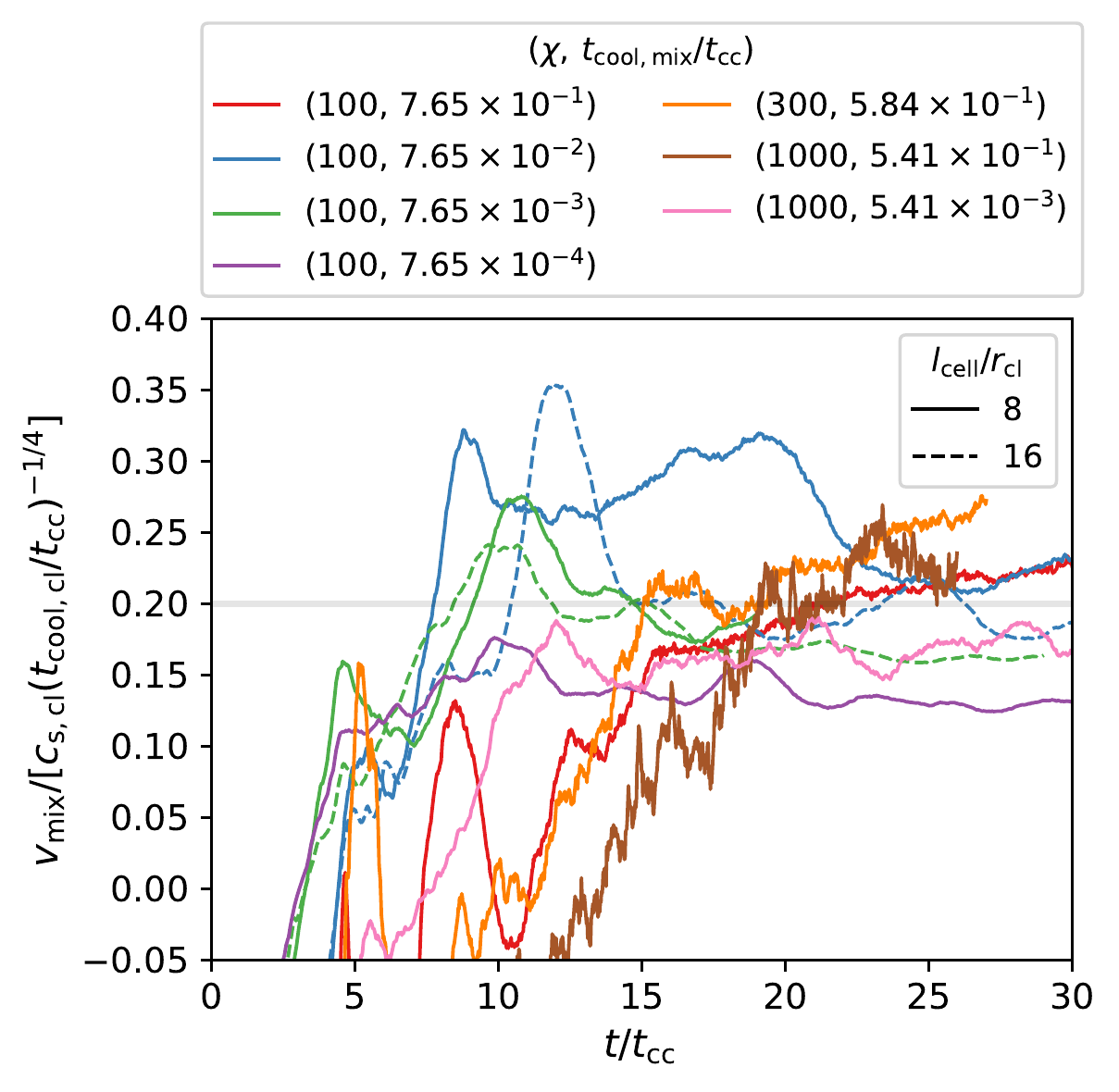}
  \caption{Evolution of the derived mixing velocity $\hat{v}_{\rm mix} \equiv \dot m_{1/3} / (A \rho_{\mathrm{wind}})$, normalized to the expected result $c_{\rm s,cl} (t_{\rm cool,cl}/t_{\rm cc})^{-1/4}$. After the initial transient (when the cloud is growing its tail), it settles  to $\hat{v}_{\rm mix} \sim 0.2 c_{\rm s,cl} (t_{\rm cool,cl}/t_{\rm cc})^{-1/4}$, to within a  factor  of 2.}
  \label{fig:mass_growth_velmix}
\end{figure}

\begin{figure}
  \centering
  \includegraphics[width=\linewidth]{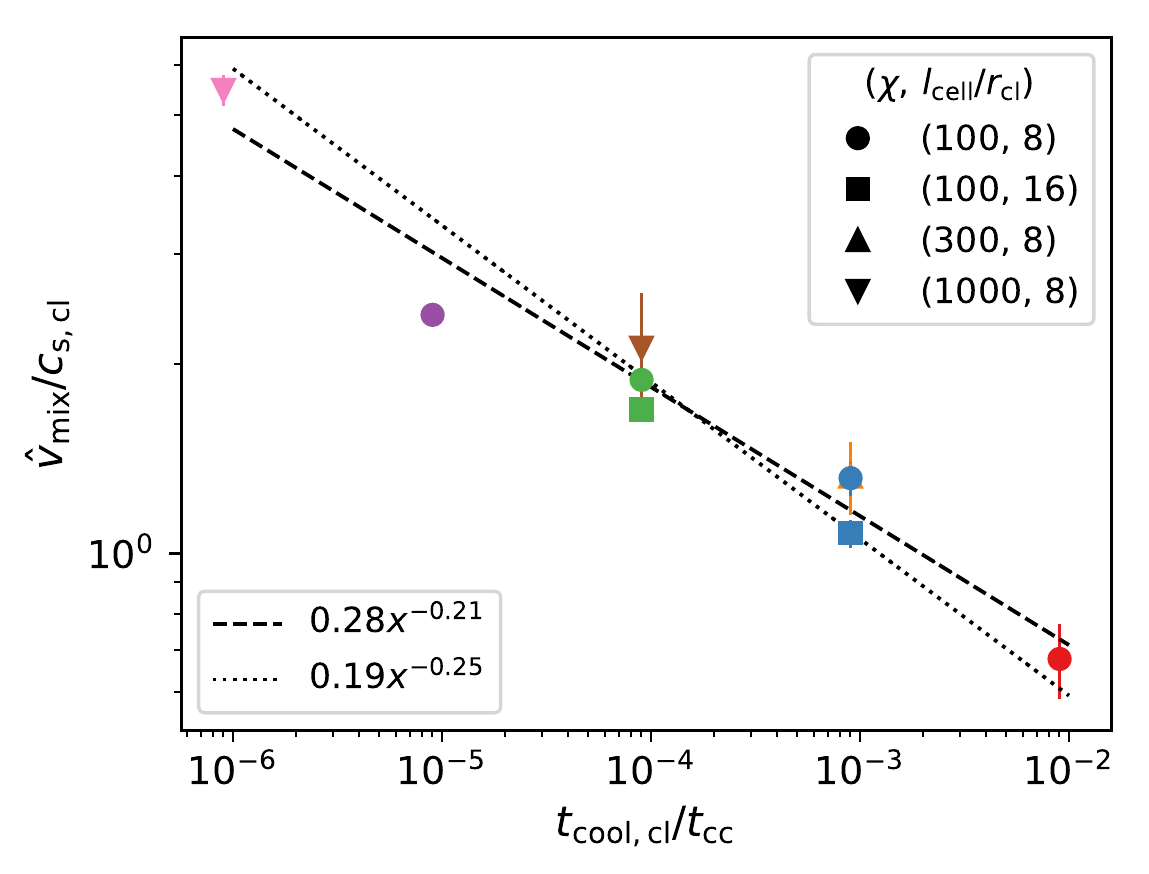}
  \caption{Derived mixing velocity averaged over $t\in [17,30]\,\tcc$ versus cooling time with the data points denoting the median, and the error bars the $16$th and $84$th percentiles. The two lines show a fit to the data points with the slope free and fixed to the theoretical expectation $\hat{v}_{\rm mix} \propto t_{\rm cool,cl}^{-1/4}$ (\textit{dashed} and \textit{dotted line}, respectively).}
  \label{fig:vmix_vs_tcool}
\end{figure}

\subsubsection{Characteristic mixing velocity}
\label{sec:vmix}

Using the measured surface area (at $T\sim T_{\mathrm{mix}}$), we can derive the mixing velocity via $v_{\mathrm{mix}}\sim \dot m / (A \rho_{\mathrm{wind}})$. Fig.~\ref{fig:mass_growth_velmix} shows this derived mixing velocity (using $A$ calculated numerically at the $T=T_{\mathrm{mix}}$ threshold).  
After the initial transient (when the cloud is growing its tail; $t\gtrsim 15\tcc$), $v_{\mathrm{mix}} \sim 0.2 c_{\mathrm{s,\cl}} (\tcoolrat{cl})^{-1/4}$ (marked with a gray line), as derived in \S~\ref{sec:analytics}. Thus, in Eq.~\eqref{eq:vmix_analytic}, $\alpha^{1/2}\approx 0.2$, or $ \alpha \approx 0.04$.  

In Fig.~\ref{fig:vmix_vs_tcool}, we show the derived mixing velocity $\hat{v}_{\rm mix} = \dot{m}_{1/3}/(A \rho_{\rm mix})$ as a function of $t_{\rm cool,cl}/t_{\rm cc}$. We find that  $v_{\mathrm{mix}} \propto (\tcoolrat{cl})^{-1/4}$, as in Eq.~\eqref{eq:vmix_analytic}. Note that the $\tcc$ dependence here is really a dependence on $t_{\mathrm{sc}}$ as discussed in \S~\ref{sec:analytics} and \S~\ref{sec:mass_growth_cause}. Plugging in numerical values into the fit obtained and using $\tcool{mix} \sim \chi \tcool{cl}$, we obtain for the mixing velocity
\begin{equation}
  \label{eq:vmix_numerical}
  v_{\mathrm{mix}}\sim 7\kms T_{\cl,4}^{1/2}\left( \frac{r_\cl}{r_{\mathrm{cl,crit}}}\right)^{1/4}\left( \frac{\chi}{100} \right)^{-1/4}
\end{equation}
which in conjunction with Eq.~\eqref{eq:rcrit} and $A \sim \chi r_{\rm cl}^{2}$ can be used to compute mass growth rates.

\section{Results: Dependence of the Mass Growth on Setup}
\label{sec:result_robustness}

In this section, we vary the simulation setup compared to the previous section and study the dependence of the mass growth on these altered conditions.
We investigate the mass growth when the wind varies (e.g., according to the CC85 solution) in \S~\ref{sec:adiabatic_wind}. We also study how  a less idealized initial cloud configuration (\S~\ref{sec:nosphere}), a higher Mach number wind (\S~\ref{sec:large_M}) and magnetic fields (\S~\ref{sec:magnetic_fields}) change our findings. We then turn to numerical effects, and show how our results depend on dimensionality (\S~\ref{sec:appendix_2d}), and numerical resolution (\S~\ref{sec:shattering}).

\subsection{Cold gas in an expanding wind}
\label{sec:adiabatic_wind}

\begin{figure}
  \centering
  \includegraphics[width=.95\linewidth]{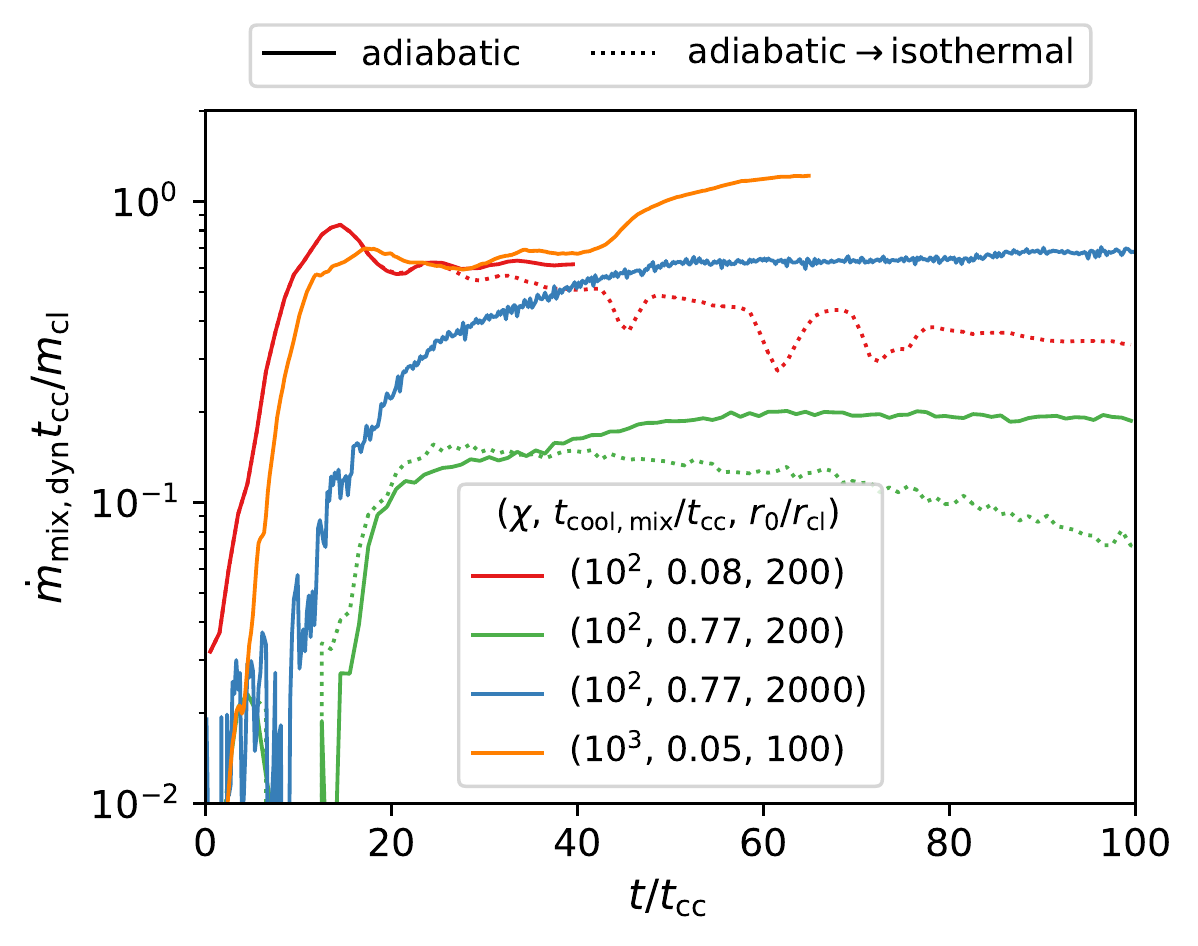}
  \caption{Temporal evolution of the mass growth in an expanding wind. The solid and dotted curves show the solutions in an adiabatically expanding wind and one where the background profile becomes isothermal for $r>r_0$, respectively. Mass growth is clearly slower in the isothermal regime.}
  \label{fig:mdot_expanding}
\end{figure}

\begin{figure}
  \centering
  \includegraphics[width=.95\linewidth]{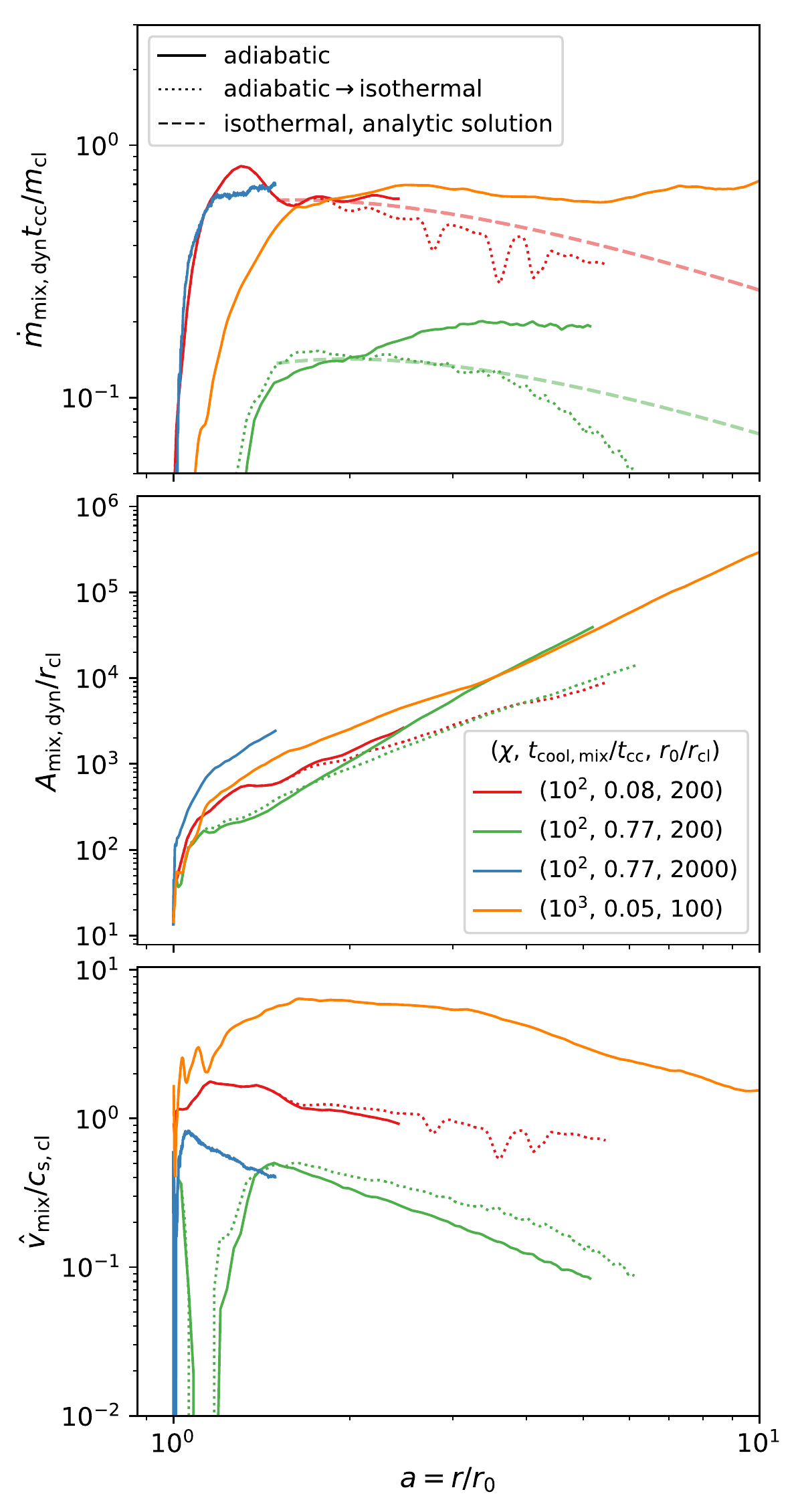}
  \caption{Mass growth, cold gas surface area, and inferred mixing speed (\textit{from top to bottom panel}) versus the `expansion factor' in an expanding wind. The solid and dotted curves show the solutions in an adiabatically expanding wind and one where the wind changes to isothermal. The thick dashed curves in the upper panel show the analytic solution for $\dot m(a)$ in an isothermal wind.  The reduction in mass growth in an isothermal wind is  primarily due to the reduced surface area of the cloud due to reduced cloud expansion.}
  \label{fig:expanding_vs_a}
\end{figure}

Figure~\ref{fig:mdot_expanding} shows the mass growth for three different runs in an adiabatically expanding wind (see \S~\ref{sec:method}). We used two different overdensities $\chi=(100,1000)$, and three different wind launch positions $r_0/r_{\rm cl}=(100,200,2000)$, and defined the cold gas to be $T < T_{\rm mix, dyn} = (T_{\rm cl} T_{\rm w,i})^{1/2} a^{-2/3}$ where $a(t)=r(t) / r_0$ is the scale factor of the simulation box, and $T_{\rm w,i}$ is the initial wind temperature at the launch radius $r_{0}$. This temperature scaling obeys the Chevalier-Clegg solution. 

The solid lines in Fig.~\ref{fig:mdot_expanding} show the mass growth in an adiabatically expanding wind. All the curves flatten out after some time as predicted in \S~\ref{sec:analytic_wind}. However, a lower value of $r_0$ leads to an overall lower $\dot m$ plateau. The choice of $r_0$ sets the scale height of the wind: a lower value corresponds to faster expansion and a more rapid drop in density with radius. In the more  rapidly expanding scenario, entrainment takes longer and more cold gas is lost to mechanical mixing during this phase. This leads to a lower surface area, and thus, a lower $\dot m$ in the entrained phase. This can be seen in the upper panel of Fig.~\ref{fig:expanding_vs_a}. In the lower panel of Fig.~\ref{fig:expanding_vs_a}, we show the derived mixing speed, which falls off with increasing radius as expected.

Both Figs.~\ref{fig:mdot_expanding} and \ref{fig:expanding_vs_a} also show simulations where we change the wind profile from adiabatic to isothermal for $r > r_{\mathrm{break}}= 1.5 r_0$ (dotted lines). An isothermal profile (i.e., where the wind entropy increases with radius) leads indeed to a slower mass growth as discussed in \S~\ref{sec:analytic_wind}. We also show the analytic solution to Eq.~\eqref{eq:mdot_entropy} for an isothermal wind profile with the boundary conditions $m(r_{\mathrm{break}})$ and $\dot m(r_{\mathrm{break}})$ fixed by the numerical results. It is a relatively good fit. 
From Fig.~\ref{fig:expanding_vs_a} it is clear that the reduced mass growth in an  isothermal wind is due to the reduced cloud surface area (since  the wind pressure falls more slowly with radius compared to an adiabatic wind). The mixing speed does not change significantly between the isothermal and adiabatic case.

\begin{figure}
  \centering
  \includegraphics[width=\linewidth]{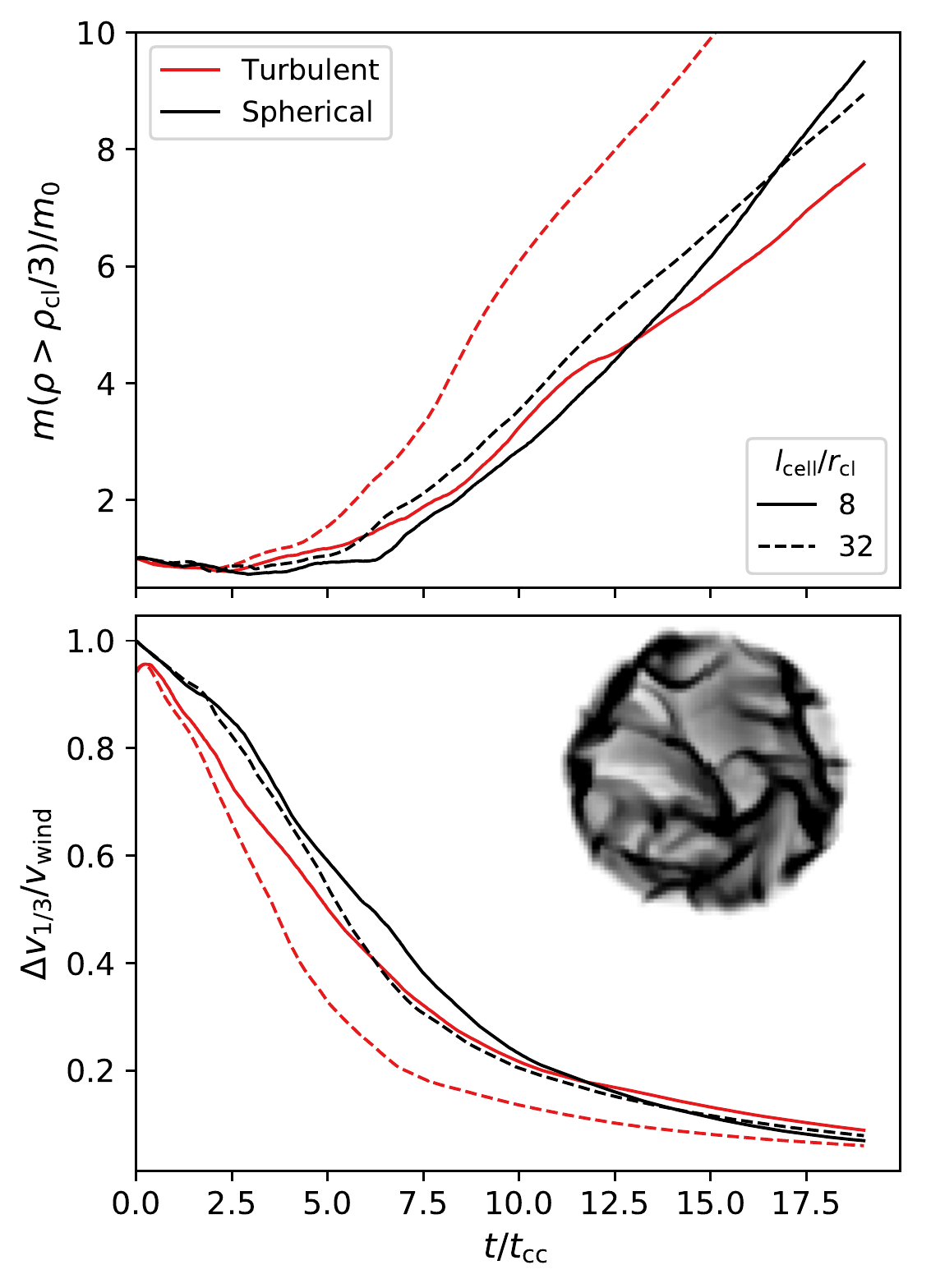}
  \caption{Mass and velocity evolution (\textit{upper} and \textit{lower panel}, respectively) of a spherical, and turbulent setup with different resolution. The \textit{inset} in the lower panel shows a density slice of the turbulent initial conditions.}
  \label{fig:evolution_turbulent_ic}
\end{figure}

\begin{figure}
  \centering
  \includegraphics[width=\linewidth]{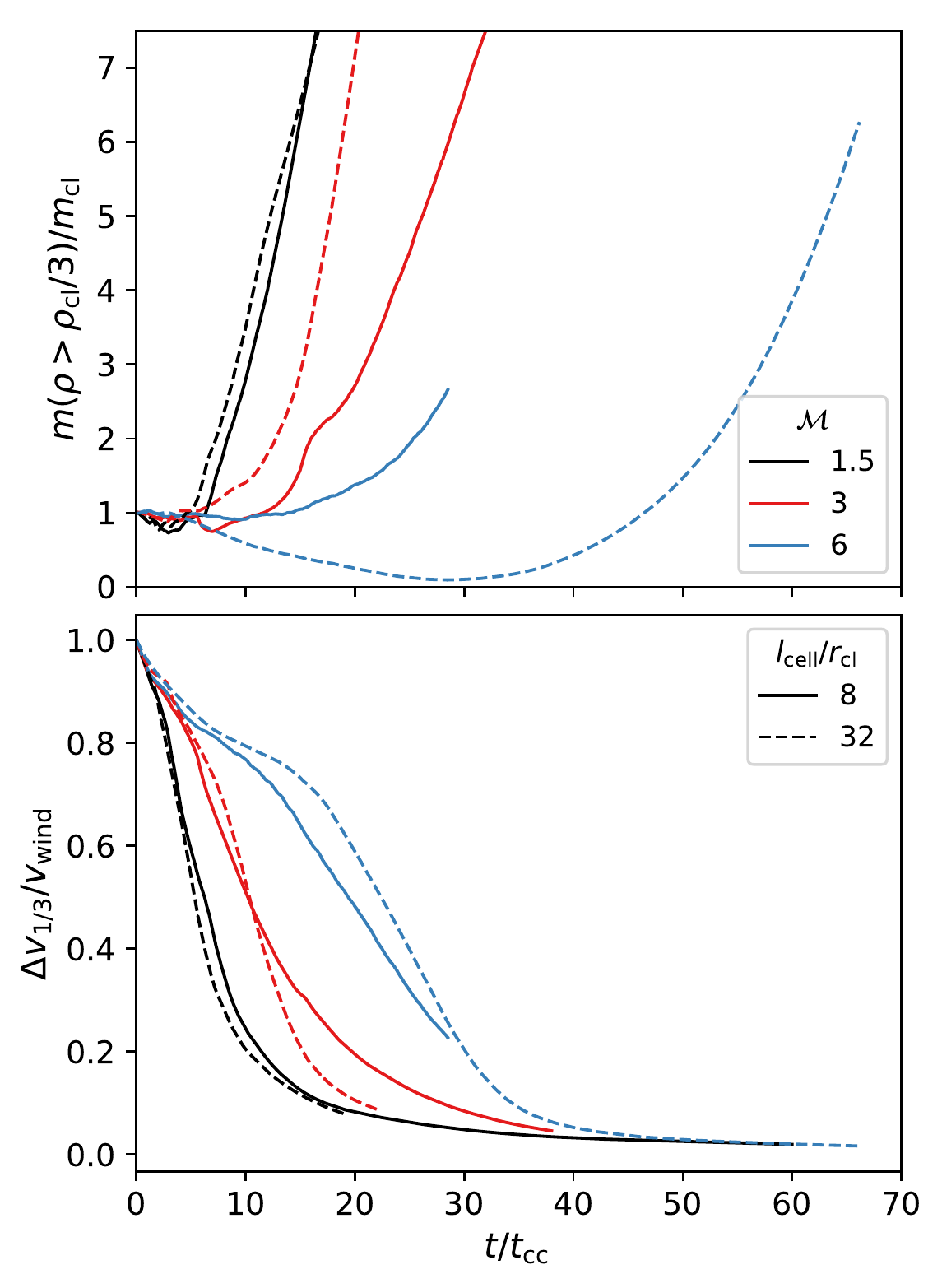}
  \caption{Mass (\textit{upper panel}) and velocity evolution (\textit{lower panel}) with different Mach numbers, and resolution.}
  \label{fig:mach}
\end{figure}

\subsection{Non-spherical setup}
\label{sec:nosphere}

The cloud geometry potentially influences its destruction and acceleration timescale \citep[see, e.g., the recent work by][for a detailed study of this effect]{Banda-Barragan2019}. To assess the impact of cloud geometry, we compare our clouds with spherically symmetric initial conditions against a set of simulations using a turbulent cloud. To do this, we extracted a spherical region out of a turbulent $128^3$-cell box which was stirred with a Mach number of $\mathcal{M}_{\mathrm{w}}=3$ and a correlation length of $1/20$ of the box size. We then scaled the density so that the mean density contrast is given by $\chi$.
A similar procedure was also followed by \citet{Schneider2016} and \citet{Liang2018} to generate non-spherical initial conditions.

Figure~\ref{fig:evolution_turbulent_ic} shows the mass and velocity evolution in the upper and lower panel, respectively. In the inset of the lower panel we also show the initial conditions of the cloud. Our findings show that neither the mass growth nor the velocity evolution are heavily affected by the altered initial conditions. In fact, the mass growth is slightly larger in the turbulent setup. This is due to the enlarged (initial) surface area leading to faster growth. Because of this enlarged growth rate, the corresponding momentum transfer is also larger, leading to a (slightly) faster entrainment in the turbulent setup.

At first sight, these results appear to disagree with the study of \citet{Schneider2016} who find a faster disintegration of the turbulent cloud compared to the spherically symmetric cloud. However, they study the $\tcoolrat{mix} > 1$ regime, corresponding to mass loss rather than growth. Thus, in their case the larger surface area of the turbulent cloud implies it undergoes faster destruction, just as in our case it grows faster.

\subsection{Higher Mach number flows}
\label{sec:large_M}

Figure~\ref{fig:mach} shows the mass and velocity evolution of the cold gas if we increase the flow Mach number from our default choice of $\mathcal{M}_{\mathrm{w}}=1.5$. The stronger shock creates a higher pressure confining wind, which compresses the cloud considerably in the direction orthogonal to the flow direction, increasing the cloud overdensity. The higher overdensities $\chi$ lead to longer cloud-crushing ($\tcc \propto \chi^{1/2}$) and entrainment ($t_{\rm drag} \propto \chi$) times. Growth still occurs, but it is delayed as a result of the longer entrainment time, as growth is suppressed during the entrainment phase. This is consistent with our previous wind tunnel simulations of higher overdensity gas (see Fig. \ref{fig:mass_growth_new_setup_multiplot}). 

These results are consistent with previous work. \citet{Scannapieco2015a}, who operated in a regime where clouds do not survive, found $t_{\rm  cc}  \propto \sqrt{1 + \mathcal{M}_{\mathrm{w}}}$ in a supersonic wind. Our results suggest that the the entrainment time is $t_{\rm drag} \propto  1+\mathcal{M}_{\mathrm{w}}$ in a supersonic wind. If we identify $\chi \propto  ({1 +\mathcal{M}_{\mathrm{w}}})$, these results agree\footnote{See \citet{Scannapieco2015a} for more discussion of the Mach number scalings of the shock. Note that while the normal shock at the head of the cloud increases the pressure there by a factor of $\sim 1+ \mathcal{M}_{\mathrm{w}}^{2}$, the downstream portion of the cloud sees an oblique shock  where the  pressure only rises as  $1+\mathcal{M}_{\mathrm{w}}$.}. However, higher resolution studies in the $t_{\mathrm{cool,mix}}/\tcc < 1$ regime with a larger set of $(\mathcal{M}_{\mathrm{w}},\,\chi)$ pairs are required in order to address this issue in detail. The compressed cloud requires higher resolution, and our simulations with $\mathcal{M}_{\mathrm{w}}> 1.5$ are not yet numerically converged, so no firm conclusions can be drawn at this point. 

\subsection{Magnetic fields}
\label{sec:magnetic_fields}

\begin{figure}
  \centering
  \includegraphics[width=.95\linewidth]{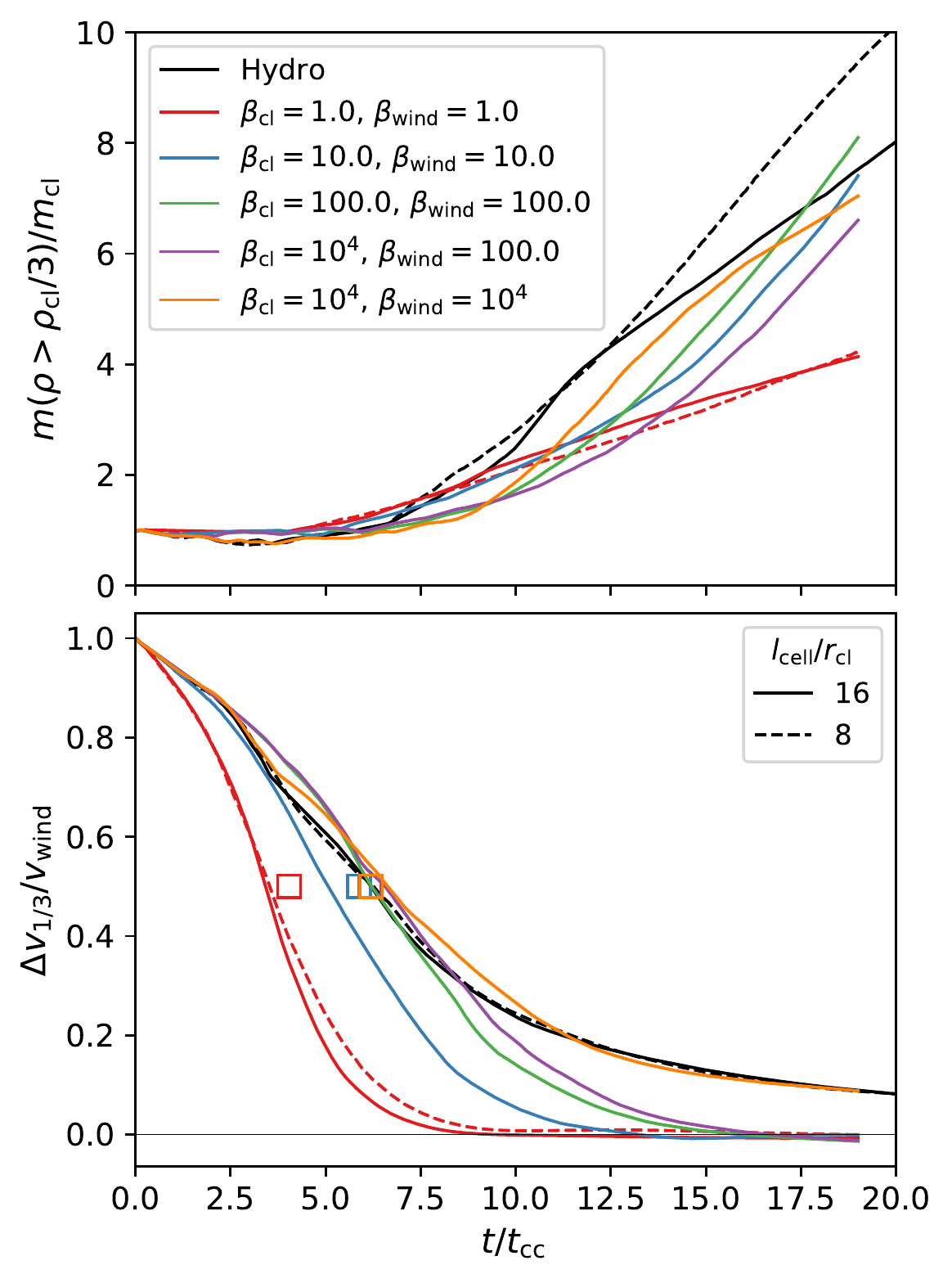}
  \caption{Evolution of the cold gas mass and the velocity difference (upper and lower panel, respectively) for simulations with $(\chi,\,\tcoolrat{mix})\sim (100,\,0.08)$ including magnetic fields. In the lower panel, we marked the $\Delta v = v_{\mathrm{wind}}/2$ point for the hydrodynamical simulation, and re-scaled it for the other runs according to Eq.~\eqref{eq:t_drag_mhd}. It corresponds well to  the  simulation results. 
    }
  \label{fig:evolution_multi_mhd}
\end{figure}

\begin{figure}
  \centering
  \includegraphics[width=\linewidth]{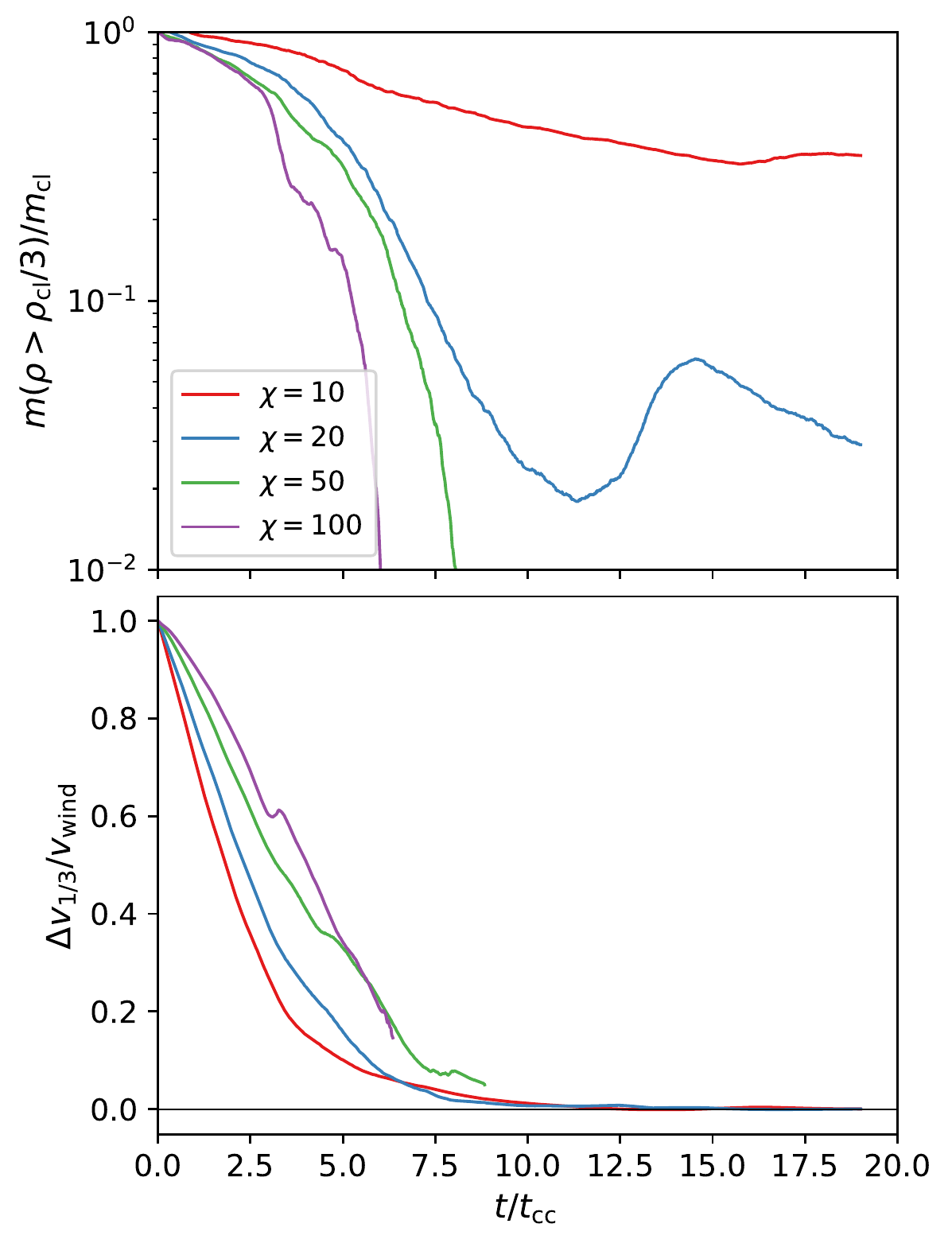}
  \caption{Mass and velocity evolution for simulations with $\beta_\cl = \beta_{\mathrm{w}} =1$ magnetic fields but without cooling. Lower overdensity clouds can be entrained, in rough agreement with Eq.~\eqref{eq:mhd_entrainment}. The upturn at $t \sim 10 t_{\rm cc}$ for the $\chi=20$ case is a transient due to gas compression, not radiative cooling.}
  \label{fig:mhd_nocool}
\end{figure}

\begin{figure*}
  \centering
  \begin{tabular}{ll}
\begin{minipage}{.8\linewidth}
\centering
    \begin{subfigure}[b]{0.5\textwidth}
      \includegraphics[width=\textwidth]{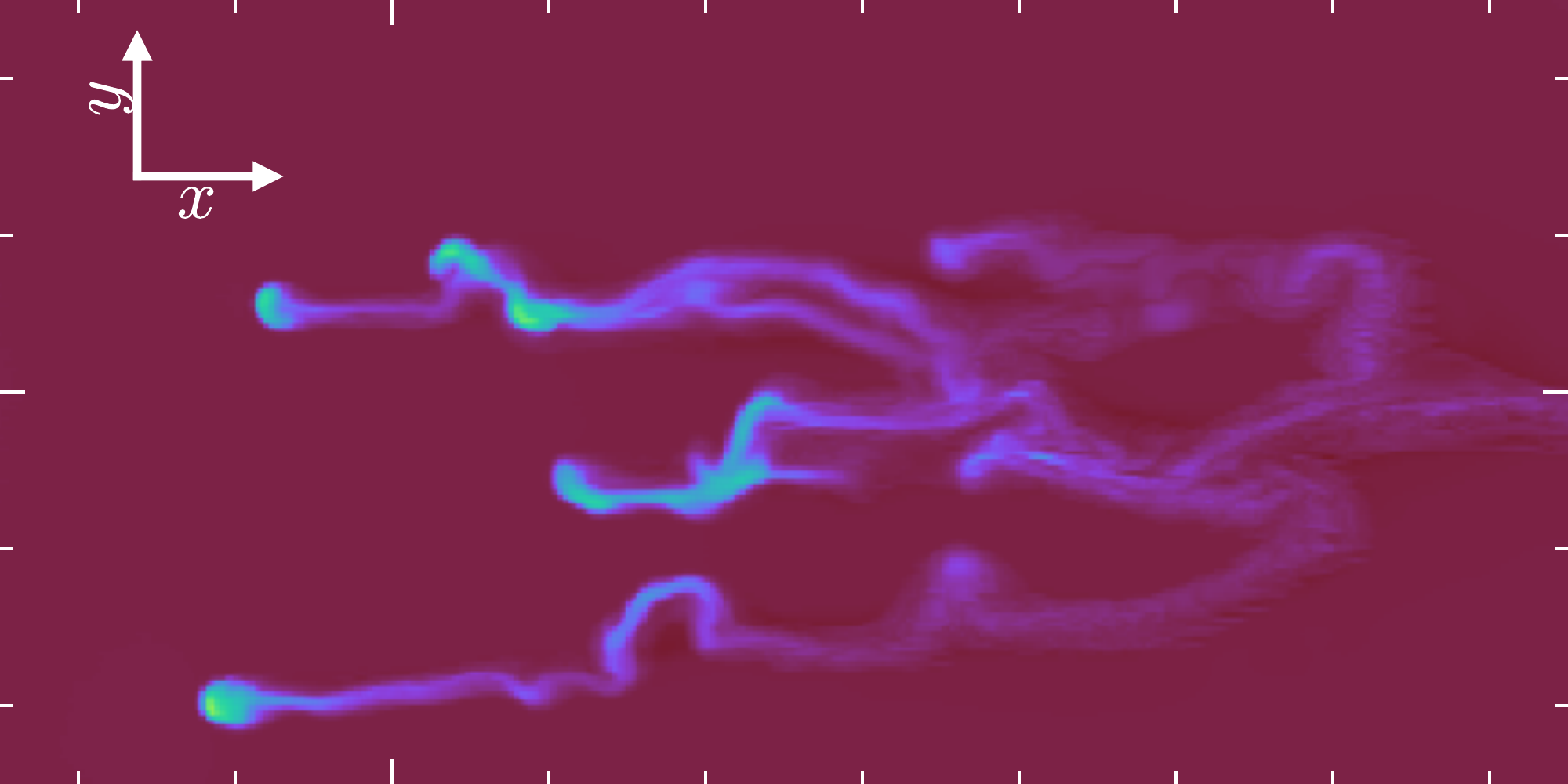}
    \end{subfigure}%
    \begin{subfigure}[b]{0.5\textwidth}
      \includegraphics[width=\textwidth]{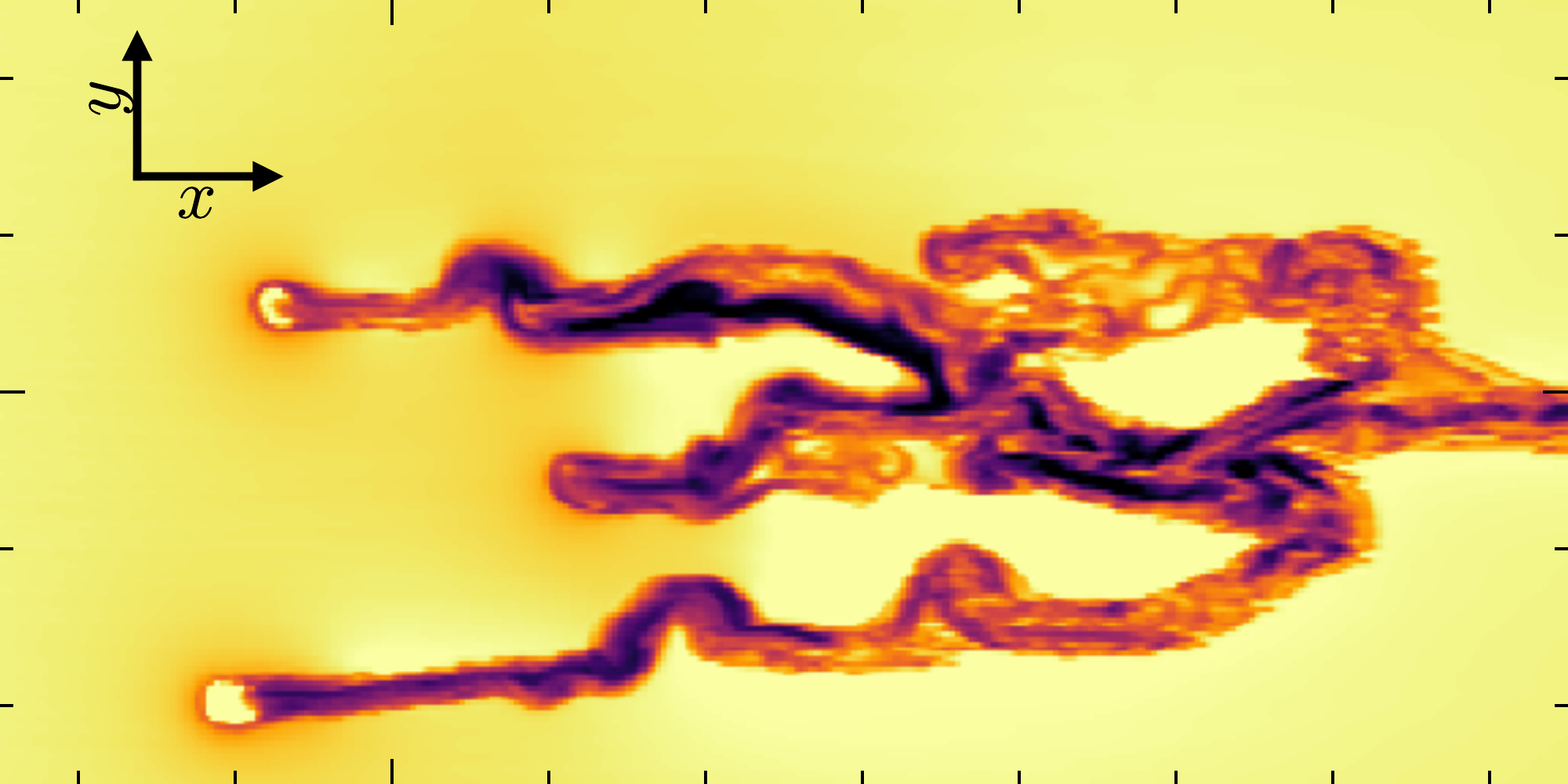}
    \end{subfigure}\\[2mm] %
    \begin{subfigure}[b]{0.5\textwidth}
      \includegraphics[width=\textwidth]{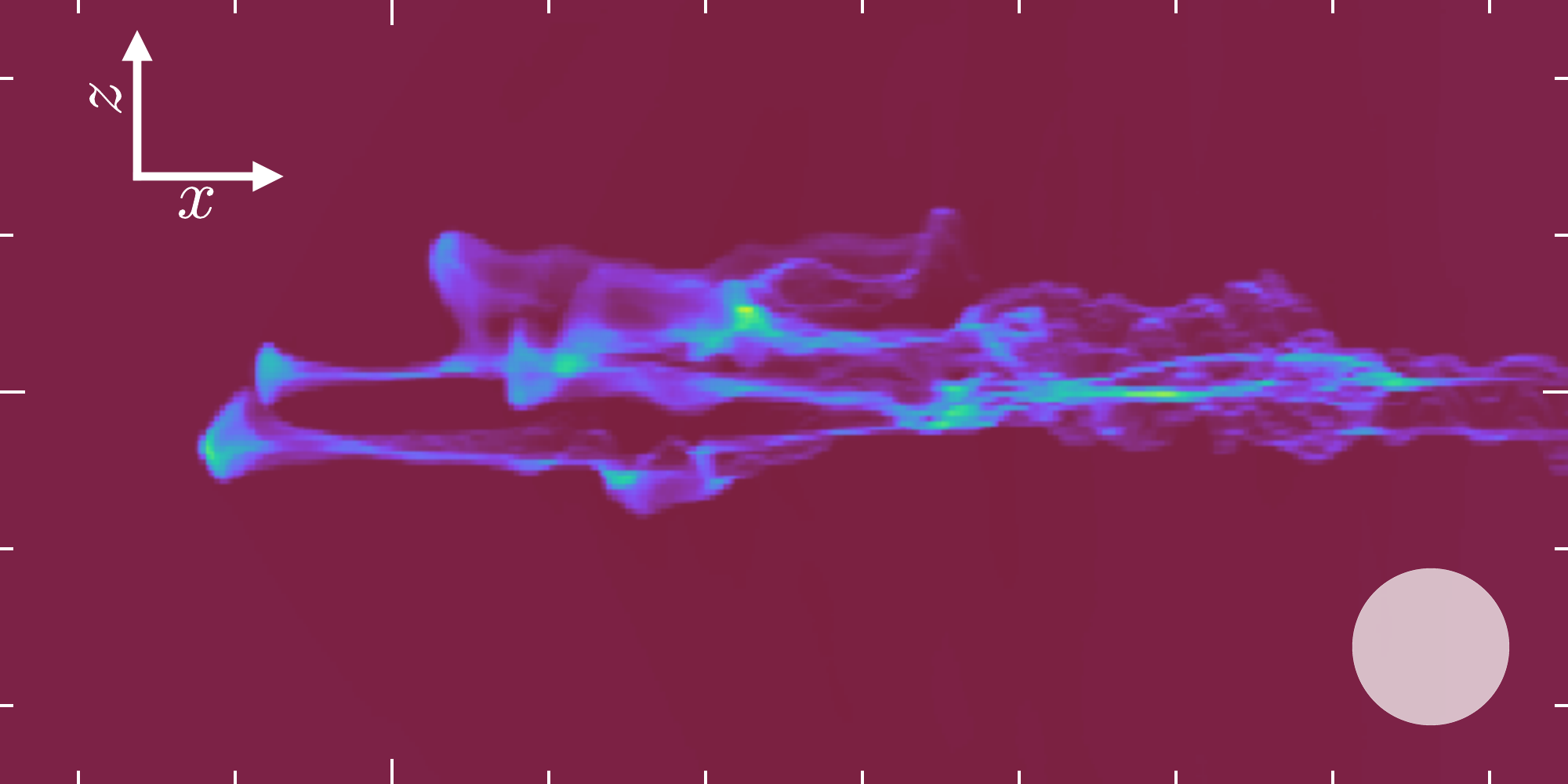}
    \end{subfigure}%
    \begin{subfigure}[b]{0.5\textwidth}
      \includegraphics[width=\textwidth]{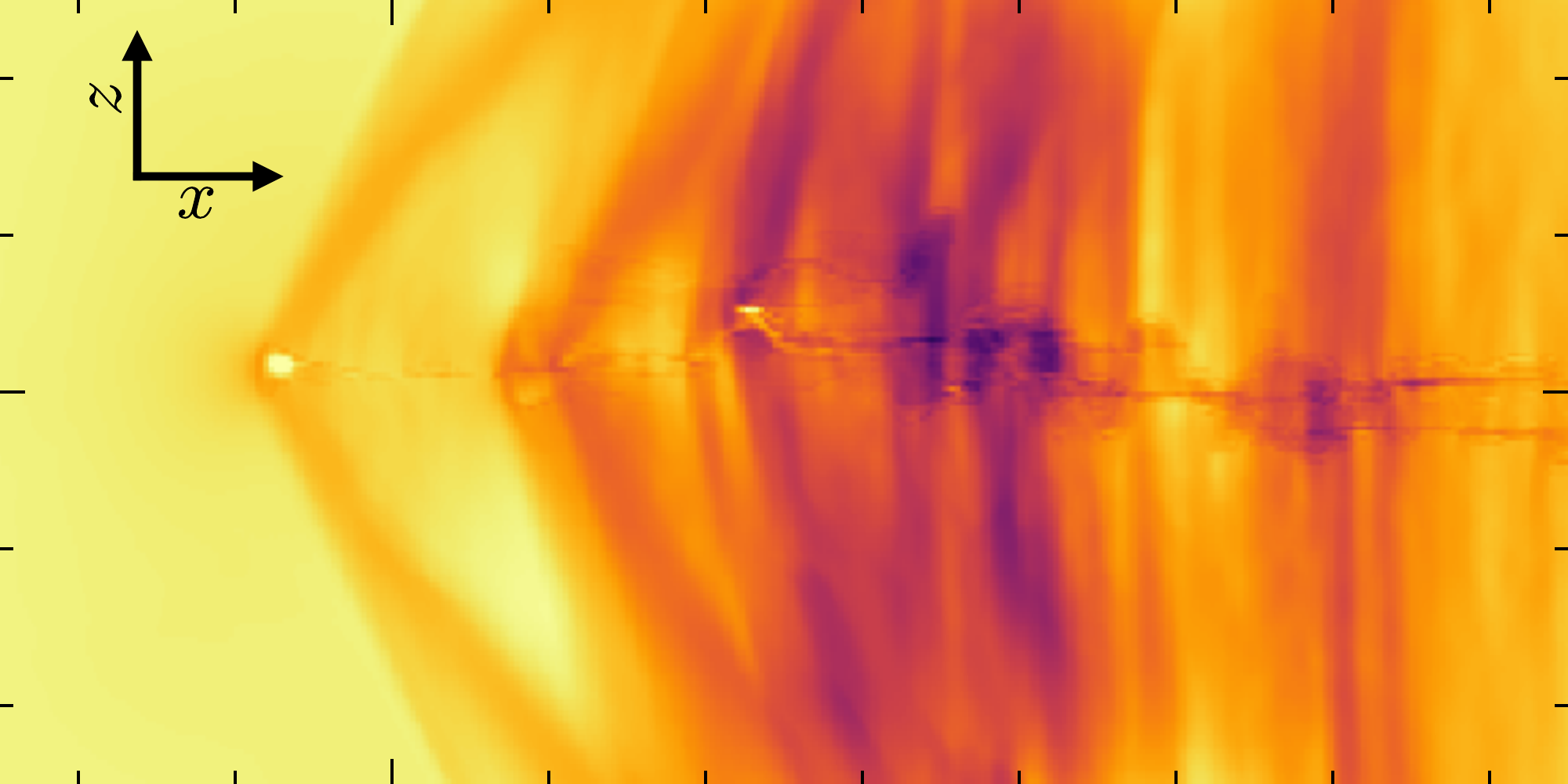}
    \end{subfigure}

\end{minipage}
&
\begin{minipage}{.2\linewidth}
  \centering
  \includegraphics[width=\textwidth]{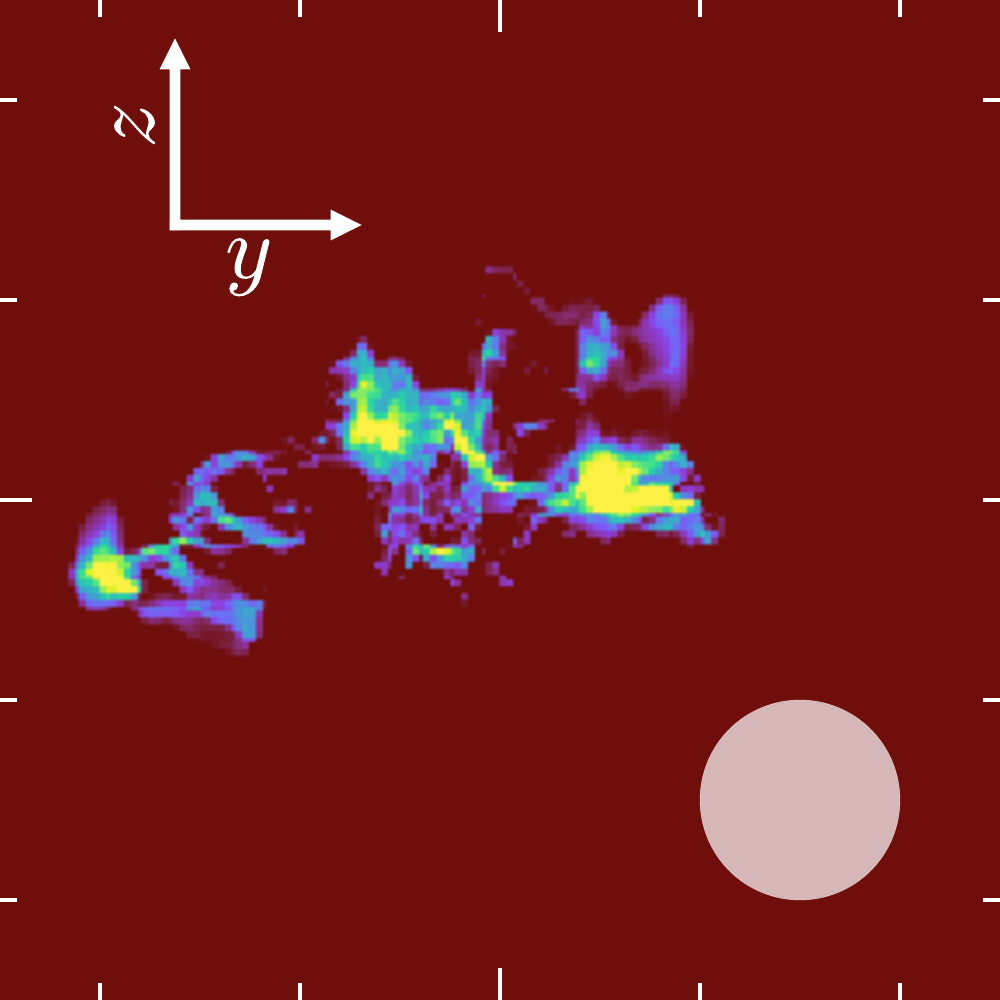}\\[1.25cm]
  \includegraphics[width=\textwidth]{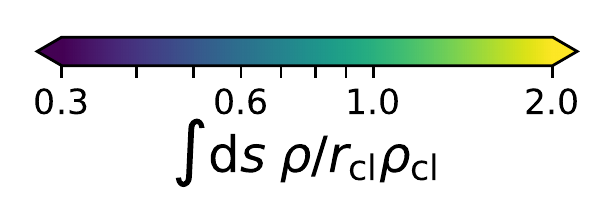}\\
  \includegraphics[width=\textwidth]{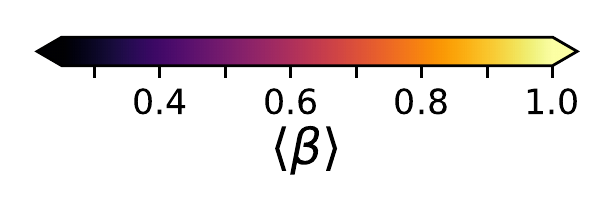}\\
\end{minipage}
\end{tabular}

  \caption{Three projections of the density (left and rightmost column; the circle in the lower left and upper right panels show the initial size of the cloud), and the average $\beta$ of a slab with $r_{\cl}$ thickness through $(0,0,0)$ (central column), where $x$  is the streamwise direction, and the $B$-field is in the y direction. Shown is the snapshot at $t=6\tcc$ of the $\beta_\cl= \beta_{\mathrm{wind}}=1$ run with $(\chi,\tcoolrat{mix})=(100,\,0.08)$.}
  \label{fig:mhd2d}
\end{figure*}

In Fig.~\ref{fig:evolution_multi_mhd} we show the mass and velocity evolution for the runs including magnetic fields. The magnetic field is orthogonal to the wind direction, so the cloud sweeps up magnetic field (and undergoes `magnetic draping') as it encounters the wind. We varied the magnetic field strength from $\beta_\cl = 1$ to $\beta_\cl = 10^4$ inside the cloud, and mostly choose $\beta_\cl = \beta_{\mathrm{wind}}$. For comparison, we show the results without magnetic fields in black. Clearly, the runs including magnetic fields show large mass growth. In fact, for all runs except the one featuring $\beta_\cl = \beta_{\mathrm{wind}} = 1$, the final mass after $\sim 20\tcc$ is comparable or even larger than the hydrodynamical simulations, to within a factor of two.\footnote{Note, that the pulsations in volume (cf. Fig.~\ref{fig:pulsations_evolution}) are somewhat suppressed compared to the pure hydro case.} This seems to be at odds with the findings of \citet{Gronnow2018} who recently claimed a significantly reduced condensation rates even for small magnetic field strengths ($\beta\sim 600$). However, these authors were studying the initial part of the evolution and followed the cloud evolution only for $t \lesssim 3\tcc$. 

In the lower panel of Fig.~\ref{fig:evolution_multi_mhd}, we show the velocity difference between the hot and the cold gas. Due to magnetic drag, the cloud acceleration is increased, leading to a faster decline of $\Delta v_{1/3}$. We expect the drag time to be reduced by \citep{Dursi2008,McCourt2015}
\begin{equation}
  \label{eq:t_drag_mhd}
  \frac{t_{\mathrm{drag}}^{\mathrm{mhd}}}{t_{\mathrm{drag}}^{\mathrm{hydro}}} = \left( 1 + \frac{2}{\beta_{\mathrm{wind}}\mathcal{M}^2} \right)^{-1}.
\end{equation}
In order to check the validity of this estimate, we rescale the time when $\Delta v_{1/3}=v_{\mathrm{wind}}/2$ in the hydro-only run according to Eq.~\eqref{eq:t_drag_mhd}, and mark the result in the corresponding color in the lower panel of Fig.~\ref{fig:evolution_multi_mhd}. It fits well to the simulation results. 

From Eq.~\eqref{eq:t_drag_mhd} and by demanding that $t_{\mathrm{drag}}^{\rm mhd} < t_{\mathrm{destruction}} \sim \xi \tcc$ (where $\xi \sim$ a few) we can show when magnetic field drag  alone will lead to entrainment, namely if:
\begin{equation}
  \label{eq:mhd_entrainment}
  \chi < \chi_{{\mathrm{crit}}} \equiv 9 \left(\frac{\xi}{3} \right)^2 \left( 1 + \frac{2}{\beta_{\mathrm{w}}\mathcal{M}} \right)^2\;.
\end{equation}
This means for $\chi > 100$ one would require very large magnetic field strengths ($\beta_{\mathrm{w}} < 1$) to accelerate clouds with $r_\cl \lesssim r_{\mathrm{cl,crit}}$.

Figure~\ref{fig:mhd_nocool} shows MHD simulations where we vary $\chi$ to determine the critical overdensity $\chi_{\rm crit}$  when entrainment by magnetic fields alone is feasible.
The simulations are adiabatic, which effectively corresponds to $r \ll r_{\rm cl,crit}$. The results agree reasonably well with our estimates: Eq.~\eqref{eq:mhd_entrainment} yields $\chi_{\mathrm{crit}}\sim 32$ for the simulated parameters, which roughly divides the cases where the cloud is destroyed and some fraction of the gas is entrained. Note, however, that the cold gas mass is still decreasing in these adiabatic simulations; they should stabilize (and potentially increase) once radiative cooling is introduced. These results suggest that in the presence of magnetic  fields, and if $\beta, \chi$ are both small, the critical scale $r_{\rm crit}= r_{\rm crit}(\beta, \chi)$ can be somewhat smaller that in the hydrodynamic case. A suite of higher resolution simulations would be needed to carefully  map out this boundary. However, given that this correction is small for dense clouds ($\chi \sim 100-1000$), and that the hydrodynamic $r_{\rm crit}$ is already very small, this is not  pressing.

\begin{figure}
  \centering
  \includegraphics[width=\linewidth]{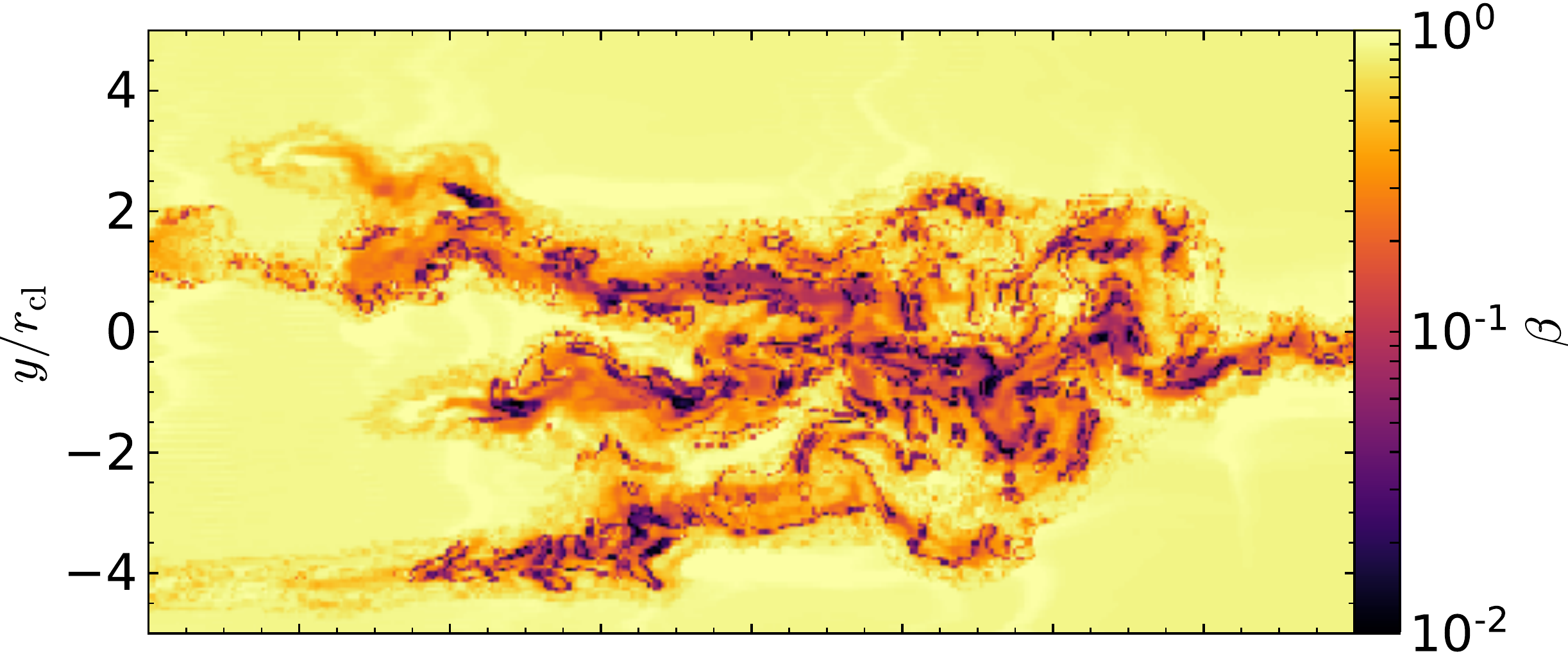}\\
  \includegraphics[width=\linewidth]{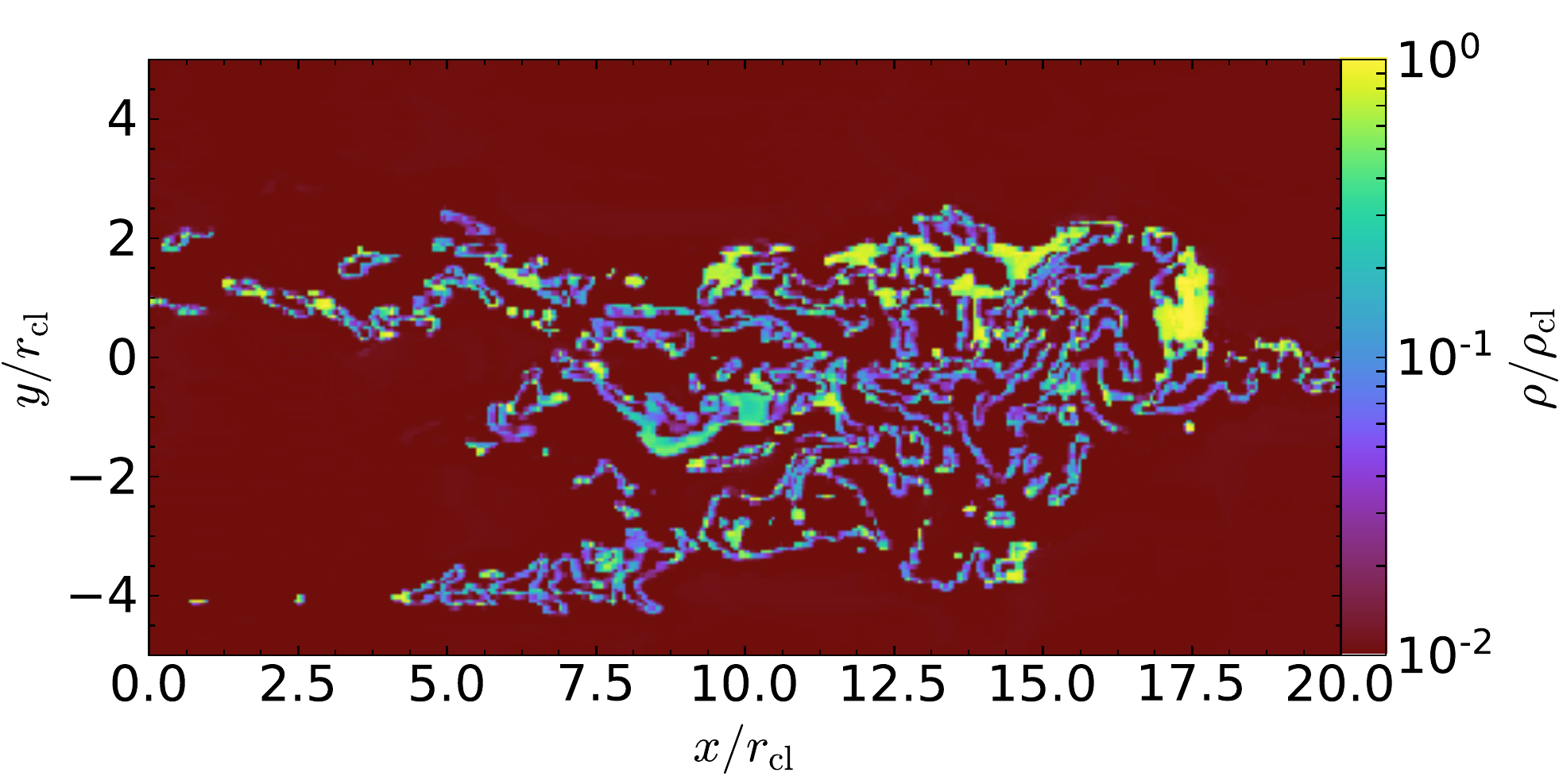}
  \caption{Plasma $\beta$ and density slice \textit{(top and bottom panel, respectively)} of the MHD simulation shown in Fig.~\ref{fig:mhd2d} at $t\sim 18\tcc$, i.e., long after the cloud has  been entrained. Note the small-scale structure in the cold gas leading to a vastly increased surface area compared to the hydro-only run, and the magnetic field is amplified due to compression. See \S~\ref{sec:magnetic_fields} for details.
  }
  \label{fig:mhd_dens_slice}
\end{figure}

Figure~\ref{fig:mhd2d} shows that the morphology of the cloud is dramatically changed due to the presence of magnetic fields. While in the pure-hydro case (cf. figure~2 in Paper I) a clear head-tail structure forms with a single, continuous tail, in the MHD case shown here the cloud broke up into three pieces which form their individual tails. The cold gas has a much stringier appearance, and is aligned with the magnetic field. The filamentary morphology can be understood from the anisotropy of gas motions once MHD forces come into play \citep{Xu2019}; it is much easier for fluid elements to slide along field lines rather than move perpendicular to it. Such cold gas morphology is widely observed in the ISM. This `stringy' structure leads to a much larger apparent surface area compared to a cloud of similar mass in a hydrodynamical simulation.
Fig.~\ref{fig:mhd2d} also shows that the cloud is 
highly magnetized due to compressional field amplification as the $\beta \sim 1$ wind gas cools, leading to $\beta\lesssim 0.1$. The complex morphology can be further seen in Fig.~\ref{fig:mhd_dens_slice} we show a slice through $z=0$ of the of the simulation with $\beta_\cl = \beta_{\mathrm{wind}} = 1$ when the cloud is entrained ($t\sim 18\tcc$). The individual tails do not coagulate but fold onto themselves instead. $\beta$ values as low as $\lesssim 10^{-2}$ can be reached due to magnetic compression.

How can we understand the similar mass entrainment rates $\dot{m}$ for the hydrodynamic and MHD cases despite the very different morphology? Is it coincidental? We have run resolution studies and (similar to the hydrodynamic case) not found any resolution dependence in the hot gas entrainment rate. As we have seen, magnetic fields  have a number of effects: (i) they enhance acceleration via magnetic drag; (ii) they suppress the KH instability via magnetic tension; (iii) they drastically change appearance by promoting a stringier, more field-aligned cold gas morphology; (iv) they provide magnetic pressure support,  thus reducing the cold gas overdensity $\chi$. The enhanced acceleration is an order unity correction for the parameters we consider (see Eq.~\eqref{eq:t_drag_mhd}),  which leads to order unity changes in the  cloud length (and hence surface area), so  (i) does not  significantly impact $\dot{m}$. Since radiative cooling, rather than the KH  instability, is responsible for drawing in hot gas, (ii) should not affect hot gas entrainment once the cloud is largely comoving with the wind (see, however, further discussion below).  While (iii) leads to much more small scale structure and greatly enhanced apparent surface area, we have already seen similar effects in  progressively higher resolution hydrodynamic simulations with  little impact on mass accretion rates; as discussed in \S~\ref{sec:shattering} what matters is the projected or `effective' surface area, which remains similar (and is of order $\sim \pi r_{\cl} l$ where $l  \sim v_{\rm wind} t_{\rm drag} \sim \chi r_{\rm  cl}$ is the cloud length after entrainment). The projected surface area would only change significantly if the cloud length changed (i.e. the drag time changed significantly) or the cloud spread out in cross-section. The latter happens only in runs  where cooling is inefficient and eventually leads to the cloud's destruction. As we have seen, cooling creates pressure gradients  which `focus' cooling gas onto the tail and keep the cross-sectional area roughly constant (cf. \citetalias{Gronke2018}). 

Point (ii) deserves more careful consideration. \citet{Ji2018} find in plane parallel simulations of the radiative KH instability that magnetic fields are shear amplified in the mixing layer to equipartition with turbulence, so that magnetic tension stabilizes the mixing layer and suppresses the hot gas entrainment rate. This is inconsistent with the discussion above. There are two alternatives: (a) $v_{\rm mix}$ {\it is} indeed suppressed, but compensated by the much larger surface area in the magnetic case\footnote{Recall that we have no independent means of measuring $v_{\rm mix}$; we can only measure $\dot{m}$ and the area $A$, and the latter quantitiy is unfortunately ambiguous.}. In this case, the close match with hydrodynamic entrainment rates is a coincidence. (b) We do not resolve the turbulent dynamo in the mixing layer. B-field amplification and thus suppression of mixing is much stronger at higher resolution, even though we do not see any resolution dependence in our current setup. In this case, mass growth would be much less. Neither of these possibilities vitiate our conclusion that clouds continue to survive in the MHD case. At the same time, we note that macroscopic clouds can have quite different behavior from a plane parallel mixing layer due to geometrical effects and the existence of a characteristic lengthscale. For instance, it is well known both from linear theory and simulation studies that magnetic fields suppress the adiabatic KH instability \citep{chandrasekhar61,miura82,jones97,ryu00}. However, adiabatic MHD cloud-crushing simulations -- which do not have to resolve a thin radiative mixing layer -- do not show cloud survival. The cloud is destroyed. 

As for (iv), Fig. \ref{fig:mhd_dens_slice} shows that a large fraction of the volume is in underdense gas, at only $\sim 10\%$ of the initial cloud density. This underdense,  low $\beta$ gas is supported by magnetic pressure, which is compressionally amplified when wind material cools and compresses into cold gas\footnote{The amplification is inconsistent with magnetic draping, where the magnetic pressure reaches equipartition with ram pressure. In our case, the strong field persists even when the cloud  is entrained.}. 
In fact, Fig. \ref{fig:mhd_dens_slice} is deceptive: while low density cold gas dominates by volume, high density cold gas dominates by mass. The mass-weighted cold gas has an overdensity comparable to the hydrodynamic case. Thus, since the size and density (and hence cooling time) of cloud material is not very different from the hydrodynamic case, it is  unsurprising that mixing velocities and mass growth rates are similar. 

At the same time, it is important to recognize that low-density, non-thermally supported gas dominates the areal covering fraction (as is evident from Fig.~\ref{fig:mhd_dens_slice}) and would dominate observations of cold gas in the CGM obtained by quasar spectroscopy. This may explain the very low cold gas densities inferred from photoionization modeling in the COS survey \citep{Werk2014}. {\it Such low-density gas is not representative of CGM gas by mass, and would lead to incorrect inferences about the abundance and physical state of cold CGM gas.} This argument also applies to other situations where the cold gas has a range of densities (e.g., due to spatially varying turbulence or cosmic ray pressure). The volume-filling low-density portion will be preferentially picked up by random lines of sight in quasar spectroscopy, which is in fact not representative of most CGM gas. In the future, this can be tested by emission line observations, which are sensitive to the dense component. 

\subsection{Comparison with $2D$ simulations}
\label{sec:appendix_2d}
\begin{figure}
  \centering
  \includegraphics[width=0.95\linewidth]{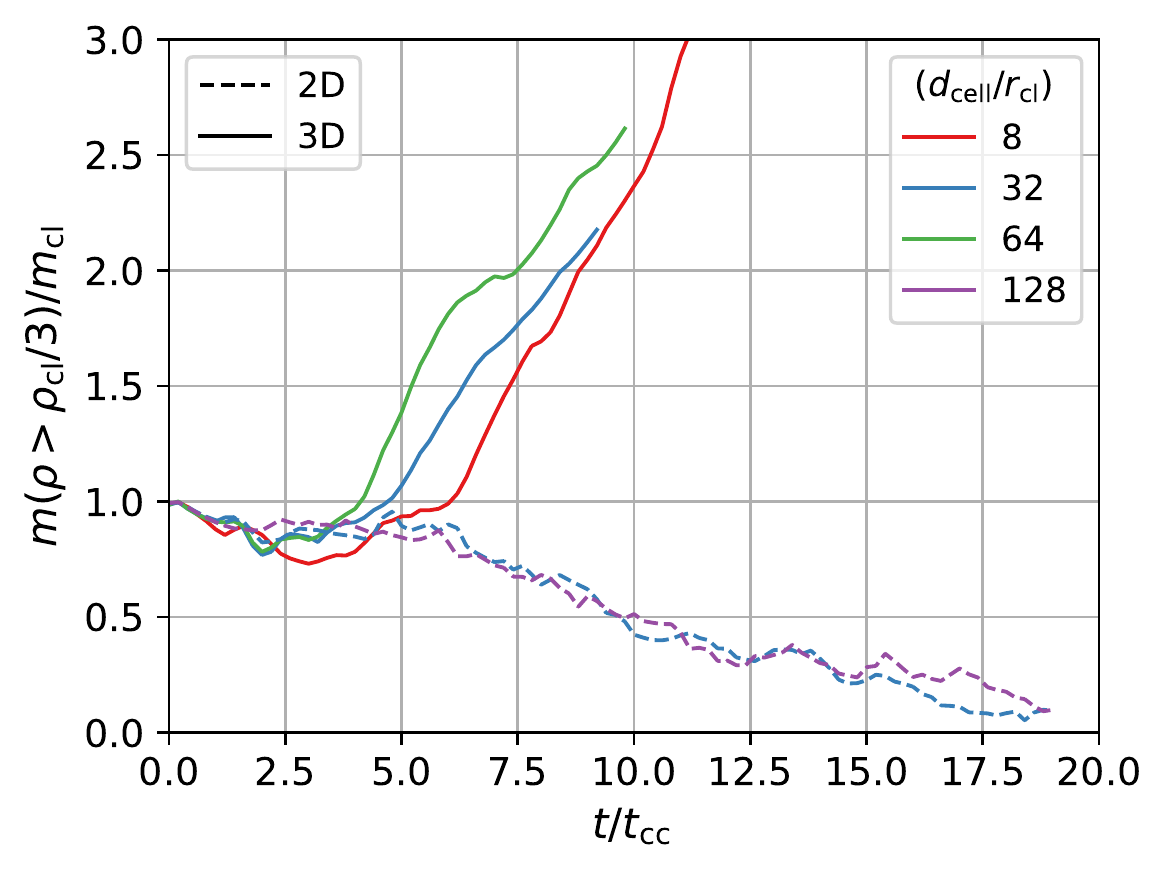}
  \caption{Mass evolution for $2D$ and $3D$ simulations with $\tcoolrat{mix}\sim 0.08$ and an overdensity of $\chi \sim 100$.}
  \label{fig:2d_comparison_mass}
\end{figure}

Figure~\ref{fig:2d_comparison_mass} shows a comparison of the cold gas evolution for 2D and 3D simulations with $\tcoolrat{mix}\sim 0.08$ and $\chi \sim 100$. The two behave quite differently: the 2D simulations show a decrease in cold gas mass and eventual cloud destruction, while our 3D setup shows the growth in mass as previously discussed. 
This highlights crucial differences between 2D and 3D simulations, likely related to the different behavior of vorticity and turbulence in 2D and 3D. It is therefore perilous to make inferences about the survival of clouds from 2D simulations. The $2$D runs also show quite different cloud morphology compared to the $3$D ones; they show more small scale structure due to the mixing and destruction of the cloud by Kelvin-Helmholtz instabilities. 

These results show that $r_{\rm cl, crit}$ (cf. Eq.~\eqref{eq:rcrit}) is different in 2D simulations, thus likely leading to mass loss in simulations where clouds should otherwise survive.  
Note, however, once $r_{\rm cl} \gg r_{\rm cl,crit}$, this distinction is less important, and  2D and 3D simulations show similar growth rates.

\begin{figure}
  \centering
  \includegraphics[width=\linewidth]{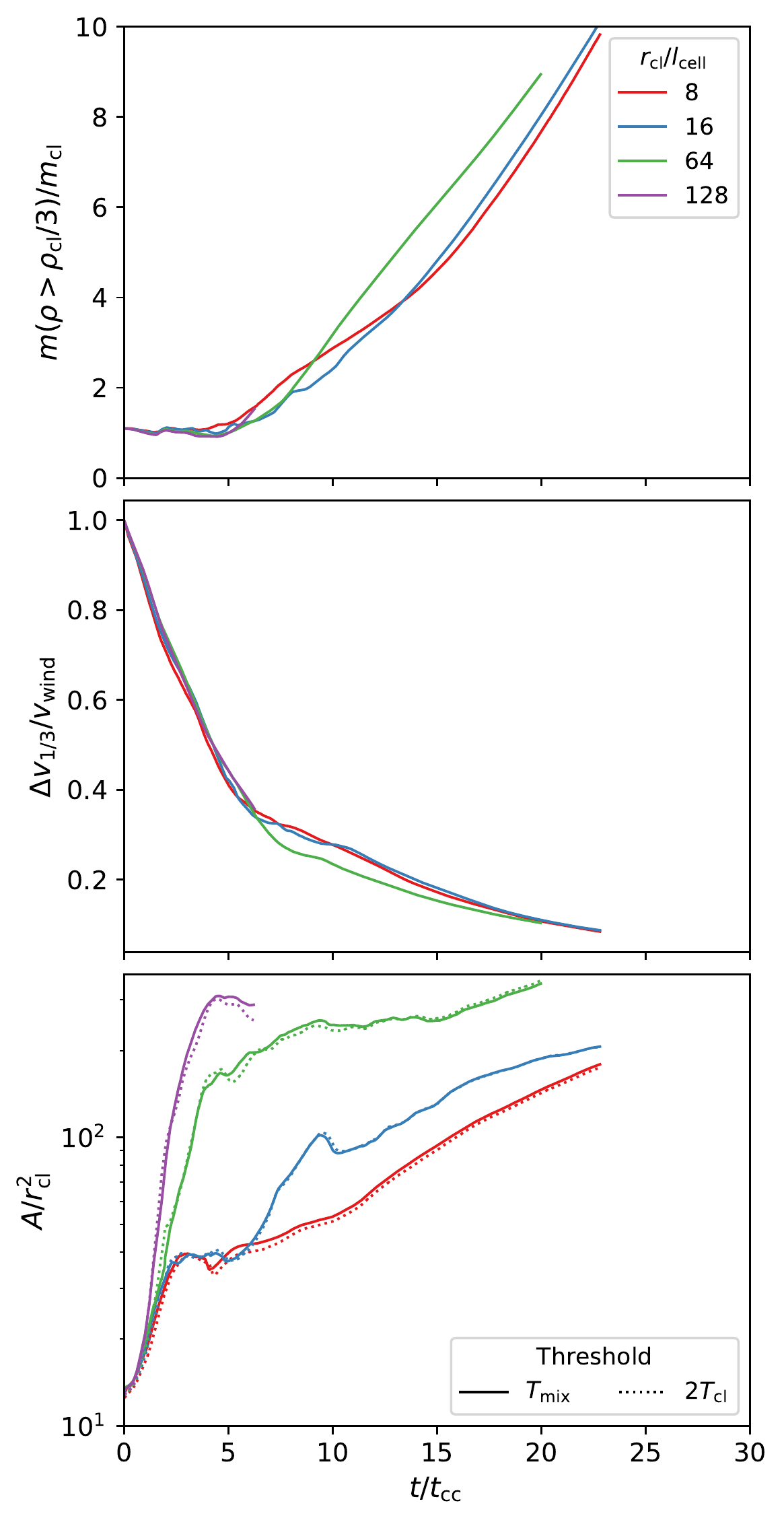}
  \vspace{-.5cm}
  \caption{Evolution of the $\chi=10$, $\tcoolrat{mix}\sim 0.45$ simulations with different resolutions where $l_{\mathrm{shatter}}$ is resolved by 3 and 6 cells respectively in the two highest resolution simulations (see \S~\ref{sec:shattering} for details).}
  \label{fig:shattering_evolution}
\end{figure}

\begin{figure*}
  \centering
  \includegraphics[width=0.4\textwidth]{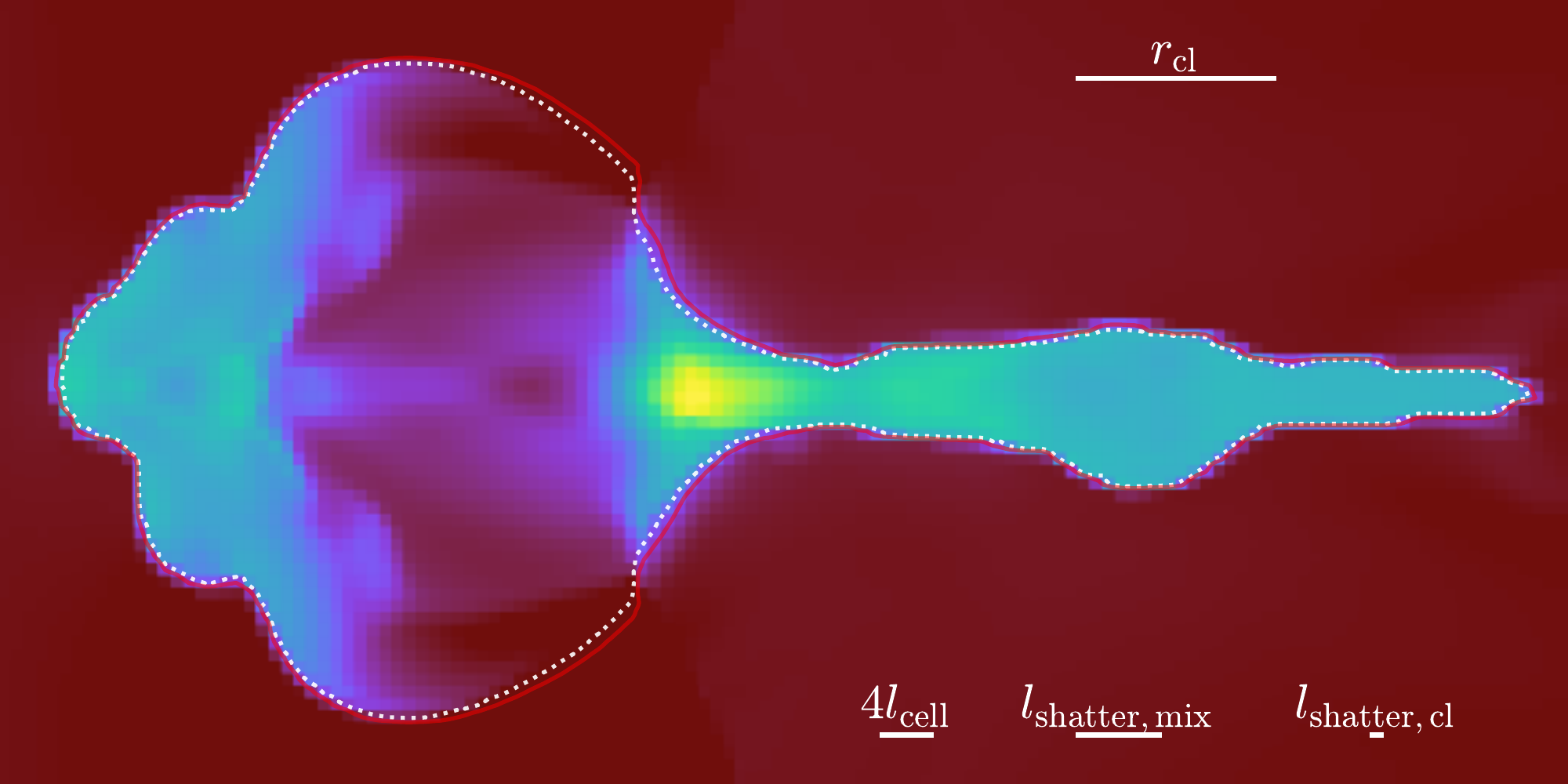}~
  \includegraphics[width=0.4\textwidth]{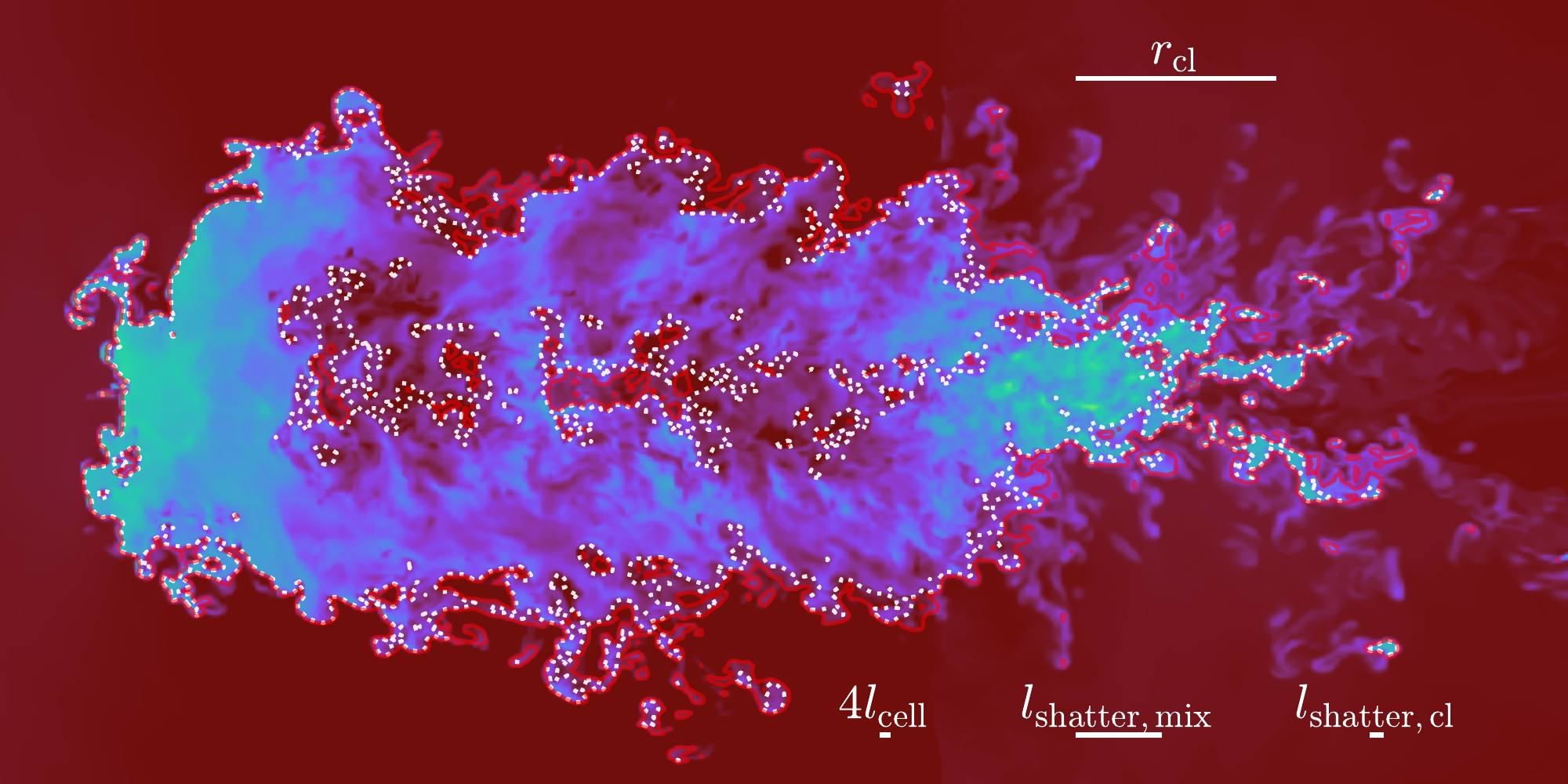}
  \includegraphics[height=3.7cm]{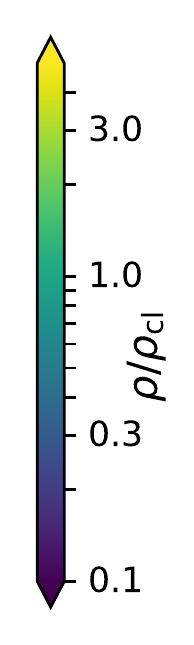}
  \caption{Density contours showing an excerpt of two runs with $\chi=10$ and $\tcoolrat{mix}\sim 0.45$ at $t = 6\tcc$. The left and right panels show simulations with resolutions $r_\cl / l_{\mathrm{cell}}=16,128$ respectively. In each panel, we show some length scales as comparison. This visualizes that $l_{\mathrm{shatter}}\equiv l_{\mathrm{shatter,cl}}$ is only resolved in the simulation shown in the right panel. The red solid and white dotted contour line show the temperature threshold $T=T_{\mathrm{mix}}$ and $T = 2T_\cl$, respectively.}
  \label{fig:plot2d_shatter}
\end{figure*}

\subsection{Numerical convergence; Connection to `Shattering'}
\label{sec:shattering}

\citet{McCourt2016} found a characteristic length scale of cold gas which is given by
\begin{equation}
  \label{eq:l_shat}
  l_{\mathrm{shatter}} = \min(c_{\mathrm{s}} t_{\mathrm{cool}}) \sim c_{\mathrm{s,cl}} t_{\mathrm{cool,cl}}\;.
\end{equation}
Note, that $l_{\mathrm{shatter}}$ is defined at the \textrm{minimum} of $c_{\mathrm{s}}t_{\mathrm{cool}}$ which sets the end of the fragmentation cascade (this differs from previous definitions of the `cooling length scale'; e.g., \citealp[][]{1999A&A...351..309H}; but see \citealp{Cornuault2018}). \citet{McCourt2016} found in their high-resolution simulations that indeed structures much larger than $\gtrsim l_{\mathrm{shatter}}$ fragment, while smaller clouds do not. Recently, other groups have claimed to confirm this results in three dimensions and with magnetic fields \citep[][]{Liang2018,Sparre2018}.

The ratio between $l_{\mathrm{shatter}}$ and our survival length scale $r_{\mathrm{cl, crit}}$ is given by (cf. equation~(3) in \citetalias{Gronke2018})
\begin{equation}
  \label{eq:rcrit_lshat_rat}
  \frac{r_{\mathrm{cl,crit}}}{l_{\mathrm{shatter}}} \approx 10 \ \mathcal{M}_{\mathrm{w}} \frac{\chi}{100} \left(
\frac{\Lambda(T_{\rm cl})/\Lambda(T_{\rm mix})}{0.1} \right)\;.
\end{equation}
This implies that in all of the simulations presented thus far, $l_{\mathrm{shatter}}$ is unresolved, and any different behavior once it is resolved is not included.

In order to check the impact of `shattering' on our results, we want to fulfill $l_{\mathrm{cell}} < l_{\mathrm{shatter}} < r_\cl$ while at the same time $r_\cl > r_{\mathrm{cl,crit}}$. We therefore ran several simulations with $\chi=10$, $\mathcal{M}_{\mathrm{w}}=1.5$, and $\tcoolrat{mix}\sim 0.45$. This means that $r_\cl / l_{\mathrm{shatter}}\sim 21$. We ran simulations with $r_{\cl} / l_{\mathrm{cell}} \sim \{8,\,16,\,64,\,128\}$. For the last two simulations $l_{\mathrm{shatter}}$ is resolved by $3$ and $6$ cells, respectively.

The overall evolution of these simulations is shown in Fig.~\ref{fig:shattering_evolution}, where we display (from top to bottom panel) the mass, velocity, and surface area of the cold gas (also see Fig.~\ref{fig:mass_growth_d100_convergence} for $\chi=100$ runs with $8$ to $64$ cells per cloud radius).
While the masses and velocities seem to have converged, the surface area is up to an order of magnitude larger in the higher resolution cases. The reason for this is increased fragmentation as illustrated in Fig.~\ref{fig:plot2d_shatter} where we contrast density slices from the $r_\cl / l_{\mathrm{cell}}\sim 16$ and $r_\cl / l_{\mathrm{cell}}\sim 128$ simulations; both at $t\sim 6\tcc$. In the higher resolution run, much more structure is visible. This also affects our measure of surface area which correspond to the contours (drawn at $T=2T_\cl$ and $T=T_{\mathrm{mix}}$) also displayed in Fig.~\ref{fig:plot2d_shatter}.

Thus, the small scale structure of the cold gas (both visually and as quantified by  the surface area) is not converged, even in these higher resolution simulations where $l_{\rm shatter}$ is resolved by 3 and 6 cells, respectively. This is perhaps unsurprising: \citet{McCourt2016} found in their 2D simulations that shattering is particularly sensitive to numerical resolution, with the fragmentation of  clumps of  $\sim 10$ cells across entirely unresolved and whose dynamics are determined by grid scale effects. The 3D simulations of \citet{Sparre2018} report results similar to  ours: while the mass survival history of cold gas as a function of time is converged, the fragmentation  of the  cloud (as measured by the number of clumps detected  by a  friends of friends  algorithm) shows much weaker convergence (see their figures 1 \& A3). The thorny  question  of numerical convergence in simulations of shattering is  beyond the scope of this paper. Our results simply hinge upon the fact that while the details of cloud morphology is not yet converged, global quantities such as the mass growth  rate do indeed  appear to be converged. One way of understanding this is to  note that cloud oscillations discussed in \S~\ref{sec:location_mass_growth} potentially introduce an effective smoothing length which averages over cloud substructure. Thus, one should not count the area  due to the `rugged coastline', but rather the effective area of a Gaussian surface encompassing the cloud (see \S~\ref{sec:convergence} for a continued discussion about numerical convergence). 

A deeper question is the physical conditions under which where a cloud breaks apart to form a `mist' (as in \citealt{McCourt2016}), or retains its monolithic identity and grows larger (as in this paper). This divide has important  consequences for the survival  and dynamics of cold gas, as well as for several observables. Both shattering and cloud growth in a wind are driven  by pressure gradients which arise from radiative cooling, but in the limits where $\delta P \sim P$ and $\delta P \ll P$ respectively. 
Shattering takes place when the cooling cloud starts out at $T \gg T_{\rm  cl}$; if the cooling time is much shorter than the sound crossing time, the cloud will be wildly out  of pressure balance ($\delta P \sim P$) and break up (see figure 4 of \citet{McCourt2016})\footnote{\citet{McCourt2016} found that `shattering' into cloudlets of size $\sim c_{\rm s} t_{\rm cool}$ also happens when a cloud is eroded by Kelvin-Helmholtz instabilities in a wind. However, downstream in the wake, cloudlets appear to have coalesced into larger structures and become entrained in the wind, similar to the  results of this work (see Fig. 6 in \citealt{McCourt2016}).}. By contrast, if a cloud is already at the temperature floor $T_{\rm cl} \sim 10^{4}$K, when cooling shuts off, then only a small volume of mixed gas is cooling and out of pressure balance with its surroundings. The small pressure fluctuations $\delta P/P \ll  1$ drive overstable sound waves; the cloud pulsates. These pulsations drive mixing and help the cloud to grow.  The interaction between `shattering', cloud coalescence and growth is an important topic for future work.

\section{Discussion}
\label{sec:discussion}

\subsection{Multiphase galactic winds}
\label{sec:disc_galactic_winds}

Galactic winds are ubiquitously observed in galaxies throughout cosmic time \citep{Veilleux2005,Rupke2018}, for instance, in NaI, SiII, HI quasar line absorption, in emission of molecular gas \citep[e.g.][]{Cicone2018}, or direct $X$-ray imaging \citep[e.g.,][]{1994ApJ...433...48V,2002ApJ...574..663M}. These multi-wavelength observations establish the multiphase nature of galactic winds.

Hot gas is thought to be accelerated via supernovae or AGN feedback. However, as discussed in the introduction, explaining the fast moving, cold components has been more challenging. 
We showed in this work that under the right conditions, entrainment and mixing of hot wind gas which subsequently cools can enable the cloud to survive, accelerate and grow. Thus, {\it both}  the original cold gas and additional hot gas which has condensed out now make up the cold phase. In this framework, the presence of dust and molecules far out in a galaxy's halo is reasonable, since the original molecular phase ejected from the galaxy (which has been observed to travel with $\gtrsim $  hundreds of \kms; e.g., \citealp{2015A&A...574A..14C}) survives. This is in contrast to scenarios  where the only source  of cold gas is condensing hot gas (see below),  which then requires survival of dust in higher temperatures or a short dust reformation timescale to explain the observations.

\subsection{The cold gas content of the CGM}
\label{sec:disc_cgm}
Absorption line measurements of bright background sources such as quasars \citep{Werk2014,Prochaska2017,Ho2017} or (stacks of) galaxies \citep{Steidel2010a,Rubin2017} paired with the knowledge of a foreground galaxies as well as direct emission maps \citep{Hennawi2015,Wisotzki2015,Battaia2018} all draw a picture of large amounts of cold gas in the circumgalactic medium of distant and galaxies \citep[also see review by][]{Tumlinson2017}.

Theoretically, the existence of this cold gas in an otherwise hot galactic halo is not well understood and cosmological simulations frequently under-predict its abundance \citep{Fielding2016,Liang2016}. In principal, four mechanisms can bring cold gas into the CGM:
\begin{enumerate}
\item Direct transport of cold gas into the CGM in outflows, as studied in this work. This can include both the central and satellite galaxies. The latter can be important contributors to the CGM, since they typically have shallow potential wells and are already at large impact parameters from the halo center \citep{Hafen2018a}.

\item Inflowing cosmological `cold flows'. In this paper, we address the survival of cold gas in the wind, but not that of the collimated supersonic cold streams, which have very different geometry, and are studied in detail by other researchers (e.g., \citealp{Mandelker2019MNRAS.484.1100M}). In cosmological simulations, these only appear in a limited halo mass and accretion rate regime, and have a relatively low covering fraction, and thus cannot be a universal explanation for ubiquitous cold CGM gas. 

\item Cooling of the hot wind. If the hot wind is expanding, strong radiative cooling and thermal instability will set in once it cools adiabatically to the peak of the cooling curve at $T \sim 10^{5}$K \citep{Wang1995,Thompson2015} which can occur at some radius $R_{\rm cool} \sim 5$kpc, given the initial properties of the wind. In this case, the clouds are born co-moving with the wind and do not suffer cloud-crushing instabilities. 
However, the growth of cold clouds as discussed here draws mass from the hot wind and reduces its density, preempting this phase. One easy way to see this is to exploit the fact that for the Chevalier-Clegg  solution, the mass growth rate is constant, $\dot{m} \sim m_{0}/t_{\mathrm{sc}, 0}$, where $m_{0}$ and $t_{\mathrm{sc},0}$ are  the initial cloud mass and sound crossing time. This implies that upon reaching the radius $R_{\rm cool}$ after a time $t_{\rm wind} \sim R_{\rm cool}/v_{\rm wind}$, the cold gas mass in the wind would have increased by a factor: 
\begin{equation}
\frac{m}{m_{0}} \sim \frac{\dot{m}t_{\rm wind}}{m_{0}} \sim 100 \left( \frac{R_{\rm cool}/r_{\cl,0}}{10^{3}} \right) \mathcal{M}_{\mathrm{w}}^{-1}  \left( \frac{\chi}{100} \right)^{-1/2}.
\label{eq:growth_factor}  
\end{equation}  
where we have used Eq.~\eqref{eq:tcc_tsc}. Thus, even if a small fraction of the wind is initially in the form of cold gas, by the  time it propagates to $R_{\rm cool}$, most of the wind would have been converted to cold gas (i.e. until $t_{\rm cool,mix}$ becomes longer than other characteristic  timescales and our  assumptions break down). While  this needs to be  verified  in galactic-scale  simulations, based on  these considerations we do not expect any sharp transition at $R_{\rm cool}$ but rather a continuous existence of cold gas -- which might ensure the survival of dust and molecules out to large radii and velocities
(see \S~\ref{sec:disc_galactic_winds} above).

\item In-situ formation via thermal instability in the hot gas. If the ratio of the cooling to free-fall time of hot gas is $t_{\rm cool}/t_{\rm ff}\lesssim 10$, hydrodynamic simulations show that thermal instability can form cold gas in  a stratified atmosphere \citep{Mccourt2012,Sharma2010}. This ratio represents the competition between radiative cooling and buoyant restoring forces which damp the instability. The criterion also depends on the background entropy gradient \citep{Voit2017}, magnetic fields \citep{Ji2018MNRAS.476..852J}, turbulence \citep{Voit2018}, all of which modulate the effects of buoyancy, or large density perturbations \citep{Choudhury2019}, which accelerates cooling. Theoretical models assume the presence of a hot halo gas component in approximate hydrostatic and thermal equilibrium. While this is justifiable for massive ellipticals, groups and clusters, it is as yet unclear whether such assumptions are justifiable at the $L_{*}$ scale and below. Note that infalling cold gas which forms via thermal instability which exceeds  $r_{\rm cl,crit}$ in size will mix and grow in mass per the model described in this work. 

\end{enumerate}

We can estimate the total cold gas deposited in the CGM through galactic winds (i.e., mechanism (i) described above) alone as:
\begin{equation}
  \label{eq:M_CGM}
  M_{\mathrm{c,CGM}}\sim \phi M_{\rm wind} \sim \phi M_{*} \eta \sim M_{*}  
\end{equation}
where 
$M_{\rm wind}$ is the total mass expelled as winds during the life of the galaxy, $M_{*}$ is the stellar mass of the galaxy, $\eta\equiv \dot M_{\mathrm{wind}} / \mathrm{SFR}$ the (cold gas) mass-loading factor of the wind, and $\phi$ is an efficiency factor. The mass loading factor is observationally constrained to be $\eta\sim 0.1 - 10$ \citep[e.g.,][]{2017MNRAS.469.4831C} for $L_{*}$ galaxies, which is also broadly consistent with numerical simulations of supernovae driven winds \citep[e.g.][]{Li2016}. From Eq.~\eqref{eq:growth_factor}, the rapid growth rates imply that we expect the efficiency factor $\phi = M_{\rm c,CGM}/M_{\rm wind}$ to be of order unity, unless a large fraction of the winds are below the escape velocity and stall before they reach large radii, or a large fraction of clouds fulfills $r_\cl < r_{\mathrm{cl, crit}}$ and does not survive. Powerful outflows above the escape velocity are frequently observed \citep{Veilleux2005,Tripp2011,Rudie2019}, although during more quiescent phases the wind simply drives a galactic fountain. While this needs to be considered in more detail, in this crude estimate, the cold gas mass in the CGM is of order the stellar mass in the galaxy, in good agreement with observations. We discuss numerical challenges in larger scale simulations which can address this shortly in \S\ref{sec:convergence}.

\subsection{Fueling of star-formation}
\label{sec:disc_fueling}

Star-forming galaxies typically have short gas depletion times ($\sim 1$  Gyr) and thus require a continual supply of fresh gas.  In the Milky Way, this inflow must come in the form of low-metallicity ($Z < 0.1 \, Z_{\odot}$) gas, to satisfy constraints from disk stellar metallicities and  chemical evolution models \citep{schonrich09,kubryk13}. At the same time, we see infall in the form of `high-velocity' and `intermediate-velocity'  clouds (HVCs and IVCs; \citealt{Putman2012}) with relatively low metallicities, as well as a galactic fountain with  continuous circulation of material  between the  disk and corona \citep{Shapiro1976a, Fraternali2008}. Fountain-driven accretion could supply the disk with gas for star formation, and explain the observed kinematics of extra-planar gas \citep{Armillotta2016,Fraternali2017}.

Our simulations give credence to the idea that star formation in the disk exerts a form of positive feedback: cold gas thrown  up into the  halo `comes back with interest', by mixing with low metallicity halo gas which cools and increases the cold gas mass. During this process, cold gas also exchanges angular momentum with coronal gas,
which  links fountain circulation to the observable kinematics of coronal gas. 

In a galactic wind, the cold gas eventually co-moves with the hot wind. By contrast, fountain  clouds always see a time-varying wind in their frame, which  is maximized near the  disk and minimized at apocenter. There is no asymptotic state; the problem is time and position dependent. The velocity profile during outflow and infall would be exactly symmetric for a ballistic particle, but the cloud's initial acceleration and mass growth during its journey, and hydrodynamic drag forces during infall break this symmetry. In particular, the latter implies that the cloud reaches a terminal velocity\footnote{Magnetic tension can also oppose gravity and prevent/slow down smaller clumps from falling, as found in MHD simulations of thermal instability \citep{Ji2018MNRAS.476..852J}. Magnetic field swept up during the infall of the cloud can be significantly amplified so that such support is possible.} $v_{\rm term} \sim (r_{\rm cl} g \chi/3 )^{1/2}$ which is less than the outflowing wind velocity, $v_{\rm wind}$, which must at the very minimum be $ v_{\rm wind} > (2 g h)^{1/2}$, where $h$ is the height the cloud reaches at turnaround. Thus, $v_{\rm wind}/v_{\rm term} > (6 \, h/\chi r_{\rm cl})^{1/2} > 1$, where we typically have $h \gg \chi r_{\rm cl}$ (note that $\chi r_{\rm cl}$ is the lengthscale over which the cloud is entrained).  Thus, if a cloud is sufficiently large to survive the outbound journey, $r > r_{\rm crit}$, it should survive the return journey. However, the details of its mass growth depend on the coronal density profile and the cloud trajectory. We will investigate this in future work.

\subsection{Cold gas in the halo of the Milky Way}
\label{sec:discussion_milky_way}

Given its promixity, the Milky Way is an ideal laboratory to observe the effects of wind-cloud interactions. 
It is inhabited by a population of cold gas clouds which can be, for instance, observed using the $21$cm spin-flip transition of neutral hydrogen. These `high-velocity' and `intermediate-velocity' clouds (HVCs and IVCs, respectively) have column densities of $N_{\mathrm{HI}}\sim 10^{18-20}\cm^{-2}$, radii of $\lesssim 50\,$pc, and move with line-of-sight velocities of $\lesssim 400\kms$ \citep[e.g.,][]{Putman2002,DiTeodoro2018}. Their observed line-widths of $\sim 10-30\kms$ correspond to temperatures and corresponding pressures of $T\sim 10^4\,$K and $\log P/k_{\mathrm{B}}\sim 10^3\,\mathrm{K}\,\cm^{-3}$, respectively \citep{Wolfire1995,Hsu2011}. 
The properties of the hot surrounding medium (probed by X-ray measurements) of these clouds is somewhat uncertain, but coronal temperatures of $T\sim 10^{6}\,$K in pressure balance with the cold clouds is usually assumed \citep{Putman2012}.

These values correspond to cloud overdensities of $\chi\sim 100 - 1000$ in a $\mathcal{M}_{\mathrm{w}}\sim 3$ wind. The observed head-tail morphology \citep{Bruns2000,Putman2011} supports the picture of cloud-wind interaction. However, it also raises the question of their survival \citep[e.g.,][]{DiTeodoro2018}.

The observed HVC properties correspond to a critical radius of cloud survival of $r_{\mathrm{cl, crit}}\sim 6\,$pc. We predict that only HVCs larger than this radius can survive, and thus we should see a drop in the observed size distribution for $r_{\cl} \lesssim r_{\mathrm{cl,crit}}$. Unfortunately, even the very highest resolution \HI observations reach only a limit of $\sim 2.5'$ corresponding to $r_{\mathrm{cl,min}}\sim 6\,\mathrm{pc} (d / 8\,\mathrm{kpc})$ \citep{McClure-Griffiths2012} -- up to which no cutoff in the cloud radius distribution is observed \citep{DiTeodoro2018}. Alternatively, we predict that that no clouds with total column $N_{\rm H} < 10^{18} {\rm cm^{-2}}$ should survive. 
Another prediction of this study is that if a cloud does survive, it will accumulate additional mass during its trajectory, which should lead to an overall larger mass of HVCs approaching the disk than outflowing (assuming that most clouds do not achieve escape velocity).

An interesting case of cold gas survival within our own Galaxy is the leading arm (LA) of the Magellanic stream, which is an ensemble of cold gas clouds continuing the Magellanic stream beyond the Small- and Large Magellanic clouds \citep{Putman1998,DOnghia2016}.
\citet{Pardy2018a} simulated the SMC-LMC interaction prior of entering the Milky Way's halo and showed convincingly that the Magellanic stream as well as the LA can be reproduced using an initial nine-to-one mass ratio of the satellite galaxies. However, they did not include any halo gas in their calculations. Recently \citet{Tepper-Garcia2019} re-visited the problem, including halo gas. They found (using several halo models -- including magnetized ones) that the hydrodynamical instabilities destroy the LA during its $\sim 1\,$Gyr passage through the halo. However, the resolution of these simulations is much larger than the critical scale\footnote{Since the ambient density is somewhat certain, this propagates to uncertainty in $r_{\mathrm{cl,crit}}$.} for cold gas $r_{\mathrm{cl,crit}}\sim 10-50\,$pc. An alternative theory for the existence of the LA invokes `forerunners', that is, satellite galaxies preceding the trajectory of the SMC/LMC with the LA consisting of their stripped gas \citep{2014MNRAS.442.2419Y}. Following the survival of cold gas clumps stripped from such systems might similarly require higher resolution.

\subsection{Numerical convergence and larger scale simulations} 
\label{sec:convergence} 

The rapid convergence  in  cloud mass accretion  rate $\dot{m}$ in our simulations might appear surprising, particularly when  other  aspects (such as cold gas morphology  and  surface area) are clearly not converged. We obtain results  converged in $\dot{m}$ even when $r_{\rm cloud}/l_{\rm cell} =8$ (see Fig. \ref{fig:mass_growth_d100_convergence}), far below what is customarily considered acceptable in cloud-crushing simulations \citep[usually $\gtrsim 100$ cells per cloud radius are customary]{Klein1994}. This is likely related to the fact that our  cloud stays intact  in  one large piece and grows in volume, whereas higher resolution is needed to follow fragment dynamics when the cloud breaks apart. 

If a cloud {\it does} fragment into  small pieces,  then the only characteristic lengthscale is $c_{\rm s} t_{\rm cool}$ (evaluated at its minimum at $T \sim 10^{4}$K).  If this is not resolved, cloudlets will mix in at the grid scale \citep{McCourt2016}. Even if the cloud does not fragment,  one  might reasonably assume that the interface between hot and cold phases, $H \sim \alpha (c_{\rm s} t_{\rm cool} r_{\rm cl})^{1/2}$, 
must be resolved for the  total luminosity of the cloud  (and the  mass accretion rate) to be converged  -- low resolution should lead to `overcooling' due to an overabundance of gas at intermediate temperatures. For instance, it has been suggested that simulations of thermal instability will not converge unless explicit thermal conduction is included and the Field length (which plays a similar role to the lengthscale $H$) is resolved \citep{Koyama2004}. The fronts are  clearly not resolved in our  simulations, given that the boundary layer is one cell thick (Fig. \ref{fig:2d_emissivty_slice} and \ref{fig:cdf_Tmax}), and yet our simulations are converged in both luminosity and mass accretion  rate. The fact that $\dot{m}$ can be correctly computed even in low-resolution simulations is also seen in simulations of cloud destruction \citep[e.g.][]{Sparre2018}. Importantly, our mass entrainment rates are in good agreement with high resolution simulations of a single turbulent mixing layer \citep{Ji2018}, where the cooling length is resolved by many cells.

This  issue clearly deserves  more study. We believe the key reason for convergence  is the  fact that large pressure imbalances never develop, $\delta P/P \ll 1$, so that  cooling is quasi-isobaric. As long as this remains true, the cloud does not `shatter' but remains in approximate pressure balance with its surroundings, and the relevant lengthscale which  needs to be resolved is not $c_{\rm s} t_{\rm cool}$ or even  the diffusion length $H \sim \alpha (c_{\rm s} t_{\rm cool} r_{\rm cl})^{1/2}$, but the cloud radius $r_{\rm cl}$. Another way of stating this is that the cloud sound crossing time dictates the growth time (Eq.~\eqref{eq:growth_rate_cc}), and this is well resolved in our simulations. Any cloud with $r  > r_{\rm cl,crit}$  which  is initially at (or close to) the temperature floor $T_{\rm min}$ will stay intact in quasi-isobaric pressure balance, which only requires that cloud dynamics be resolved. If the cloud is initially at some $T  \gg T_{\rm min}$, large pressure imbalances  will develop that cause the cloud to `shatter'  into pieces of size $\sim c_{\rm s} t_{\rm cool}$, which  then require high resolution. If the cloud has $r < r_{\rm crit}$, it will break up due  to hydrodynamic instabilities, which eventually erode and mix away.

At present, both cosmological and galaxy scale simulations are numerically unconverged  with respect to the properties of  their CGM, in particular the cold gas mass which is required to make comparisons against observations of line absorption and emission \citep{VandeVoort2018,Hummels2018,Peeples2018,Suresh2018}. As numerical resolution increases, so does the cold gas mass \citep[see also][]{Faucher-Giguere2016}. One potential way out of this dilemma is to construct a sub-grid model for cold gas, treating as a second fluid which exchanges mass, momentum and energy with the hot gas. This inevitably throws away information, and a viable model has yet to be elucidated. A more attractive approach would be to demonstrate that a physical scale exists which, if resolved, would  enable mass convergence, even if other properties such as cold gas morphology are  not necessarily resolved. Our results suggest  that $r_{\rm cl,crit}$  is  indeed  such a scale, and that it only needs to be resolved with a few cells,  rather than  the $\sim 100$ or more cells need to resolve systems far from pressure balance (e.g. \citealp{McCourt2016}).  This has very promising implications for direct resolution of cold gas in galaxy-scale simulations, since $r_{\rm cl,crit}$ is at least an order of magnitude larger than $c_{\rm s} t_{\rm cool}$ (cf. Eq.~\eqref{eq:l_shat}).

While $r_{\rm cl,crit}$ is $\sim$pc scale at the launch radius, because $r_{\rm cl,crit}\propto P^{-1}$, it  declines rapidly with radius  and becomes several hundred pc in the CGM. This is not too far from the resolution of current simulations, which reach a resolution of $\sim 500\,$comoving pc throughout the CGM (e.g., \citealt{Hummels2018}). Such requirements are well suited to current simulation setups, which have high resolution close to the galaxy and progressively lower resolution in the  low  density CGM. Equivalently, since $r_{\rm cl,crit} \propto 1/n$, if one resolves the cloud column density
\begin{equation}
N_{\rm crit} \approx 10^{18} \ {\rm cm^{-2}} 
\mathcal{M}\frac{\chi}{100} \left(
\frac{\Lambda(T_{4})/\Lambda(T_{\rm mix})}{0.1} \right)
\end{equation}
by at least several cells throughout the wind and the CGM, we would predict that the simulations should approach convergence. 
Our results suggest  that rather than  ``forced (uniform) resolution'' or tying resolution to density, it would be most efficient to resolve a fixed column density, or equivalently to tie spatial resolution to pressure. 

The approach to convergence depends on the mass spectrum of  clouds,  and the fraction of cloud mass at small scales. It may  not even  be necessary to resolve  down to $r_{\rm cl,crit}$, since cold clouds are only `seeds' for radiative cooling and condensation, and a sufficient mass in larger clouds will lead to the same asymptotic efficiency $\dot{m}_{\rm cold} \sim \dot{m}_{\rm wind}$. Note that smaller clouds with $r < r_{\rm cl,crit}$ can still drain momentum and energy from the hot gas. However, this should be a relatively short phase (given the short cloud-crushing times at these small lengthscales). An important unknown is  the efficiency with which smaller clouds can coagulate and clump to form  larger clouds (see  Fig. 6 of \citealp{McCourt2016} for dynamically driven coagulation, and \citealp{Waters2019} for cooling driven coagulation), and the necessity of resolving such dynamics. 

Other physical processes could also alter mass growth and resolution requirements -- for instance, turbulence. A turbulent velocity field is quite different from the significantly more ordered velocity field  of a wind. While the shear between a cloud and the wind becomes progressively smaller as the cloud comoves with the wind, there is always significant shear in a chaotic velocity field. Turbulence also provides an effectively adiabatic form of pressure support which reduces pressure gradients and thus entrainment due to cooling. This is an important avenue for future work.

Overall, our results are consistent with the lack of convergence in cold gas mass in the current generation of simulations, but offers hope that convergence and direct resolution of condensation may not be too far off.

\section{Conclusions}
\label{sec:conclusion}

Most studies of $T \sim  10^{4}$K gas clouds exposed to a hot wind have found the clouds to be eroded by cloud crushing and Kelvin-Helmholz instabilities, even in the presence of radiative cooling. In \citetalias{Gronke2018}, we found in 3D hydrodynamic simulations that cold clouds can not only survive in a hot wind: they can grow in mass, by more than an order of magnitude. These results are not incompatible with previous simulations: one needs both a sufficiently long simulation box, even in a comoving frame (to capture mass growth in the tail), and a sufficiently large initial cloud (so that the cloud crushing time is longer than the cooling time), to witness this growth. \citet{Gronke2018} formulated a quantitative criterion for cloud survival and growth,  $t_{\mathrm{cool,mix}}\lesssim \tcc$, where $t_{\mathrm{cool,mix}}$ is the cooling time of the {\it mixed} gas (which is equivalent to $r_\cl\gsim r_{\mathrm{cl,crit}}$, where $r_{\mathrm{cl,crit}}$ is a critical cloud size in Eq.~\eqref{eq:rcrit}).  In this paper, we have studied the mass growth in much greater detail. Our results are as follows: 

\begin{itemize}
\item{{\bf Entrainment mechanism.} Clouds grow by mixing with hot gas; the mixed gas sits at the peak of the cooling curve, and quickly cools radiatively. One might expect that the mixing is primarily due to the Kelvin-Helmholtz instability, and that mass growth peaks when velocity shear is strongest, fading as the cloud becomes entrained and $\Delta v$ falls. Instead, we find the opposite trend: mass growth {\it rises} as $\Delta v$ falls; it peaks and continues even when the cloud is almost fully entrained\footnote{Note, however, that in a fully static setup, we find much lower  mass  growth rates. Clearly some perturbations are necessary to set off the thermal instability responsible for mixing and growth.} (see Fig. \ref{fig:mass_growth_new_setup_multiplot}). Consistent with the findings of \citet{Ji2018}, who simulate a single radiative mixing layer at high resolution, we find that radiative cooling, rather than hydrodynamic instabilities, dominate gas mixing, by  creating pressure gradients which siphon hot gas into the mixing layer; this becomes a self-sustaining process. Rather than the entrainment of the cold cloud in the wind, we should speak of the entrainment of cooling hot gas onto the cloud. As in an inelastic collision, this gas then imparts its mass and momentum to the cloud, growing and accelerating it.}

\item{{\bf Mass growth rates.} The mass growth rate can be written as $\dot{m} \sim A \rho_{\rm w} v_{\rm mix}$, where the $A$ is the surface area of the cloud, $v_{\rm mix}$ is the characteristic velocity of the mixing process, and $\rho_{\rm w}$ is the wind density. After an initial transient where the cloud surface area grows rapidly due to the development of a tail, surface area roughly scales as $A \sim (m/\rho_{\rm cl})^{2/3}$. On the other hand, the mixing velocity is of order the cloud sound speed $v_{\rm mix} \sim c_{\rm s,cl}$, scaling as $v_{\rm mix} \propto t_{\rm cool}^{-1/4}$, as also seen in the high resolution simulations of \citet{Ji2018}. This allows one to develop analytic scalings for mass growth, and implies mass growth rates of $\dot m \sim m_{\mathrm{cl, i}}/ t_{\rm sc}$, where $m_{\mathrm{cl},i}$ is the initial cloud mass and $t_{\rm sc}$ the sound-crossing time of the cloud.}  

\item{{\bf Magnetic Fields.} Magnetic fields significantly change cloud morphology, leading to a much `stringier', elongated appearance aligned with $B$-fields. They also speed up cloud acceleration, due to magnetic drag enhancing momentum transfer from the wind. 
Magnetic tension also suppresses Kelvin-Helmholtz mixing. 
Despite this, mass growth rates are within a factor of $\sim 2$ of hydrodynamic mass growth rates, for plasma $\beta \equiv P_{\rm gas}/P_{\rm B} \gtrsim 1$. Compressional amplification of entrained wind material can lead to magnetically dominated clouds supported by magnetic pressure where $\rho_{\rm cl}/\rho_{\rm hot} \ll T_{\rm hot}/T_{\rm cl}$, which could potentially explain unexpectedly low-density cloud material inferred from COS observations \citep{Werk2014}.}  

\item{{\bf Background wind.} We considered three modifications to our time-steady hot ${\mathcal M}_{\mathrm{w}} = 1.5$ wind: (i) an adiabatically expanding hot wind which obeys the Chevalier-Clegg (CC85) solution. Here, the mass growth rate per unit area of the cloud decreases with radius, since the density of the hot wind falls and the cooling time of mixed gas increases. At the same time, the area of the cloud increases due to the decreased pressure of the confining hot gas, with the rather remarkable consequence that $\dot{m}$ is independent of radius, as can be understood from analytic scalings. (ii) An isothermal background wind for which the mass growth eventually ceases in accordance with scaling arguments. (iii) Higher Mach number flows. In this case, the post-shock cloud compression leads to higher cloud overdensities and longer acceleration times. Here, the cloud still eventually grows, though we caution that convergence in the high Mach number case is poor and requires further investigation.}

\item{{\bf Initial cloud geometry.} A spherical cloud is a rather idealized setup. To test the influence of initial cloud geometry, we generate a fractal cloud by extracting a portion of a simulation stirred with supersonic turbulence, similar to previous studies \citep[e.g.][]{Schneider2016,Liang2018}. We find somewhat higher mass growth compared to a spherical cloud, due to the larger surface area-- the converse of simulations in the $r_{\rm crit} < r_{\rm cl,crit}$ regime, which find more rapid cloud destruction with a fractal cloud.
}

\item{{\bf Convergence.} Our 3D simulations, which range from $8-128$ cells per cloud radius (typical for cloud-crushing simulations), are not converged with respect to cloud morphology. Higher resolution reveals progressively more substructure, which can be quantified by the rapid growth of surface area with resolution. One might only expect convergence if the lengthscale $c_{\rm s} t_{\rm cool}$, evaluated at the cloud temperature, is well-resolved \citep{McCourt2016}. However, our mass growth rates are remarkably well-converged with respect to numerical resolution, and show no appreciable difference in simulations where $c_{\rm s} t_{\rm cool}$ is resolved. This is despite the fact that the cloud boundary (which harbors most of the emission; see Fig. \ref{fig:2d_emissivty_slice} and \ref{fig:cdf_Tmax}) is numerically unresolved. Our mass entrainment rates are in good agreement with high resolution simulations of a single turbulent mixing layer \citep{Ji2018}, where the cooling length is resolved by many cells.
We speculate that turbulence and cloud pulsations introduce an effective smoothing length, but this needs to studied further.
The fact that only $\sim 8$ cells per $r_{\mathrm{cl,crit}}$ are necessary in order to achieve convergence in the mass growth gives hope that larger  cosmological simulations can reliably predict cold gas observables in the near future.}

\item{{\bf Implications.} Overall, our results point toward remarkably robust cold cloud acceleration and mass growth if clouds exceed a critical size, even if a host of other features are varied. The ability of radiative cooling to overcome the effect of hydrodynamic instabilities can explain observations of high-velocity clouds in the Milky Way, and cold gas outflowing in galactic winds at high velocity. Even more remarkably, the {\it growth} of such clouds by entraining and cooling hot gas from the wind could contribute to the cold gas content of the CGM (up to a mass comparable to the stellar mass of the host galaxy; Eq.~\ref{eq:M_CGM}), and fuel star formation in the disk as cold gas which is recycled in a galactic fountain grows and accretes mass from the halo before falling back onto the disk. These striking implications clearly require more work.}  

\end{itemize}

Many important questions remain, amongst these the role of thermal conduction, turbulence, the nature of cloud growth in a fully galactic/cosmological setting, the role of cloud pulsations, a better understanding of numerical convergence, and cloud growth during infall. We plan to tackle these issues in future work.

\section*{Acknowledgments}
We thank C. Cicone, D. Fielding, Y.-F. Jiang, Y. Li, C. Liang, P. Sharma, E. Schneider, B. Tan, R. Wissing, and Y. Zheng for interesting discussions.
We also wish to thank Suoqing Ji for providing the turbulent initial conditions for \S\ref{sec:nosphere}.
This research made use of \texttt{yt} \citep{2011ApJS..192....9T}, \texttt{matplotlib} \citep{Hunter:2007}, \texttt{numpy} \citep{van2011numpy}, and \texttt{scipy} \citep{jones_scipy_2001} whose communities we also thank for development \& support.
We thank the referee Evan Schneider for a very  thorough and close reading of the paper, which prodded us to clarify a number of points. We acknowledge support from NASA grant NNX17AK58G, HST theory grant HST-AR-15039.003-A, XSEDE grant TG-AST180036, and the Texas Advanced Computing Center (TACC) of the University of Texas at Austin.
MG was supported by NASA through the NASA Hubble
Fellowship grant \#HST-HF2-51409 awarded by the Space Telescope Science
Institute, which is operated by the Association of Universities for
Research in Astronomy, Inc., for NASA, under contract NAS5-26555. A portion of this work was performed at the Aspen Center for Physics, which is supported by National Science Foundation grant PHY-1607611. 

\appendix

\bibliographystyle{mnras}
\bibliography{refs}

\appendix

\section{Near static setups}
\label{sec:static_setups} 

\begin{figure}
  \centering
  \includegraphics[width=\linewidth]{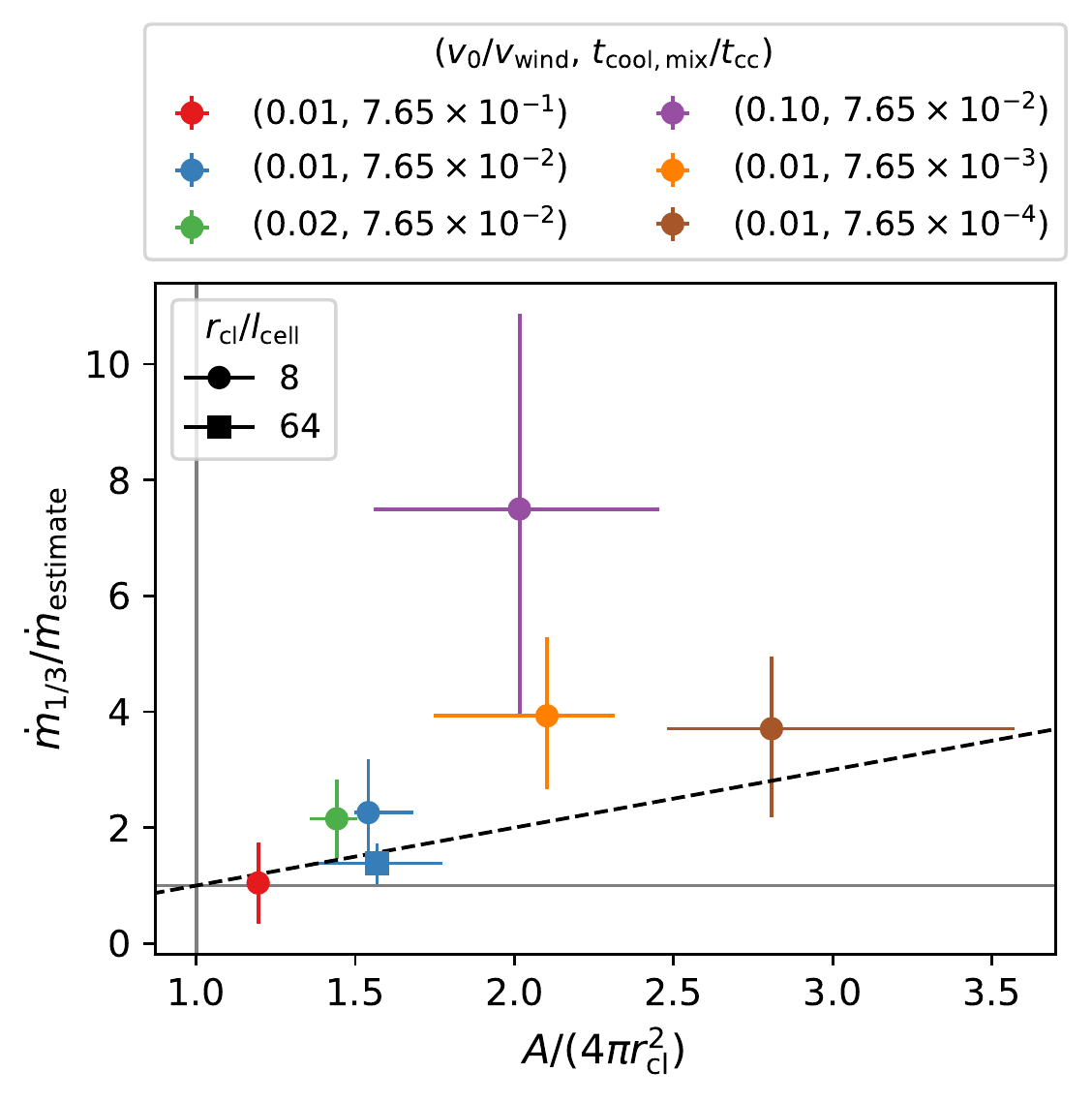}
  \caption{Time averaged mass growth rate versus the measured area for a `near-static' setup. Unlike the case of a wind, a tail does not develop and the surface area is of order the initial value, $A = 4 \pi r_{\rm cl}^{2}$. Circles and error bars symbolizes the $50$th percentile, and the differences to the $16$th and $84$th percentiles, respectively. Note that we normalized the growth rate by the simple estimate $\dot m_{\mathrm{estimate}}\sim r_\cl^2 \rho_{\mathrm{wind}} c_{\mathrm{s,cl}} (\tcoolrat{mix})^{-1/4}$. The black dashed line shows the identity function, i.e., if the mass growth scaled linearly with $A$. We varied the initial wind speed $v_0$, and the cloud radius (expressed through \tcoolrat{mix}).}
  \label{fig:mass_growth_little_wind}
\end{figure}

\begin{figure}
  \centering
  \includegraphics[width=\linewidth]{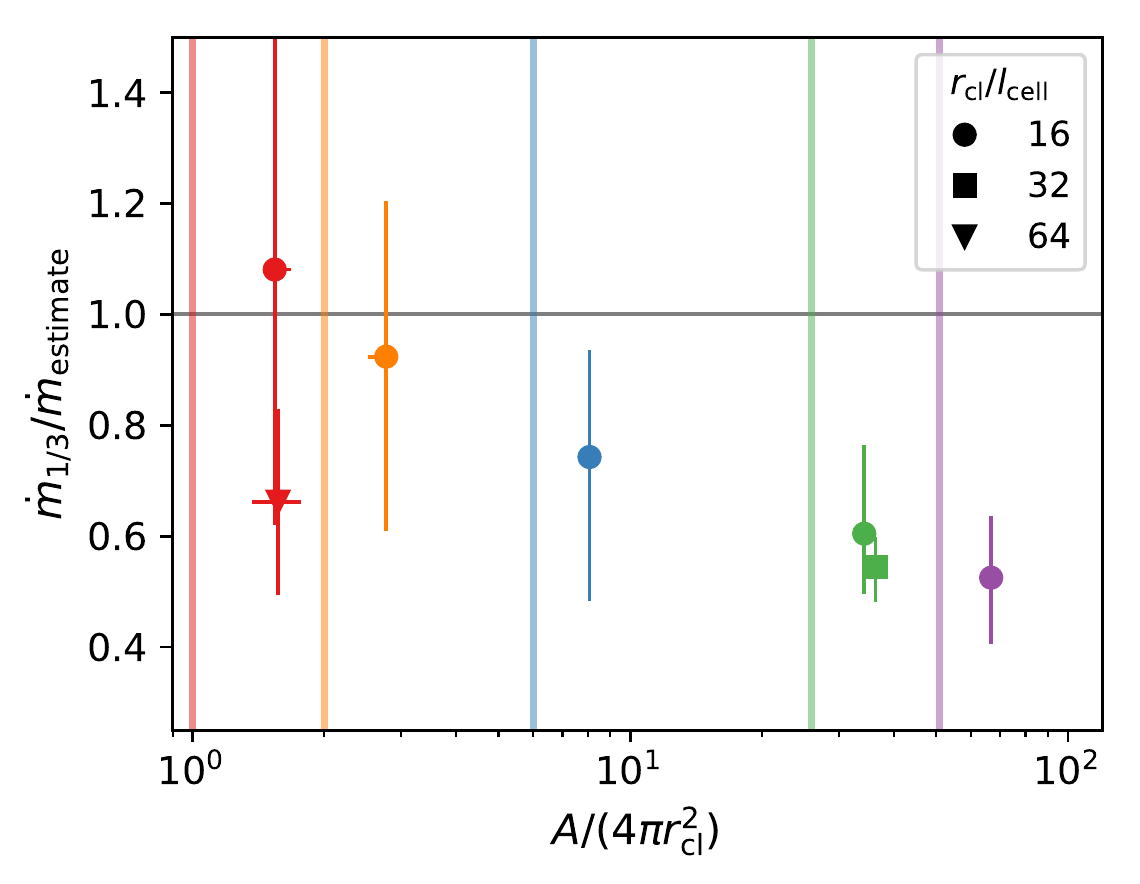}
  \caption{Time averaged mass growth rate versus the measured area for a `near-static' setup with $v_0=0.01 v_{\mathrm{wind}}$, $r_\cl$ corresponding to $\tcoolrat{mix}\sim 0.08$, and elongated clouds along the wind direction. Circles and error bars symbolizes the $50$th percentile, and the differences to the $16$th and $84$th percentiles, respectively. Note that we normalized the growth rate by the simple estimate $\dot m_{\mathrm{estimate}}\sim A_\cl \rho_{\mathrm{wind}} c_{\mathrm{s,cl}}$. The vertical lines in the corresponding colors represent the clouds' initial surface areas.}
  \label{fig:mass_growth_ellipsoids}
\end{figure}

A potential issue is that while the mass growth presented in the upper right panel of Fig.~\ref{fig:mass_growth_new_setup_multiplot} seems to flatten for $\Delta v_{1/3} < v_{\mathrm{wind}} / 10$, we could not probe the regime of much smaller $\Delta v_{1/3}$ due to computational constraints. To overcome this, we performed `near-static' experiments where we initialized the flow to $v_0$ with $v_0 / v_{\mathrm{wind}} \ll 1$. Note that setting $v_0 = 0$ does not allow the formation of a realistic mixing layer, and thus we chose values of $v_0\sim 0.01 v_{\mathrm{wind}}$. Fig.~\ref{fig:mass_growth_little_wind} shows the mass growth rate in this near-static setup versus the cold gas surface $A$. We averaged both quantities in the time range $5 < t/\tcc < 15$ (when the mass growth is time-steady; note that this happens earlier in the near static setup, since the transient associated with entrainment and the development of the tail is not present), and present the median and $16$th and $84$th percentiles in Fig.~\ref{fig:mass_growth_d100_convergence}. Note that the cause of the large spread in $\dot m_{1/3}(t)$ even after $t>5\tcc$ is the oscillation in mass growth due to the non-dynamic initial conditions which lead to rather large but symmetric errors.

We normalize $\dot m_{1/3}$ in Fig.~\ref{fig:mass_growth_little_wind} to an analytic estimated based in the considerations in Sec.~\ref{sec:analytics} $\dot m_{\mathrm{estimate}}\sim r_{\cl}^2 \rho_{\mathrm{wind}} c_{\mathrm{s,cl}} (\tcoolrat{mix})^{-1/4}$, i.e., by setting $A\sim r_\cl^2$ and $v_{\mathrm{mix}} \sim c_{s,\cl} (\tcoolrat{mix})^{-1/4}$ in Eq.~\eqref{eq:mdot_general}. That this is an over-simplification can be seen from the mismatch in Fig.~\ref{fig:mass_growth_little_wind}. However, when correcting for area in the estimate to the measured value (indicated by the dashed, black line) the estimate improves.

Clearly, a spherical cloud placed in a hot wind with small velocity does not correspond to the later stages of our cloud crushing setup, where a tail has formed leading to a elongated cold cloud geometry. To simulate this, we place also `cigar shaped' clouds with varying length $l_{\mathrm{cl}}$ in a wind with velocity $v_0 = 0.01 v_{\mathrm{wind}}$. Fig.~\ref{fig:mass_growth_ellipsoids} shows the resulting mass growth normalized by $\dot m_{\mathrm{estimate}}\sim A_\cl \rho_{\mathrm{wind}} c_{\mathrm{s,cl}}\sim 2 \pi r_\cl (l_\cl + 2 r_\cl) \rho_{\mathrm{wind}} c_{\mathrm{s,cl}}$, and as in the previous figure averaged over $t\in [5,\,15]\,\tcc$. The vertical lines in Fig.~\ref{fig:mass_growth_ellipsoids} show $A_\cl$ and, thus, the offset denotes how much the clouds expand. At fixed $t_{\rm cool,mix}/t_{\rm cc} \sim 0.08$, our ansatz is very stable to changes in the surface area: over a factor of $\sim 50$ in surface area, $\dot{m}_{1/3}/\dot{m}_{\rm estimate}$ varies only by a factor of 2.\\

\bsp	
\label{lastpage}

\end{document}